\documentclass[11pt,epsfig]{article} 
\usepackage[utf8]{inputenc}
\usepackage[top=1in, left=0.95in, bottom=1.1in, right=0.95in]{geometry}
\usepackage[colorlinks=true,
linkcolor=blue,
urlcolor=red,
citecolor=red]{hyperref}
\usepackage[vcentermath,enableskew]{youngtab}
\usepackage{ytableau}
\usepackage{tikz}
\usepackage{young}
\usepackage{tabularx}
\usepackage{diagbox}
\usepackage[toc]{appendix}
\usepackage{longtable}
\usepackage{enumerate}
\usepackage{float}
\usepackage{subfig}
\usepackage{xcolor}

\usepackage{cite}

\let\counterwithin\relax
\usepackage{chngcntr}
\usepackage{amssymb, amsmath,mathrsfs}
\usepackage{multicol,multirow}

\usepackage{graphics}
\usepackage{graphicx}
\usepackage{epsf}
\usepackage{epsfig}
\usepackage{float}
\usepackage{makecell}

\usepackage{color}
\usepackage{xcolor}
\usepackage{simplewick}
\usepackage{amsmath}
\usepackage{amsfonts}
\usepackage{makeidx} 
\usepackage[section]{placeins}
\newcommand{\bea}{\begin{eqnarray}}
\newcommand{\eea}{\end{eqnarray}}

\def\baselinestretch{1.16}

\usepackage[normalem]{ulem} 

\begin{document}
\begin{center}



{\Large \textbf  {Constructive Heavy Particle Effective Theory with Nonlinear Poincar\'{e} Symmetry}}\\[10mm]

Yong-Kang Li$^{a, b}$\footnote{liyongkang@itp.ac.cn}, Yi-Ning Wang$^{a, b}$\footnote{wangyining@itp.ac.cn}, Jiang-Hao Yu$^{a, b, c, d}$\footnote{jhyu@itp.ac.cn}\\[10mm]

\noindent 
$^a${\em \small CAS Key Laboratory of Theoretical Physics, Institute of Theoretical Physics, Chinese Academy of Sciences,    \\ Beijing 100190, P. R. China}  \\
$^b${\em \small School of Physical Sciences, University of Chinese Academy of Sciences,   Beijing 100049, P.R. China}   \\
$^c${\em \small School of Fundamental Physics and Mathematical Sciences, Hangzhou Institute for Advanced Study, UCAS, Hangzhou 310024, China} \\
$^d${\em \small International Centre for Theoretical Physics Asia-Pacific, Beijing/Hangzhou, China}\\[10mm]

\date{\today}   
          
\end{center}

\begin{abstract}

We develop a constructive heavy particle effective theory (HPET) through the nonlinear realization of the  spontaneously broken Poincar\'{e} symmetry $R^{3,1} \rtimes SO(3,1) \rightarrow R^{3,1} \rtimes SO(3)$. 
Starting from the heavy one-particle state, we find the nonlinear boost transformation indicates the shift symmetry in the coset construction, corresponding to the reparameterization invariance. 
Using the little group Wigner rotation, we obtain the nonlinear boost transformation for corresponding heavy field, recovering the Foldy-Wouthuysen transformation. 
At the operator level, since interaction terms would modify the nonlinear transformation, we propose a most general parametrization on the boost transformation only based on symmetry. The nonlinear boost transformation relates different Wilson coefficients of the HPET operators, providing a bottom-up approach of constructing the independent HPET operators, and generalizing the top-down HPET operators beyond the tree-level integrating out. Utilizing the HPET as example, we obtain additional constraints for the boost transformation as well as the additional variation $\delta \mathcal{L}$ at the $1/m^3$.




\end{abstract}

\newpage
\setcounter{tocdepth}{3}
\setcounter{secnumdepth}{3}

\tableofcontents

\setcounter{footnote}{0}

\def\baselinestretch{1.5}
\counterwithin{equation}{section}

\newpage
\section{Introduction}

It is well known that spacetime symmetry plays an essential role in physics. According to Weinberg's statement~\cite{Weinberg:1996kw}, any quantum theory satisfying Lorentz invariance and cluster decomposition would look like a quantum field theory at sufficiently low energy. In his textbook~\cite{Weinberg:1995mt}, a constructive quantum field theory is established by systematically constructing particle states, quantum fields, and Lagrangian under the Poincar\'{e} symmetry. The relativistic one-particle states are defined as the unitary irreducible representations (irreps.) of the Poincar\'{e} group $R^{3,1}\rtimes SO(3,1)$, transforming under the Wigner's little group rotation through the induced representations. On the other hand,  relativistic quantum fields are classified as finite-dimensional irreducible representations of the homogeneous Lorentz group $SO(3,1)$, connecting to the one-particle state via the Poincar\'{e}-Lorentz relation of the wave function. Finally, after converting the little group indices to the Lorentz indices, the Lorentz invariant Lagrangian can be built from the local fields only. This procedure constitutes a constructive framework for the quantum field theory.

The foundational role of the spacetime symmetry is further exemplified in a range of well-established effective field theories. Notable instances include the familiar Heavy Quark Effective Theory (HQET)~\cite{Caswell:1985ui,Isgur:1989vq,Eichten:1989zv,Georgi:1990um,Grinstein:1990mj,Luke:1992cs,Eichten:1990vp,Manohar:1997qy,Manohar:2000dt}, 
the Non-relativistic Quantum Chromodynamics  (NRQCD)~\cite{Caswell:1985ui,Bodwin:1994jh,Manohar:1997qy,Manohar:2000dt,Brambilla:2003nt},
the Non-relativistic Quantum Electrodynamics (NRQED)~\cite{Caswell:1985ui,Ragusa:1993rm,Kinoshita:1995mt,Hill:2011wy,Hill:2012rh},
nucleon contact interactions~\cite{Weinberg:1990rz,Weinberg:1991um,Weinberg:1992yk,Ordonez:1995rz,Kaplan:1998we,vanKolck:1999mw,Bedaque:2002mn,Entem:2003ft,Epelbaum:2004fk,Epelbaum:2005pn,Epelbaum:2008ga,Girlanda:2010ya, Girlanda:2011fh, Xiao:2018jot}, effective theories for Dark Matter~\cite{Fan:2010gt,Hill:2011be,Fitzpatrick:2012ix,Cirelli:2013ufw,Bishara:2016hek,Bishara:2017pfq,Roszkowski:2017nbc,DelNobile:2021wmp}, and the more recent Heavy Black Hole Effective Theory~\cite{Bern:2019nnu,Bern:2019crd,Damgaard:2019lfh}. These NR heavy systems have the $SO(3)$ as their manifest (homogeneous) symmetry group.
Conventionally, HQET is derived from the Quantum Chromodynamics (QCD) following the top-down approach, describing the behavior of a hadron with a single heavy quark (the bottom quark $b$ or the charm quark $c$), where the momentum of the heavy quark is $p^{\mu}=mv^{\mu}+k^{\mu}$. Since the heavy quark mass $m \gg \Lambda_{\text{QCD}}$, these two energy scales are separated. On the one hand, this heavy system, characterized by the heavy quark velocity $v^{\mu}$ (with $v^2=1$), remains sufficiently stable and does not decay into light relativistic particles. The majority of the energy is consistently contained within the heavy quark mass in the rest frame or the label momentum $mv^{\mu}$ in a general frame.
On the other hand, the light components such as the sea quarks and gluons (often referred to as ``brown muck'') carry small momentum on the order of $ \Lambda_{\text{QCD}}$. Minor fluctuations $k^{\mu}$ arising from the soft QCD interactions are also of the same order. 
Thus, dividing the relativistic fields into the heavy and light modes according to $v^{\mu}$, the effective theory is obtained by integrating out the antiparticle degrees of freedom (d.o.f.)~\cite{Eichten:1989zv,Bodwin:1994jh,Hill:2013hoa,Damgaard:2019lfh,Fickinger:2016rfd,Kobach:2018pie}, with the power expansion according to $k^{\mu}/m$.
As for the HPET which we discuss in this work, while sharing structural similarities with the HQET, it exhibits a distinct gauge symmetry structure: the HPET incorporates a U(1) gauge group corresponding to the electromagnetic interactions, whereas the HQET is governed by the SU(3) non-Abelian gauge symmetry inherent to the QCD.
This fundamental difference leads to their respective applicability – the HPET describes the systems dominated by long-range electromagnetic forces, while the HQET specifically addresses the heavy quark dynamics under the strong interactions.
While the conventional integrating out approach in the HQET and HPET maintains manifest relativistic covariance and reveals the relations among the Wilson coefficients, it suffers from ultraviolet (UV) dependence – different UV completions (e.g., QCD or low-energy effective theories) yield distinct coefficients, with discrepancies also arising between the tree-level and loop-level calculations. Furthermore, a non-relativistic (NR) operator basis is required when implementing this matching procedure.

For the heavy and NR particle, the Lagrangian is usually formulated by the NR field transforming under the $R^{3,1}\rtimes SO(3)$ symmetry. Given that the NR field is almost defined as the Fourier transformation of the particle d.o.f., with the anti-particle d.o.f. integrated out, a hidden Poincar\'{e} symmetry should persist even if it is formulated in the NR formalism.
From the bottom-up view, this hidden symmetry can be viewed as the spontaneous symmetry breaking (SSB) of the Poincar\'{e} group, which is nonlinearly realized by the boost. With the emergence of the shift symmetry, this also manifests as the reparameterization invariance (RPI) in the NR formalism~\cite{Luke:1992cs, Chen:1993np, Finkemeier:1997re}.
Based on the symmetry breaking pattern, a Callan-Coleman-Wess-Zumino (CCWZ) coset construction~\cite{PhysRev.177.2239,PhysRev.177.2247} can be built and leads to an effective field theory description.
As for spacetime symmetry, various symmetry breaking patterns are widely discussed and classified in different areas, such as Refs.~\cite{Low:2001bw, Goon:2012dy, Nicolis:2013lma, Watanabe:2013iia, Nicolis:2015sra, Watanabe:2014fva,Beekman:2019pmi,Hidaka:2012ym,Brauner:2014aha,Watanabe:2019xul,Naegels:2021ivf}.
%
Specifically, considering the SSB pattern with the broken boost symmetry, for sufficiently low-energy physical processes involving a single heavy particle, the description of the system typically becomes frame-dependent, and is usually confined to a certain frame such as the rest frame, with suppressed particle-antiparticle pair production due to the huge mass gap constraints.
In this case, the particle number conservation emerges with the physical ground state $|\Omega\rangle$ coinciding with a scalar single-particle state carrying momentum. This ground state has a non-vanishing vacuum expectation value
(VEV) $\langle \Omega|P^{\mu}|\Omega\rangle=mv^{\mu}$ that is proportional to the heavy mass $m$, whereas the timelike vector $v^{\mu}$ represents either the particle's four-velocity or the preferred frame itself.
The spacetime symmetry group consequently undergoes spontaneous breaking from the Poincar\'{e} group $R^{3,1} \rtimes SO(3,1)$ to its subgroup $R^{3,1}\rtimes SO(3)$, with the surviving $SO(3)$ corresponding to the little group preserving $v^{\mu}$. The broken boosts become nonlinearly realized, equivalent to a shift $v^{\mu}\longrightarrow v^{\mu}+\delta v^{\mu}$.

This work adopts a bottom-up approach, analogous to Weinberg's paradigm, applied to the systems with spontaneously broken Poincar\'{e} symmetry. Using HPET as a concrete example, we develop the framework from the heavy one-particle state construction, to the heavy field, and finally to Lagrangian formulation.
\begin{itemize}
\item For the heavy one-particle state $|v,\vec k,\sigma\rangle$, the transformation under the boost $B(q)$ defined in Eq.~\eqref{eq:Bqdef} is
\begin{equation}
U(B(q))|v,\vec k,\sigma\rangle=\sum_{\sigma^{\prime}}D^{(s)}_{\sigma^{\prime}\sigma}(W[B(q),p])|v,\vec k^{\prime},\sigma^{\prime}\rangle,
\end{equation}
where $W[B(q),p]\equiv U(L(\vec k^{\prime}))^{-1}U(B(q))U(L(\vec k))$ is the little group transformation of $v^{\mu}$ and given by Eq.~(\ref{Dsmatrix}), and $L(\vec k)$ is the standard Lorentz boost. From the symmetry perspective, the coset construction not only naturally gives this boost transformation, but also incorporates the shift symmetries connecting different momentum $|k|$ states, which precisely corresponds to the RPI.

\item The one-particle state defines a two-component heavy field $N_{I}(x)$. Here $I$ is the little group index of $v^{\mu}$. Under the boost transformation $B(q)$, the heavy field transforms nonlinearly
\begin{equation}
     U(B(q))^{-1} N(x)U(B(q))=e^{i\vec q\cdot\vec x}\left(1+\frac{\vec q}{m}\cdot\vec {\mathcal{K}}\right)N(x')=\left(1+\frac{\vec q}{m}\cdot\vec {\mathcal{K}}_x\right)N(x'),
\end{equation}
such that the corresponding  the boost generator of the heavy field $N(x)$ is
\begin{equation}\label{eq:fnabla-intr}
 \vec {\mathcal{K}}=-i\frac{\vec\sigma}{2}f(\vec\nabla),\quad\text{where}\quad f(\vec\nabla)=\frac{-i\vec\sigma\cdot\vec\nabla}{2m+i\partial_t}.
\end{equation} 
where $x'=B(q)^{-1}x$.
Identifying the contribution from the anti-particle, we build the connection between the heavy field and the Dirac field
\begin{equation}
    \Psi(x)=e^{-imt}\left[\begin{array}{c}
          N(x)  \\
         \tilde N_I(x) 
    \end{array}\right] = e^{-imt}\left[\begin{array}{c}
          N(x)  \\
         f(\vec\nabla)  N(x) 
    \end{array}\right].
\end{equation}
where $\tilde N_I(x)$ transforms under the boost as
\begin{equation}
    U(B(q))^{-1}\tilde N(x)U(B(q))=e^{i\vec q\cdot\vec x}\left(\frac{\vec \sigma\cdot\vec q}{2m}+f(\vec\nabla)\right)N(x').
\end{equation}
This identity gives the consistency condition Eq.~\eqref{eq:zetacc1}. 
Accordingly we obtain that the Dirac field transforms linearly under the boost transformation.

\item Given the heavy field under the $SO(3)$ symmetry, it is ready to construct the most general operators in HPET. Different from other effective theory, the Wilson coefficients of these HPET operators are not independent due to the underlying Lorentz symmetry. These relations can be obtained from the boost transformation of the heavy field. To derive relationships among the Wilson coefficients, there are two methods: 
\begin{enumerate}

    \item Bottom-up approach: The variation of the NR Lagrangian $\delta\mathcal{L}$ under the Lorentz boost is investigated. Utilizing the invariance $\delta\mathcal{L}=0$, relations among the Wilson coefficients are obtained. Specifically, the nonlinear boost transformation of the heavy field in the interaction case is needed:
\begin{equation}
U(B(q))^{-1} N(x)U(B(q))=e^{i\vec{q}\cdot\vec{x}}\left(1+\frac{\vec{q}}{m}\cdot\vec {\mathcal{K}}\right)N(x'),\quad\text{where}\quad \vec {\mathcal{K}}=-i\frac{\vec\sigma}{2}f(\frac{D_{\mu}}{m}),
\end{equation}
while this generally constructed $f(\frac{D_{\mu}}{m})$ is constrained by the consistency condition:
\begin{equation}\label{eq:introcc}
f(\frac{B(q)_{\mu}^{\nu} D_{\nu}}{m})e^{i\vec q\cdot\vec x}\left(1+\frac{\vec\sigma\cdot\vec q}{2m}f(\frac{D_{\mu}}{m})\right)N(x')=e^{i\vec q\cdot\vec x}\left(\frac{\vec \sigma\cdot\vec q}{2m}+f(\frac{D_{\mu}}{m})\right)N(x').
\end{equation}

\item Top-down approach: A relativistic UV completion is matched to the NR Lagrangian, such that all Wilson coefficients are determined directly by the Lorentz symmetry. The anti-particle component are parameterized as
\begin{equation}
    \tilde N(x)=f(\frac{D_{\mu}}{m})N(x),
\end{equation}
where $f(\frac{D_{\mu}}{m})$  again satisfies the consistency condition Eq.~\eqref{eq:introcc}. 
In this approach, the Dirac field is expanded as 
\begin{equation}
    \Psi(x)=e^{-imt}\left[\begin{array}{c}
          N(x)  \\
         f(\frac{D_{\mu}}{m})  N(x) 
    \end{array}\right],
\end{equation}
in the UV Lagrangian, and then the matching results can be obtained.

\end{enumerate}


\end{itemize}

Having setup a systematic conceptual framework, we realize the practical difference caused by different definition of the $f(D^\mu)$ from the past works. 
For the free field, the Eq.~\eqref{eq:fnabla-intr} gives
\begin{equation}
    f(\vec\nabla)=\frac{-i\vec\sigma\cdot\vec \nabla}{2m+i\partial_t}=\left(-\frac{i\vec\sigma\cdot\vec\nabla}{2m}+\frac{i\vec\sigma\cdot\vec\nabla i\partial_t}{4m^2}+...\right),
\end{equation}
while in the gauge interaction, the naive replacement in Ref.~\cite{Heinonen:2012km} is  used 
\begin{equation}\label{eq:fdnaive-intro}
    f(D)=\left(-\frac{i\vec{\sigma}\cdot\vec{D}}{2m}+\frac{i\vec\sigma\cdot\vec DiD_t}{4m^2}+...\right),
\end{equation}
while the tree-level integrating out Quantum Electrodynamics (QED) gives
\begin{equation}\label{eq:treefd-intro}
    f(D)=\left(-\frac{i\vec{\sigma}\cdot\vec{D}}{2m}-\frac{\{D_t,\vec{\sigma}\cdot{\vec{D}}\}}{8m^2}-\frac{ig\vec{\sigma}\cdot\vec{E}}{8m^2}+...\right).
\end{equation}
However, in this work, a generally constructed  $f(\frac{D_{\mu}}{m})$ constrained by the consistency condition Eq.~\eqref{eq:introcc} is utilized
\begin{equation}\label{eq:fd-intro}
   f(\frac{D_{\mu}}{m})= -\frac{i\vec{\sigma}\cdot\vec{D}}{2m}+c^{(2)}_A\frac{\{D_t,\vec{\sigma}\cdot{\vec{D}}\}}{8m^2}+c^{(2)}_B\frac{ig\vec{\sigma}\cdot\vec{E}}{8m^2}+\mathcal{O}(\frac{1}{m^3}),
\end{equation}
satisfying the consistency condition Eq.~\eqref{eq:introcc}. 
To confront these challenges when gauge fields exist, the covariant boost for the heavy field is proposed with the consistency conditions of $f(\frac{D_{\mu}}{m})$ according to the Lorentz symmetry. Specifically,
the NR heavy field and other building blocks combine as an expansion of the relativistic field, which is rigorously constrained by both discrete spacetime symmetries (space inversion $P$ and time reversal $T$) and the linear transformation properties of the relativistic field under the continuous spacetime symmetries (the rotational and Lorentz boost transformations).
Utilizing this expansion, both the bottom-up and top-down approaches can be implemented and those problems mentioned above are addressed, demonstrated through application to the HPET with the relations among the Wilson coefficients derived.
The following results are obtained
\begin{itemize}

\item In the bottom-up approach, the   little group transformations encounters obstructions when determining the gauge field-dependent terms in the nonlinear  boost~\cite{Heinonen:2012km,Hill:2012rh}, where the derivatives $\partial_{\mu}$ are naively replaced  with $D_{\mu}$, as in Eq.~\eqref{eq:fdnaive-intro}. In this work, we use the boost transformation with the gauge field-dependent term properly determined by the consistency condition as in Eq.~\eqref{eq:fd-intro}.

\item In the bottom-up approach, the general parameterized boost generator $\vec{\mathcal{K}}_x=\vec{\mathcal{K}}+m\vec x$ constrained by the boost commutator is discussed in Refs.~\cite{Brambilla:2003nt,Berwein:2018fos}. In this work, we construct the $f(\frac{D_{\mu}}{m})$ instead, and the corresponding consistency condition is equivalent to the Lorentz boost commutator as shown in subsection~\ref{bottomupcompare}, with fewer free parameters we use.

\item In the top-down approach, the tree-level integrating out of the    relativistic Lagrangian is discussed in Ref.~\cite{Manohar:1997qy}, as in Eq.~\eqref{eq:treefd-intro}. In this work, we start with the most general relativistic effective  Lagrangian to incorporate the possible UV completion information. Using the general expansion with $f(\frac{D_{\mu}}{m})$ in Eq.~(\ref{gexp}) and \eqref{fform2}, the matching results extend beyond  the tree-level.

\end{itemize}

This paper is organized as follows.
In section \ref{sec:general ex}, from the perspective of SSB, we construct the one-particle states of the HPET as well as its transformation, and use the coset construction to derive the reparameterization of the label velocity. In section \ref{sec4}, we derive the heavy field and its boost transformation based on the corresponding particle state and the wave function, extending to the covariant form in the gauge interaction. In section \ref{sec:b-HQET} and \ref{ahqet}, several bottom-up and top-down methods to constrain the Wilson coefficients in the HPET from the underlying Lorentz symmetry are discussed, respectively.
In appendix \ref{ap1}, the calculation details of the boost generator for the little group transformations are provided. In appendix \ref{ap2}, the Goldstone theorem is briefly reviewed. In appendix \ref{app3}, the detailed  boost transformation of the heavy field as well as the commutator of the boost generator are discussed.
\section{Heavy Particle State}\label{sec:general ex}
In the relativistic quantum field theory, fields and particle states represent two essential but distinct frameworks for describing physical systems. Usually the fields serve as the foundational building blocks for constructing the relativistic Lagrangians, while the particle states are indispensable when describing the S-matrix by analyzing the scattering amplitudes. Transformation properties of these two descriptions under the spacetime symmetry group differ. A quantum field $\phi_{\alpha}(x)$ carrying the Lorentz indices $\{\alpha\}$ transforms under a finite-dimensional linear representation of the Lorentz group
\begin{equation}\label{relativisticphilambda}
   \phi^{\prime}_{\alpha}(x^{\prime})\equiv U(\Lambda)^{-1}\phi_{\alpha}(x)U(\Lambda) = D_{\alpha}^{\beta}(\Lambda) \phi_{\beta}(\Lambda^{-1}x),
\end{equation}
   where $D(\Lambda)$ is a non-unitary representation matrix and $\Lambda$ denotes a Lorentz transformation. 
On the other hand, one-particle states $|p,\sigma\rangle$ carrying the momentum $p^\mu$ and the little group indices $\{\sigma\}$ (e.g., spin/helicity) transform under the infinite-dimensional unitary irreducible representations of the Poincar\'{e} group. This transformation is implemented through the creation operator $a^\dagger_{p,\sigma}$ as 

\begin{equation}
     U(\Lambda,b) a^\dagger_{p,\sigma} U^{-1}(\Lambda,b) = e^{i\Lambda p\cdot b}\sum_{\sigma'} D^{(s)}_{\sigma'\sigma}(W(\Lambda,p)) a^\dagger_{\Lambda p,\sigma'},
\end{equation}
where the little group transformation $W(\Lambda,p)$~\cite{Wigner:1939cj} keeps the momentum $p^\mu$ invariant. Both transformations are clear in the relativistic quantum field theory and they are related by the wave function. 
For the constructive quantum field theory (QFT), however, the one-particle states are the starting point. In constructing the HPET which is a NR theory after the spacetime SSB, we utilize the NR one-particle states, and the realization of the Lorentz symmetry is nonlinear. 


For a quantum system with given symmetry group, the Hilbert space is spanned by the one-particle states. The Hilbert space and particle states of the usual relativistic theory and HPET are discussed in subsection \ref{relativistichilbert} and \ref{hilbertspace}, respectively.  
Utilizing the coset description, the boost transformation of the states as well as the equivalence of the reparameterization invariance and boost are shown in subsection \ref{coset}. Besides, the absence of Goldstone Boson in this boost breaking single-particle system is discussed in subsection \ref{missingGB}.

\subsection{Relativistic One-Particle State}
\label{relativistichilbert}

Before we discuss the non-relativistic theory, let us briefly review the relativistic ones, which refer to those that have the Poincar\'{e} group as the spacetime symmetry. The generators of the Poincar\'{e} group $G= R^{3,1}\rtimes SO(3,1)$ are
\begin{equation}
\left\{
\begin{array}{ll}
     \hat P^{\mu}, &\text{Translations}, \\
     \\
         \hat J^i,&\text{Rotations}, \\
         \\
       \hat K^i,&\text{Boosts},
\end{array}
  \right. 
\end{equation}
where $\hat J^i=\frac{1}{2}\epsilon^{ijk}\hat J^{jk}, \hat K^i=\hat J^{i0} $. The translation generator is $\hat P^{\mu}=(\hat P^0, \hat P^i)$, where $\hat P^0$ and $\hat P^i$ are the time and spatial translation generators, respectively. Note that we use the metric tensor with the signature $(+,-,-,-)$ throughout this work. These generators define the Lie algebra 
\begin{equation}
    \begin{array}{lll}
      \lbrack\hat P^{\mu},\hat P^{\nu}\rbrack&=&0,\\
      \\
       \lbrack\hat J^{\alpha\beta},\hat P^{\mu}\rbrack&=&-i\left(g^{\mu\alpha}\hat P^{\beta}-g^{\mu\beta}\hat P^{\alpha}\right),\\
       \\
       \lbrack\hat J^{\alpha\beta},\hat J^{\mu\nu}\rbrack&=&-i\left((g^{\alpha\mu}\hat J^{\beta\nu}-g^{\beta\mu}\hat J^{\alpha\nu})-(g^{\alpha\nu}\hat J^{\beta\mu}-g^{\beta\nu}\hat J^{\alpha\mu})\right),
    \end{array}
\end{equation}
or equivalently
\begin{equation}
\begin{array}{llllllll}
     \lbrack\hat J^i,\hat J^j\rbrack&=&i\epsilon^{ijk}J^k,\quad \lbrack\hat J^i,\hat K^j\rbrack&=&i\epsilon^{ijk}K^k,\quad &\lbrack\hat K^i,\hat K^j\rbrack&=&-i\epsilon^{ijk}J^k,\\
     \\  
           \lbrack\hat J^i,\hat P^j\rbrack&=&i\epsilon^{ijk}P^k,\quad  \lbrack\hat K^i,\hat P^0\rbrack&=&iP^i,\quad&  \lbrack\hat K^i,\hat P^j\rbrack&=&iP^0\delta^{ij},
            
\end{array}    
\end{equation}
while other commutators vanish. An element in the Poincar\'{e} group $g \in G$ can be written as $(a,\Lambda)$ where $a \in R^{3,1}$ and $\Lambda \in SO(3,1)$. Note that, we can rewrite the generators of the rotation and boost as
\begin{equation}\label{eq:twosu2lrgen}
\begin{array}{lll}
      \hat A^i\equiv(\hat J^i+i\hat K^i),\quad  \hat B^i\equiv(\hat J^i-i\hat K^i),
\end{array}
\end{equation}
such that they satisfies the Lie algebra of the $SU(2)_L\times SU(2)_R$ group
\begin{equation}
         \lbrack\hat A^i, \hat A^j\rbrack=i\epsilon^{ijk}\hat A^k, \quad
         \lbrack\hat B^i, \hat B^j\rbrack=i\epsilon^{ijk}\hat B^k,  \quad
         \lbrack\hat A^i, \hat B^j\rbrack=0.  
\end{equation}

The basic principle of the quantum mechanics tells us the physical states are those state vectors $|\Phi\rangle$ that span the Hilbert space $\mathcal{H}$. These states satisfy the interpretations of probability, and for a symmetry operator $U(g)$ we have
\begin{equation}
    \langle\Psi|\Phi\rangle=\langle\Psi|U^{\dagger}(g)U(g)|\Phi\rangle.
\end{equation}
As a result, the symmetry operator $U(g)$ that keeps vacuum invariant $U(g)|0\rangle=|0\rangle$ should also be unitary.
With regard to the $U(g)$, a unitary irreducible representation (UIR) of the symmetry group $G$ in the Hilbert space $\mathcal{H}$ spanned by $|\Phi\rangle$ can be obtained.
As for non-compact group like the Poincar\'{e} group, the UIR is infinite-dimensional, using the induced representation.

Firstly, the state is labeled by eigenvalues. The first Casimir operator is $\hat P^{\mu}\hat  P_{\mu}$, whose eigenvalue is mass $m$ of the particle. The second Casimir operator $\hat W^{\mu}\hat W_{\mu}$ gives eigenvalue $-m^2s(s+1)$ and $s$ is utilized as another label, where $\hat W^{\mu}=-\frac{1}{2}\epsilon^{\mu\nu\rho\sigma}\hat J_{\nu\rho}\hat P_{\sigma}$ is the Pauli-Lubanski vector.
The state is further specified by the eigenvalues $p^{\mu}$ of $\hat P^{\mu}$, where $\hat P^{\mu}$ forms an Abelian invariant subgroup. The remaining quantum number is denoted as $\xi$.
Therefore, we write a state in the Hilbert space as $|m,s;p,\xi\rangle$. To determine the unspecified label $\xi$, notice that on the same orbit $p^2=m^2$, different points $p^{\mu}$ will have the little Hilbert space $\mathcal{H}_{p}$ that is isomorphic to each other, i.e., $\mathcal{H}_{p}\cong \mathcal{H}_{\Lambda p}$, and this degree of freedom belongs to the little group that keeps the momentum $p^{\mu}$ invariant.

Secondly, the one-particle state is classified by the orbit. %
Applying the Lorentz transformation, all the four-momentum $p^{\mu}$ with the same $m$ could be obtained, known as an orbit. The $m=0$ is the light-like orbit and the $m>0$ is the time-like orbit, as we know that massless and massive particles are different UIRs. 
\begin{itemize}
    \item For the time-like orbit, $m>0$, a standard momentum $k^{\mu}=(m,0,0,0)$ is chosen, and the little group is of course the $SO(3)$. A standard Lorentz transformation $L(p)$ is demanded such that arbitrary $|p,\sigma\rangle$ is obtained by 
    \begin{equation}
        |p,\sigma\rangle=U(L(p))|k,\sigma\rangle,
    \end{equation} where $\sigma=-s,...,s$, which is the little group index and remains the same under the standard Lorentz boost. The UIR of the Poincar\'{e} group is induced by the $2s+1$-dimensional UIR of this $SO(3)$ rotation group. Under translation $a$ and Lorentz transformation $\Lambda$, it reads 
    \begin{equation}
        \begin{array}{lll}
            U(a,\Lambda)|p,\sigma\rangle&=&U(L(\Lambda p))U(L(\Lambda p))^{-1}U(a,\Lambda)U(L(p))|k,\sigma\rangle\\
            \\
&=&e^{i\Lambda pa}U(L(\Lambda p))\sum_{\sigma^{\prime}}D_{\sigma^{\prime}\sigma}^{(s)}(W[\Lambda,p])|k,\sigma^{\prime}\rangle
\\
            \\
            &=&e^{i\Lambda pa}\sum_{\sigma^{\prime}}D_{\sigma^{\prime}\sigma}^{(s)}(W[\Lambda,p])|\Lambda p,\sigma^{\prime}\rangle,
        \end{array} 
    \end{equation}
where $W[\Lambda,p]=L(\Lambda p)^{-1}\Lambda L(p)$ is the Wigner rotation for the little group, and the rotation matrix is
$D_{\sigma^{\prime}\sigma}^{(s)}(W[\Lambda,p])$ $=\langle k,\sigma^{\prime}|U(W[\Lambda,p])|k,\sigma\rangle$.

\item For the light-like orbit, $m=0$, a standard momentum could be  $k^{\mu}=(E,0,0,E)$ and the physical little group is $SO(2)$. A general helicity state is defined as \begin{equation}
    |p,\lambda\rangle=U(R_p)U(L_z(p))|k,\lambda\rangle,
\end{equation} where $\lambda$ is a fixed number corresponding to the positive or negative helicity. Related by the Parity, we have $\lambda=\pm s$. And the $R_p$ rotates the direction along z-axis into the direction along $\vec p$.  The UIR of the Poincar\'{e} group is thus induced as 
\begin{equation}
    \begin{array}{lll}
         U(a,\Lambda)|p,\lambda\rangle&=& e^{i\Lambda pa} U(R_{\Lambda p}L_z(\Lambda p)) (U(R_{\Lambda p}L_z(\Lambda p)))^{-1}U(\Lambda)U(R_pL_z(p))|k,\lambda\rangle\\
         \\
         &=&e^{i\Lambda pa}e^{-i\lambda\theta(p,\Lambda p)}|\Lambda p,\lambda\rangle,
    \end{array}
\end{equation}
where $\theta(p,\Lambda p)$ is determined by the little group element $W[\Lambda,p]$ $=$
$(R_{\Lambda p}L_z(\Lambda p))^{-1}\Lambda R_pL_z(p)$.
\end{itemize}

\subsection{Heavy One-Particle State}\label{hilbertspace}


%
%
In the HPET, a time-like four-velocity $v^{\mu}$ is introduced, obeying $v^2=1$, representing the reference frame where the kinematics of the heavy particle is NR. The momentum $p^{\mu}$ of the heavy particle is then written by
\begin{equation}\label{pequalmvk}
    p^{\mu}=mv^{\mu}+k^{\mu},
\end{equation}
where the residual momentum $k^{\mu}$ dynamically evolves and the mass of particle $m\gg | k|$. Different $v^{\mu}$ split the mass shell of a particle into separate individual small areas of size $|k|$,
whereas the residual momentum $k^{\mu}$ fluctuates in each small area. Under the decomposition Eq.~\eqref{pequalmvk}, the on-shell condition $p^2 = m^2$ is equivalent to $2mv\cdot k + k^2 = 0$.

In general, an arbitrary four-vector, such as $k^{\mu}$ is written in terms of the $v^{\mu}$ as
\begin{equation}
    k^{\mu}=(v\cdot k)v^{\mu}+k_{\perp}^{\mu},
\end{equation}
such that the perpendicular component $k_{\perp}^{\mu}$ is orthogonal to the velocity $v\cdot k_{\perp}=0$. In the rest frame $v^{\mu}=(1,0,0,0)=(1,\vec 0)$, the perpendicular component reduces to the spatial vector $k_{\perp}^{\mu}=(0,\vec k)$, while the $v\cdot k$ reduces to the time-component $(v\cdot k)v^{\mu}= (k^0,\vec 0)$.
Therefore, the perpendicular component of the residual momentum is
\begin{equation}
\begin{array}{lll}
   k_{\perp}^{\mu}&=& k^{\mu}-(v\cdot k)v^{\mu}\\
 
   &=&p^{\mu}-(v\cdot p)v^{\mu},
\end{array}
\end{equation}  
where in the second line we have used $v^2=1$ and $v\cdot k_{\perp}=0$. In the rest frame, this is $k_{\perp}^{\mu}=(0,\vec k)=(0,\vec p)$.

Although the HPET is Poincar\'{e} invariant, the ground state $|\Omega\rangle$ breaks the boost but conserves the translations and rotations. For this ground state, the momentum operator $\hat P^{\mu}$ has the non-vanishing expectation value 
\begin{equation}\label{VEV}
    \langle \Omega|\hat P^{\mu}|\Omega\rangle\equiv mv^{\mu}.
\end{equation}
However, the residual momentum operator 
\begin{equation}\label{eq:k0hat}
    \hat k^{\mu}=\hat P^{\mu}-\langle\Omega|\hat P^{\mu}|\Omega\rangle,
\end{equation}
has the vanishing VEV $\langle \Omega|\hat k^{\mu}|\Omega\rangle=0$, and its eigenvalue is small. 
Thus it is the ground state of $\hat k^{\mu}$ --- the residual Hamiltonian and residual 3-momentum.
The interpretation of this ground state $|\Omega\rangle$ is that, 
there exists a frame with velocity $v^{\mu}$ in which this free heavy particle is always at rest. This reference has no dynamical property and the only variable is the $v^{\mu}$ determined by the heavy particle. This velocity remains a constant while $|p^{\mu}|$ is proportional to $m$. 
The spacetime symmetry spontaneously breaks,
\begin{equation}
    R^{3,1}\rtimes SO(3,1) \longrightarrow R^{3,1}\rtimes SO(3),
\end{equation}
leaving the ground state $|\Omega\rangle$ invariant.

Without loss of generality, we could choose $v^{\mu}=(1,0,0,0)$. We also introduce the three spatial unit polarization vectors $(n^i)_{\mu}=\delta^i_{\mu},~ i=1,2,3$, such that $v\cdot n=0$.
The generators of the unbroken subgroup $R^{3,1}\rtimes SO(3)$ can be written out of $\hat P^{\mu}$ and $\hat J^{\mu\nu}$ in the Poincar\'{e} group as
\begin{equation}
\left\{
\begin{array}{ccll}
  \hat H_{v}&=&v\cdot \hat P-m& ,\quad\text{Time translation},\\
  \\
    \hat P_{v}^i&=&n^i\cdot \hat P&,\quad \text{Spatial translations}, \\
    \\
    \hat J_{v}^i&=&\frac{1}{2}(n^i)^{\mu}\epsilon_{\mu\nu\rho\sigma}\hat J^{\nu\rho}v^{\sigma}&,\quad\text{Rotations}.\\
\end{array}
\right.
\end{equation}
The Lie algebra of the unbroken subgroup is
\begin{equation}
    \begin{array}{lll}
         \left[\hat H_{v},  \hat P_{v}^i\right]&=&0,\\
         \\
         \left[\hat H_{v}, \hat J_{v}^i\right]&=&0, \\
         \\
    \left[\hat P_{v}^i,\hat P_{v}^j\right]&=&0,\\
    \\
     \left[\hat J_{v}^i,\hat P_{v}^j\right]&=&i\epsilon^{ijk}(\hat P_{v})^k,\\
     \\
      \left[\hat J_{v}^i,\hat J_{v}^j\right]&=&i\epsilon^{ijk}\hat J_{v}^k.
    \end{array}
\end{equation}
For the rest frame, we can identify the generator of the rotations as
\begin{equation}
    \hat J^i=\frac{1}{2}(\hat A^i+\hat B^i),
\end{equation}
where $\hat A^i$ and $\hat B^i$ are defined in Eq.~\eqref{eq:twosu2lrgen}.

Comparing to the Poincar\'{e} group, the coset generator is,
\begin{equation}
   \hat {K}_v^i=v_{\mu}(n^i)_{ \nu} \hat J^{\mu\nu},\quad\text{Boost},
\end{equation}
and in the rest frame we denote it as $\hat K^i=\hat K^i_v=\hat J^{0i}$.

\paragraph{UIR by the Induced Representation}

Each heavy system we considered has a given VEV~\eqref{VEV} with fixed four-vector $v^{\mu}$. All different $v^{\mu}$ label different systems, and the time-like orbit $v^2=1$ is separated into different individual pieces, where each Hilbert space is labeled by the corresponding $v^{\mu}$. 
Therefore, we denote the state as $|v,k^i,\xi\rangle$, 
with the eigen equation of the translation generators:

\begin{equation}
    \begin{array}{rlr}
         \hat H_v|v,k^i,\xi\rangle&=&k^0|v,k^i,\xi\rangle,  \\
         \\
         \hat{\vec{P}}_{v}^i~|v,k^i,\xi\rangle&=&\vec k~|v,k^i,\xi\rangle.
    
    \end{array}
\end{equation}
Here $\xi$ denotes the undetermined indices for the remaining subspace. In the rest frame, the eigenvalue of $\hat H_v$ is $v\cdot k=k^0$, and the eigenvalue of $\hat{\vec{P}}$ is the perpendicular component of the residual momentum $\vec k$, since $k^{\mu}_{\perp}=(0,\vec k)$. Depending on the equation of motion (EOM), $k^0$ is the function $|\vec k|$, and this EOM relates to the broken UV symmetry. For both Galilean group and Poincar\'{e} group, we have $k^0=\frac{1}{2m}\vec k^2$ at the leading order. Thus $k^0=0$ when $|\vec k|=0$.

The UIR of the unbroken subgroup $R^{3,1}\rtimes SO(3)$ can be obtained as the usual induced representation procedure according to the little group. The translation $U(a)$ acts on the states just as the relativistic case and will not be repeated. In the Poincar\'{e} group $R^{3,1}\rtimes SO(3,1)$, the little group is $SO(3)$ for massive particles, and different momentum $p^{\mu}$ on the mass shell $p^2=m^2$ are related by the Lorentz boost. However, in the broken spacetime symmetry $R^{3,1}\rtimes SO(3)$, different $|\vec k|$ belong to individual irreducible representations and are not related, as a consequence of the absence of boost. Besides, the little groups are different for $|\vec k|=0 $ and $|\vec k|>0$:
\begin{itemize}
  
    \item For $|\vec k|=0$, the little group is $SO(3)$.
    
    For the rotation $U(R)=\exp\{i\vec \theta\cdot \hat{\vec{J}}_v\}\in SO(3)$ with the rotation angle $\vec \theta$, the state transforms as  
    \begin{equation}
        U(R)|v,0,\sigma\rangle=\sum_{\sigma^{\prime}}D_{\sigma^{\prime}\sigma}^{(s)}(R)|v,0,\sigma^{\prime}\rangle,
    \end{equation}
where the little group index is the spin index $\sigma$, and $D^{(s)}_{\sigma^{\prime}\sigma}(R)\equiv \langle v,0,\sigma^{\prime}|U(R)|v,0,\sigma\rangle$ is the rotation matrix. It also gives the representation of the translations, with vanishing eigenvalue of $\hat H_v$ and $\hat{\vec{P}}_v$ due to $|\vec k|=0$.

\item For $|\vec k|>0$, the little group is $SO(2)$.  

For the subspace considered, we characterize it by a standard momentum $\vec k_0=(0,0,|\vec k|)$. The subgroup $SO(2)$, that leaves $\vec k_0$ invariant is the little group of $\vec k_0$. In this situation, the little group transformation is $U(R_3(\omega))=\exp\{-i\omega \hat{J}_v^3\}\in SO(2)$, and the particle state transforms as
\begin{equation}
    U(R_3(\omega))|v,\vec k_0,h\rangle=e^{-ih\omega}|v,\vec k_0,h\rangle.
\end{equation}
Here $h$ is the eigenvalue of $\hat J_v^3$, due to $\hat J_v^3|\vec k|=\hat{\vec{J}}_v\cdot\vec k_0$, and 
$
   \hat{\vec{J}}_v\cdot\hat{\vec {P}}_v|v,\vec k_0,h\rangle=|\vec k| h~|v,\vec k_0,h\rangle,
$
we know that $h$ is the helicity. In this orbit with the same $|\vec k|$ and $h$, other directions of momentum can be obtained by a standard rotation $U(R(\phi,\theta))=\exp\{-i\phi\hat{J}^3_v\}\exp\{-i\theta \hat{J}^2_v\}$,
    \begin{equation}
        |v,\vec{k},h\rangle= U(R(\phi,\theta))|v,\vec k_0,h\rangle,
    \end{equation}
    and thus the irreducible representation of the spacetime symmetry $SO(3)$ is induced by 
    \begin{equation}\label{irepkbig0}
\begin{array}{ll}
     U(R)|v,\vec k,h\rangle & =U(R)U(R(\phi,\theta))|v,\vec k_0,h\rangle \\
     \\
     & =U(R(\phi^{\prime},\theta^{\prime}))U(R_3(\omega))|v,\vec k_0,h\rangle\\
     \\
     &=e^{-ih\omega}|v,R\vec k,h\rangle,
\end{array}
\end{equation}
where $ R\vec k=R(\phi^{\prime},\theta^{\prime})\vec k_0$ and $R_3(\omega)=R^{-1}(\phi^{\prime},\theta^{\prime})~R~R(\phi,\theta)$.

\end{itemize}

In the UIR of $R^{1,3}\rtimes SO(3)$, each $|k|$ belongs to a different irreducible representation, which is apparently not the same as
the relativistic one-particle states $|p,\sigma\rangle$ that transform according to the UIR of Poincar\'{e} group.
The above classification of orbits has important physical interpretation and it is discussed in subsection \ref{missingGB}.

\paragraph{The breaking boost}
Since the spacetime SSB occurs, there is the hidden Lorentz boost to relate different $|\vec k|$ in the heavy particle states. %
First take the standard Lorentz boost $L(\vec{k})$, which boosts the particle from the rest to a small momentum $\vec{k}$. Then under a general Lorentz transformation $\Lambda$, one obtains $p^{\mu}\rightarrow \Lambda^{\mu}_{~\nu} p^{\nu}$ while $v^\mu$ unchanged. Given that $k^{\prime\mu}=\Lambda^{\mu}_{~\nu}p^{\mu}-mv^{\mu}$, the residual momentum $k^{\mu}$ would transform as
\begin{equation}\label{klorentz}
    k^{\mu}\longrightarrow k^{\prime\mu}=\Lambda^{\mu}_{~\nu}k^{\nu}+m\Lambda^{\mu}_{~\nu}v^{\nu}-mv^{\mu}.
\end{equation}

The general form of the Lorentz boost from $u^\mu$ to $w^\mu$ could be parameterized as 
\begin{equation}\label{lambdawv}
\Lambda(w,u)\equiv\text{exp}\left(-iJ_{\alpha\beta}w^{\alpha} u^{\beta}\theta_w\right),
\end{equation}
where $J_{\alpha\beta}$ is the generator of Lorentz group, and $\theta_w\equiv\frac{\sinh^{-1}\left(\sqrt{(w\cdot u)^2-1}\right)}{ \sqrt{(w\cdot u)^2-1}}$, such that $w=\Lambda(w,u) u$. In the vector representation, the Lorentz generator is $(J^{\rho\sigma})^{\mu}_{~\nu}=i(g^{\rho\mu}\delta^{\sigma}_{~\nu}-g^{\sigma\mu}\delta^{\rho}_{~\nu})$.

First, let us apply Eq.~\eqref{lambdawv} to the standard Lorentz boost $L(\vec{k})$, we identify $u^\mu = (1,0,0,0)$ and $w^\mu = \frac{p^{\mu}}{m}\equiv(1+\frac{k^0}{m},\frac{k^i}{m})$. The standard Lorentz boost $L(\vec k)$ takes the form
\begin{equation}
    L(\vec k)\equiv \exp\left(-iJ_{i0}\frac{k^{i}}{m}\theta_k\right)=\text{exp}\left(-i\hat{\vec{K}}\cdot\vec{\eta}_k\right),
\end{equation}
where the boost generator 
\begin{equation}
    (K^i)^{\mu}_{~\nu}=(J^{0i})^{\mu}_{~\nu}=i\left(g^{0\mu}\delta^i_{\nu}-g^{i\mu}\delta^0_{\nu}\right),
\end{equation}
and the rapidity $\vec \eta_k$ is
\begin{equation}
    \eta^i_k\equiv \frac{k^i}{m}\theta_k=k^i~ \frac{\sinh^{-1}|\frac{\vec k}{m}|}{|\vec k |}.
\end{equation}

\begin{figure}
    \centering
    \includegraphics[width=0.5\linewidth]{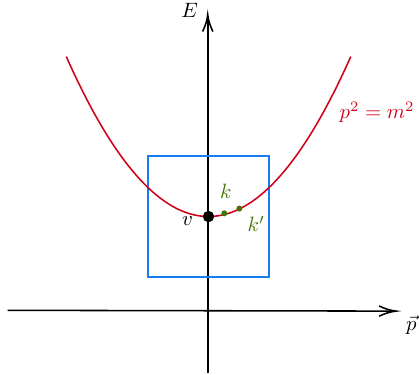}
   \caption{This diagram illustrates the  boost transformation of the residual momentum from $k^{\mu}$ to $k^{\prime\mu}$.  The red curve denotes the mass-shell $p^2 = m^2$. The black dot marks the reference momentum $m v^{\mu}$, while the blue box centered around it represents the validity region of the Heavy Particle Effective Theory for this specific velocity $v^{\mu}$. Each small green dot represent a residual momentum. Under a boost, the residual momentum transforms as $k^{\mu} \to k^{\prime\mu}$, whereas the reference velocity $v^{\mu}$ remains unchanged.  }
   \label{fig:deltak}
\end{figure}

Then under a general Lorentz transformation $\Lambda$, 
the total and residual momenta transform
\begin{equation}
\delta_{\Lambda}:\quad\left\{\begin{array}{lll}
&v^{\mu}\longrightarrow v^{\mu},
\\
\\
  & k^{\mu}\longrightarrow \Lambda k^{\mu}+m\Lambda v^{\mu}-mv^{\mu},
  \\
  \\
 &p^{\mu}\longrightarrow \Lambda p^{\mu}. 
   \end{array}  \right.
\end{equation}
And this  boost transformation is illustrated in Fig.~\ref{fig:deltak}. Let us focus on an infinitely small boost transformation, according to appendix~\ref{ap1}, it can be parametrized by the standard boost with three parameters as
\begin{equation}\label{eq:Bqdef}
     B(q)^{\mu}_{~\nu} \equiv\Lambda(v-\frac{q}{m},v) = \delta^{\mu}_{~\nu}+\frac{1}{m}(v^{\mu}q_{\nu}-q^{\mu}v_{\nu}) + \mathcal{O}(q^2),
\end{equation}
with $v^2=(v-q/m)^2=1$.
%
%
%
Accordingly, the residual momentum $k^{\mu}$ transforms as
\begin{equation}
\begin{array}{lll}
         k^{\mu}\longrightarrow  k^{\prime\mu}&=&B(q)^{\mu}_{~\nu}k^{\nu}+mB(q)^{\mu}_{~\nu}v^{\nu}-mv^{\mu}
         \\
         \\
         &=&k^{\mu}-q^{\mu}+\frac{v^{\mu}(q\cdot k)-q^{\mu}(v\cdot k)}{m}, 
\end{array}
\end{equation}
and the momentum $p$ transforms as
\bea
         p^{\mu}\longrightarrow  p^{\prime\mu} = p^{\mu}+\frac{v^{\mu}(q\cdot p)-q^{\mu}(v\cdot p)}{m}.
\eea

The heavy one-particle state should be also boosted, and we obtain the heavy one-particle states from the $|\vec k|=0$ mode defined as
\begin{equation}\label{unknownboost1}
        |v,\vec k,\sigma\rangle\equiv U(L(\vec k))|v,0,\sigma\rangle.
    \end{equation}
On the other hand, we have obtained the heavy one-particle states $|v,\vec k, h\rangle$ by the induced representation for the non-vanishing residual momentum $|\vec k|>0$. For the momentum $\vec k$, there are $2s+1$ components according to Eq.~\eqref{unknownboost1}, where $\sigma=-s,...,+s$. Correspondingly, there should also be $2s+1$ values of helicity $h=-s,...,+s$, such that
\begin{equation}\label{sigmahconver}
|v,\vec k,\sigma\rangle=\bigoplus_{h=-s}^{h=+s}|v,\vec k,h\rangle.
\end{equation}
Thus the states are still inside the Hilbert space spanned by the UIR of $R^{1,3}\rtimes SO(3)$. However, we use $|v,\vec k,\sigma\rangle$ to represent the momentum $\vec k$ particle with the hidden Lorentz symmetry, since it reveals the spin structure of the massive particle by carrying the spin index of the $SO(3)$ little group.

The underlying Lorentz covariance demands that the boost generator $\vec K$ obeys the Lorentz group algebra, and the orbit in the Hilbert space is $p^2=(mv+k)^2=m^2$, which is equivalent to $k^2+2mv\cdot k=0$. Similar states could be found in Ref.~\cite{Dugan:1991ak}.
One could also choose to recover the Galilean group symmetry by imposing the corresponding algebra and restoring the EOM $k^0-\frac{\vec k^2}{2m}=\text{constant}$.

The heavy one-particle state $|v,\vec k,\sigma\rangle$ transforms nonlinearly under the Lorentz boost.
For the state with general residual momentum, it corresponds to a boost transformation $U(L(\vec k))$ in the coset as Eq.~\eqref{unknownboost1}. Under an additional Lorentz transformation $\Lambda$, with the decomposition 
\begin{equation}
U(\Lambda)U(L(\vec k))|v,0,\sigma\rangle=U(L(\vec k^{\prime}))W[\Lambda,p]|v,0,\sigma\rangle,
\end{equation}
where $W[\Lambda,p]\equiv U(L(\vec k^{\prime}))^{-1}U(\Lambda)U(L(\vec k))$ is the little group rotation, we then find that
the heavy one-particle state $|v,\vec k,\sigma\rangle$ related to the coset element $U(L(\vec k))$ is transformed as
\begin{equation}\label{boostofstate}
    U(\Lambda)|v,\vec k,\sigma\rangle=\sum_{\sigma^{\prime}}D^{(s)}_{\sigma^{\prime}\sigma}(W[\Lambda,p])|v,\vec k^{\prime},\sigma^{\prime}\rangle,
\end{equation}
with $D^{(s)}_{\sigma^{\prime}\sigma}(W[\Lambda,p])\equiv\langle v,0,\sigma^{\prime}|W[\Lambda,p]|v,0,\sigma\rangle.$
By selecting one special Lorentz transformation, such as $\Lambda=B(q)$ defined in Eq.~\eqref{eq:Bqdef}, this little group transformation $W[\Lambda,p]$ can be directly calculated from its definition to the first order of $q/m$,
\begin{align}
    W[B(q),p]&=1-\frac{i}{2}\frac{1}{m(m+v\cdot p)}(q^{\alpha}k_{\perp}^{\beta}-k^{\alpha}_{\perp}q^{\beta})J_{\alpha\beta},
\end{align}
with details exhibited in appendix \ref{ap1}.
The perpendicular component $k_{\perp}^{\mu}\equiv k^{\mu}-v^{\mu}(v\cdot k)$ is defined previously. 
In the spinor representation $J^{ij}=\frac{1}{2}\epsilon^{ijk}\sigma^k$, the transformation matrix $D_{\sigma^{\prime}\sigma}^{(\frac{1}{2})}$ in Eq.~\eqref{boostofstate} is 
\begin{equation}\label{Dsmatrix}
    D^{(\frac{1}{2})}(W[B(q),p])=1+i\frac{1}{m\left(2m+v\cdot k\right)}q^ik_{\perp}^{j}\epsilon^{ijk}\frac{\sigma^k}{2} \equiv 1 +i \frac{\vec q}{m}\cdot \vec K.
\end{equation}
Here we identify the boost generator as
\begin{equation}\label{kforstate}
  	\vec{K}=\frac{-\frac{\vec \sigma}{2}\times\vec{k}_{\perp}}{2m+v\cdot k}.
  \end{equation}
Without explicit Lorentz covariant wave function, the exact dependence on momentum $\vec k_{\perp}$ translates into a dependence on derivatives of heavy fields in the coordinates space. Consequently, the boost transformation of fields is nonlinear.

Finally, the vacuum satisfies $a_{v,0}^{\sigma}|\Omega\rangle=0$, and the heavy one-particle state is created by $|v,0,\sigma\rangle=a^{\dagger\sigma}_{v,0}|\Omega\rangle$. The creation operator with the general residual momentum is 
\begin{equation}\label{unknownboost2}
a^{\dagger\sigma}_{v,k}=U(L(\vec k))a^{\dagger\sigma}_{v,0}U(L(\vec k))^{-1},
\end{equation}
such that the heavy one-particle state with the momentum $\vec k$ is
\begin{equation}\label{adgstate}
    |v,\vec k,\sigma\rangle=a_{v,k}^{\dagger\sigma}|\Omega\rangle.
\end{equation}
The commutator is
\begin{equation}\label{quantizationA}
        \{a_{v,k}^{\sigma},a_{v^{\prime},k^{\prime}}^{\dagger\sigma^{\prime}}\}=(2\pi)^3v^0\delta_{\sigma\sigma^{\prime}}\delta_{vv^{\prime}}\delta^{(3)}(\vec k-\vec k^{\prime}).
    \end{equation}
where the velocity superselection rule~\cite{Georgi:1990um} is implemented.

\subsection{Coset Description of the State}
\label{coset}

\paragraph{Coset construction $\delta \eta$}
To keep the residual momentum $k^{\mu}$ invariant,
there is another way to reveal the underlying Lorentz symmetry of the NR effective theory with the spacetime SSB, where the $SO(3)$ is the unbroken subgroup of the $SO(3,1)$. We follow a similar procedure of the Callan-Coleman-Wess-Zumino (CCWZ) construction~\cite{PhysRev.177.2239,PhysRev.177.2247,Sun:2022ssa}, where the nonlinear transformation of the coset elements indicates the transformation properties.

Before discussing the general residual momentum, we first consider the vanishing residual momentum $k^{\mu}=0$, the three-velocity $\vec v$ is the velocity of the particle, $\vec p=m\vec v$. The corresponding rapidity $\vec{\eta}_{v}$ is defined by
\begin{equation}\label{rapidity}
\vec{\eta}_{v}=\vec v~\theta_{v}=\vec{v}~\frac{\sinh^{-1}|\vec{v}|}{|\vec v|},
\end{equation}
it connects to a coset element --- the standard boost $
    L(\vec v)=e^{i\vec K\cdot\vec \eta_{v}}$,
taking the rest to the velocity $\vec v$. Here $\vec K$ is the boost generator. In the vector representation, it reads
\begin{equation}
    (K^i)^{\mu}_{~\nu}=(J^{0i})^{\mu}_{~\nu}=i\left(g^{0\mu}\delta^i_{\nu}-g^{i\mu}\delta^0_{\nu}\right).
\end{equation}
Thus $\vec v$ labels the coset element $L(\vec v)$ in $SO(3,1)/SO(3)$.
The coset description usually decomposes a group into its subgroup and coset space, correspondingly, a general Lorentz transformation can be generated by rotation and boost. Therefore, the Lorentz transformation $\Lambda$ acts on the coset element $L(\vec v)$ as
\begin{equation}\label{cosetdecompose}
    \Lambda L(\vec v)=L(\vec v')R^{\prime},
\end{equation}
where $R'$ is the Wigner rotation
\begin{equation}
    R'=L(\vec v')^{-1}\Lambda L(\vec v)=e^{-i\vec{\eta}_{v'}\cdot \vec{K}}\Lambda e^{i\vec{\eta}_v\cdot \vec{K}}\in SO(3).
\end{equation}
Notice that, $L(\vec v')R^{\prime}$ and $L(\vec v')$ belong to the same coset space. Consequently, under the Lorentz transformation $\Lambda$, the coset element is changed as
\begin{equation}\label{etatrans}   
     L(\vec v)\longrightarrow L(\vec v^{\prime})=\Lambda L(\vec v)R^{\prime -1}.
\end{equation}
The coset element is written by $L(\vec v)=e^{i\vec K\cdot\vec \eta_{v}}$, then we consider the specific transformation in the above equation:
\begin{itemize}
    \item For the pure rotation $\Lambda=R^{\prime}=e^{i\vec\theta^{\prime}\cdot\vec J}$, where $\vec J$ is rotation generator and $\theta'$ is the angle, the $\vec \eta$ transforms as
\begin{equation}
    \vec  \eta_v\longrightarrow   \vec\eta_{v'} =R^{\prime}\vec  \eta_v.
\end{equation}
Due to the definition of rapidity Eq.~\eqref{rapidity}, we know that the velocity transforms as
\begin{equation}
    \vec v\longrightarrow\vec v'=R'\vec v.
\end{equation}
This is a linear transformation, since for the rapidity with the same magnitude $|\vec v|$ that are connected by the rotation, they belong to the same coset space.
\item For the Lorentz boost $\Lambda=B(q)$, where $\frac{q^{\mu}}{m}\equiv(0,-\frac{\vec q}{m})$ is an infinitesimal parameter. The
rapidity $\vec \eta_v$ has a shift
\begin{equation}\label{shifteta}
\begin{array}{lll}
   \vec \eta\longrightarrow \vec \eta^{\prime}&=&\frac{\vec v+\frac{\vec q}{m}\sqrt{1-\vec v^2}}{|\vec v+\frac{\vec q}{m}\sqrt{1-\vec v^2}|}\sinh^{-1}|\vec v+\frac{\vec q}{m}\sqrt{1-\vec v^2}|\\
   &=&\vec\eta+\frac{\vec q}{m}+...\ ,
\end{array}
\end{equation}
where the velocity $\vec v$ transforms nonlinearly as
\begin{equation}\label{boostv}
\begin{array}{lll}
     \vec v\longrightarrow \vec v'&=&\vec v+\frac{\vec q}{m}\sqrt{1-\vec v^2}
     \\
     &=&\vec v+\frac{\vec q}{m}+... 
\end{array}
\end{equation}
\end{itemize}

For a general residual momentum in the rest frame, $k^{\mu}$ has small spatial components. But we still have $v^2=(v+\frac{q}{m})^2=1$ for $v\cdot q=0$ and infinitesimal $\frac{q^{\mu}}{m}=(0,-\frac{\vec q}{m})$.
Therefore, the nonlinear transformation of the coset element under the Lorentz transformation $\Lambda$ in Eq.~\eqref{etatrans} indicates that, the transformation that leaves the residual momentum $k^{\mu}$ invariant while changing $v^{\mu}$ is:
\begin{equation}\label{rpivp}
\delta\eta:\left\{\begin{array}{lll}
v^{\mu}\longrightarrow v'^{\mu}=\Lambda  v^{\mu}+\frac{1}{m}\Lambda  k^{\mu}-\frac{1}{m}k^{\mu}, \\ \\k^{\mu}\longrightarrow k^{\mu},
     \\
     \\
p^{\mu}\longrightarrow p'^{\mu}=\Lambda  p^{\mu}.
\end{array}\right.
\end{equation}
Due to the fact that every coset element $L(\vec v)$ is related to a frame $v^{\mu}$, we derive the transformation of $v^{\mu}$ through the coset construction. The three-velocity $\vec v$ transforms linearly under the $SO(3)$ while nonlinearly under the coset $SO(3,1)/SO(3)$. This kind of transformation $\delta\eta $ is equivalent to the usual reparameterization $\delta v$, thus, although the introduction of $v^{\mu}$ breaks the $SO(3,1)$ symmetry to the $SO(3)$, the broken symmetry can be realized by the reparameterization.

\begin{figure}
    \centering
    \includegraphics[width=0.5\linewidth]{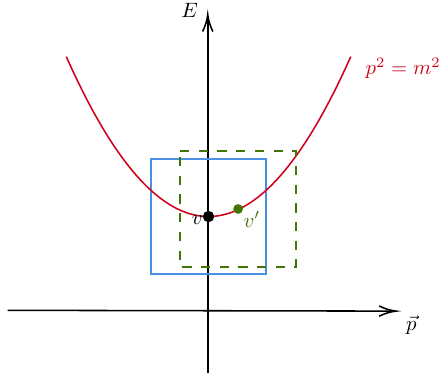}
   \caption{This figure illustrates the reparameterization  of the velocity from $v^{\mu}$ to $v^{\prime\mu}$.  The blue box, centered on the original velocity $v^{\mu}$, and the dashed green box, centered on the new velocity $v^{\prime\mu}$, represent the validity regions of the effective theory in their respective frames. Under the reparameterization, the total physical momentum $p^{\mu}$ remains invariant. }
   \label{fig:deltav}
\end{figure}

\paragraph{Reparameterization $\delta v$}

Reparameterization invariance is a symmetry arising from the redundancy in choosing the label $v^{\mu}$ for the heavy particle, under which the effective Lagrangian remains invariant~\cite{Luke:1992cs,Finkemeier:1997re,Manohar:1997qy}. 
For a given total momentum $p^{\mu}$, the reparameterization implies that the decomposition in Eq.~\eqref{pequalmvk} 
\begin{equation}
    p^{\mu}=mv^{\mu}+k^{\mu},
\end{equation}
is not unique for the pair $(v^{\mu},k^{\mu})$, 
\begin{equation}\label{pairrpi}
    (v^{\mu},~k^{\mu})\longleftrightarrow (v^{\mu}-\frac{q^{\mu}}{m},~k^{\mu}+q^{\mu}),
\end{equation}
where $\frac{q^{\mu}}{m}$ is an infinitesimal parameter satisfying $v^2=(v-\frac{q}{m})^2=1$. The reparameterization can be written as
\begin{equation}\label{lambdaprimerpi}
\delta v:\quad\left\{\begin{array}{lll}
     &  v^{\mu}\longrightarrow v^{\prime \mu}=\Lambda   v^{\mu},
     \\
     \\
&k^{\mu}\longrightarrow k'^{\mu}=k^{\mu}-m\Lambda  v^{\mu}+mv^{\mu},
     \\
     \\
     &  p^{\mu}\longrightarrow p^{\mu}.
\end{array}\right.
\end{equation}
And Eq.~\eqref{lambdaprimerpi} reduces to Eq.~\eqref{pairrpi} when $\Lambda  =B(q)$. The reparameterization is illustrated in Fig.~\ref{fig:deltav}.

\paragraph{Equivalence for $\delta_{\Lambda  }$, $\delta v $ and $\delta\eta$}
Denoting the boost transformation in the last subsection as $\delta_{\Lambda  }$, it turns out that $\delta_{\Lambda  }$ is equivalent to the reparameterization $\delta v$ at the level of invariants:
Although the boost transformation $\delta_{\Lambda  }$ only acts on the dynamic variables, while the reparameterization $\delta v$ only works for the reference frame $v^{\mu}$, we find that $\delta=\delta_{\Lambda  }+\delta v$ is at most a linear transformation, i.e., the rotation, thus the variation of singlets is trivial $\delta=0$.

At the momentum level, since the velocity $v^{\mu}$ and the residual momentum $k^{\mu}$ transform as
\begin{equation}
\delta v:\quad
\left\{\begin{array}{lll}
     &  v^{\mu}\longrightarrow \Lambda   v^{\mu},
     \\
     \\
&k^{\mu}\longrightarrow k^{\mu}-m\Lambda  v^{\mu}+mv^{\mu},
\\
\\
&p^{\mu}\longrightarrow p^{\mu},
    
\end{array}\right.
\end{equation}

\begin{equation}
\delta_{\Lambda  }:\quad\left\{\begin{array}{lll}
&v^{\mu}\longrightarrow v^{\mu},
\\
\\
  & k^{\mu}\longrightarrow \Lambda  k^{\mu}+m\Lambda  v^{\mu}-mv^{\mu},
  \\
  \\
 &p^{\mu}\longrightarrow \Lambda  p^{\mu}, 
   \end{array}  \right.
\end{equation}
the combined transformation $\delta=\delta v+\delta_{\Lambda  }$ is
\begin{equation}
\delta:\quad\left\{\begin{array}{lll}
&v^{\mu}\longrightarrow \Lambda  v^{\mu},
\\
\\
  & k^{\mu}\longrightarrow \Lambda  k^{\mu},
  \\
  \\
 &p^{\mu}\longrightarrow \Lambda  p^{\mu}.
   \end{array}  \right.
\end{equation}
As a result, all four-vectors transform covariantly under $\delta$. Therefore, the singlets, such as $mvx, kx,$ and $ px$, are Lorentz invariant under $\delta$, and the variation of singlets is $\delta=0$. For these singlets, $\delta v=-\delta_{\Lambda  }$.

Besides, the combined transformation $\delta\eta-\delta$ is
\begin{equation}
   \delta\eta-\delta:\left\{\begin{array}{lll}
      &v^{\mu}\longrightarrow v^{\mu}+\frac{1}{m}k^{\mu}-\frac{1}{m}\Lambda^{-1}k^{\mu},
\\
\\
  & k^{\mu}\longrightarrow \Lambda^{-1}k^{\mu},
  \\
  \\
 &p^{\mu}\longrightarrow p^{\mu}.
   \end{array}\right. 
\end{equation}
This corresponds to a reparameterization for
\begin{equation}
    (v^{\mu},~k^{\mu})\longleftrightarrow (v^{\mu}+\frac{1}{m}k^{\mu}-\frac{1}{m}\Lambda^{-1}k^{\mu},~\Lambda^{-1}k^{\mu}),
\end{equation}
thus we conclude that $\delta\eta$ from the coset construction is equivalent to the reparameterization $\delta v$.

At the state level, for the full Lorentz covariant theory, the relativistic one-particle state $|p,\sigma\rangle$ is already defined in the lab frame. However, in a second moving frame with velocity $v^{\mu}$, $|p,\sigma\rangle$ is boosted as a new state 
\begin{equation}
    |p_0,\sigma\rangle=U(L(\vec v))^{-1}|p,\sigma\rangle,
\end{equation}
with the momentum $p_0^{\mu}\equiv L(\vec v)^{-1}p^{\mu}$. For the state with the momentum $p_0^{\mu}\sim mv_0^{\mu}=m(1,0,0,0)$,
it can be rewritten in the heavy one-particle Hilbert space as 
\begin{equation}
    |v_0,\vec k(v,p),\sigma\rangle=|p_0,\sigma\rangle,
\end{equation}
with the residual momentum 
\begin{equation}
    k^{\mu}(v,p)\equiv p_0^{\mu}-mv_0^{\mu}=L(\vec v)^{-1}p^{\mu}-mv_0^{\mu}.
\end{equation}
In a nutshell, we build a map from $|p,\sigma\rangle$ to the heavy one-particle state $|v_0,\vec k(v,p),\sigma\rangle$, with the proper velocity $v^{\mu}\sim p^{\mu}/m$.
For a third frame with velocity $v^{\prime\mu}$ and state $|p^{\prime},\sigma^{\prime}\rangle $ obtained by:
\begin{equation}\label{Lambdav}
    \begin{cases}
         v \longrightarrow v'\equiv\Lambda   v,\\
         \\
         |p,\sigma\rangle\longrightarrow| p^{\prime},\sigma^{\prime}\rangle\equiv U(\Lambda  )| p,\sigma\rangle,
    \end{cases}
\end{equation}
these two states $|p^{\prime},\sigma^{\prime}\rangle$ and $|p,\sigma\rangle$ in the heavy one-particle Hilbert space are related by
\begin{align}\label{Wtransform}
    |v_0,\vec k(v^{\prime},p^{\prime}),\sigma^{\prime}\rangle&=U(L(\vec v'))^{-1}|p^{\prime},\sigma'\rangle\nonumber\\
    &=U(L(\vec v'))^{-1}U(\Lambda  )|p,\sigma\rangle\nonumber\\
    &=U(L(\vec v'))^{-1}U(\Lambda  )U(L(\vec v))|v_0,\vec k(v,p),\sigma\rangle\nonumber\\
    &=U(W)|v_0,\vec k(v,p),\sigma\rangle,
\end{align}
with $W\equiv L(\Lambda   \vec v)^{-1}\Lambda   L(\vec v)$ being a rotation in the $SO(3)$.

Given that $v^{\mu}$ can be treated as a spurion which transforms under  the Lorentz group like Eq.~(\ref{Lambdav}), the physical theory must be reparameterization invariant in order that the dependence on spurion can be removed consistently. An $SO(3)$ invariant theory can be made $SO(3,1)$ invariant by adding a new Lorentz covariant spurion $v^{\mu}$. Conversely, for any infinitely small transformation, the theory is Lorentz covariant as Eq.~(\ref{Wtransform}). As a result, the variation of singlet vanishes, with all relativistic building blocks including the $v^{\mu}$ transform covariantly.

At the Lagrangian level, the detailed examination is derived in subsection \ref{sec8rpi}, accordingly these two kinds of variation are equivalent. A theory is reparameterization invariant if and only if it is boost invariant. The physical interpretation of the Lorentz invariance further requires the invariance under both reparameterization and boost, and thus non-trivial relations between the Wilson coefficients of the effective theory are required.

\subsection{Missing Goldstone}
\label{missingGB}

In a spontaneously broken symmetry theory, the Goldstone modes should be considered. The naive counting of the Goldstone modes by the number of broken generators could fail for spontaneously broken space-time symmetry, and this counting 
has been improved in Refs.~\cite{Low:2001bw,Nicolis:2015sra}. 
In some situations, parts of the Goldstone modes are linear dependent of the derivatives of other modes, thus the number of Goldstone modes is reduced. This is known as the inverse-Higgs constraints~\cite{Ivanov:1975zq}.

To be specific, in the original group $G$, for the unbroken translation $P^{\mu}$ and the broken generators $Q$ and $Q'$, if 
\begin{equation}
     Q^{\prime}\subset [ P^{\mu}, Q],
\end{equation}
then the Goldstone modes of $ Q$ and $ Q'$ are dependent. As a result they generate massless excitations less than the number of the broken generators. 
However, when this description applies to the Lorentz group,
where the broken generator is the boost generator $K^i$,  we can not find any other Goldstone mode, to eliminate the mode that corresponds to the broken boost generator, since
\begin{align}
    &[P^i,  K^j]=-iP^0\delta^{ij},\nonumber\\
    &[ P^0,K^i]=-i P^i.
\end{align}
Therefore, no inverse-Higgs constraint can be imposed in the spontaneously broken Lorentz symmetry, and the number of the Goldstone modes is still three corresponding to the three broken boosts. Associating to the three spontaneously broken boosts, these Goldstone modes are gapless.

For the spontaneously broken boost symmetry, a Goldstone boson field $\vec\eta(x)$ can exist if the energy-momentum tensor $T^{\mu\nu}$ of the system satisfies
\begin{equation}
    \rho+\mathcal{P}=0,
\end{equation}
where $\rho\equiv\langle\Omega|T^{00}|\Omega\rangle$ is the energy density, $ \mathcal{P}\delta^{ij}\equiv\langle\Omega|T^{ij}|\Omega\rangle$ is the pressure. This system is called the type-I framids, with the Goldstone modes $\vec\eta(x)$ called the framons~\cite{Nicolis:2015sra}.

However, for the Fermi liquid as well as the heavy particle, they are low energy NR system and their mass density is larger than any other scale, such that
\begin{equation}
    \rho+\mathcal{P}\neq0,
\end{equation}
and there should not be any local Goldstone field $\vec\eta(x)$ when boost breaks only~\cite{Alberte:2020eil}. In fact, in the HPET there are states $|v,\vec k,0\rangle$ play the role of the Goldstone modes, but do not exist as one-particle states alone.

The Goldstone theorem suggests that for the broken generator $K^i$ which satisfies $K^i|\Omega\rangle\neq0$, there are the Goldstone modes $|n,\vec k\rangle$ created by $K^i$ satisfying the gapless dispersion relation $k^0=0$ when $\vec k=0$, where details can be found in appendix \ref{ap2}.
For the HPET, the Goldstone bosons must be picked up from the heavy one-particle Hilbert space spanned by $|v,\vec k,\sigma\rangle$. As we discussed in subsection \ref{hilbertspace}, the little group of $|v,0,\sigma\rangle$ is $SO(3)$ and it represents a massive heavy particle with spin-$s$ and the constant velocity $v^{\mu}$. 
On the other hand, the little group of $|v,\vec k,h\rangle$ with $\vec k\neq 0$ is $SO(2)$. For a scalar state $h=0$, we have the following relations
\begin{equation}
    \begin{cases}
         |v,\vec k,0\rangle=U(L(\vec k))|\Omega\rangle,\\
         \\
    k^0=\frac{\vec k^2}{2m}+\mathcal{O}(\vec k^4),
    \end{cases}
\end{equation}
thus the state $|v,\vec k,0\rangle$ serves as the Goldstone mode.

The Goldstone bosons are not really detected in these heavy particle system experimentally, 
since there is no one-particle state of the Goldstone boson and the role is played by the so-called particle-hole continuum instead, such as in the Fermi liquid~\cite{ Alberte:2020eil}. 
Similar analysis can also be carried out within the description of the heavy one-particle state.
For the vacuum $|\Omega\rangle$, we have two kinds of states:
\begin{itemize}
    \item a spin-$S$ heavy particle $|v,0,\sigma\rangle, \sigma=-s,...s,$
    \item a scalar $h=0$ Goldstone mode $|v,\vec k,0\rangle$.
\end{itemize}
In the frame with the velocity $v^{\mu}$, the heavy massive particle $|v,0,\sigma\rangle$ is not dynamical by itself other than the spin d.o.f., while the Goldstone mode $|v,\vec k,0\rangle$ carries the residual momentum $\vec k$ related to the interaction.
Nevertheless, as a spurion created by $L(\vec k)$, $|v,\vec k,0\rangle$ doesn't exist without $|v,0,\sigma\rangle$. Besides, multi-particle states such as $|v,\vec k_1,0\rangle|v,\vec k_2,0\rangle$ are ruled out.
According to the representation we introduce
\begin{equation}
|v,\vec k,\sigma\rangle=U(L(\vec k))|v,0,\sigma\rangle,
\end{equation}
the massive particle at rest $|v,0,\sigma\rangle$ absorbs the Goldstone mode $ |v,\vec k,0\rangle $ to give 
\begin{equation}
    |v,\vec k,\sigma\rangle=\left\{|v,\vec k,0\rangle,~ |v,0,\sigma\rangle\right\}.
\end{equation}
Thus, in the standard discussion, the Goldstone modes are absent and hidden in the heavy particle state $|v,\vec k, \sigma\rangle$ as long as we try to recover the Lorentz symmetry.

For general residual momentum with $|\vec k|>0$ and each helicity $h$, Eq.~\eqref{irepkbig0} gives the UIR, and the corresponding Hilbert space can be denoted as $\bigoplus_{h=-s}^{h=+s}\mathcal{H}_{|\vec k|,h}$. Due to Eq.~\eqref{sigmahconver}, we know that
\begin{equation}
    |v,\vec k,\sigma\rangle\in\bigoplus_{h=-s}^{h=+s}\mathcal{H}_{|\vec k|,h}.
\end{equation}
In fact, for the given momentum magnitude $|\vec k|$, this Hilbert space can be recognized as the tensor product 
\begin{equation}
    \bigoplus_{h=-s}^{h=+s}\mathcal{H}_{|\vec k|,h}=\mathcal{H}_{|\vec k|}\otimes \mathbf{C}^{2s+1},
\end{equation}
where $\mathbf{C}^{2s+1}$ is the $2s+1$ dimensional complex Hilbert space for $|v,0,\sigma\rangle$, and $\mathcal{H}_{|\vec k|}$ is the momentum space for the given $|\vec k|$. This is similar to the spirit of dividing the d.o.f. into two irreducible representations: the boost carrying the residual momentum, and the rest heavy particle carrying the spin.
\section{Nonlinear Boost of Heavy Field}
\label{sec4}
Based on discussion in previous section where the transformation property of the state for the NR theory is derived, we consistently find the boost transformation of the free wave functions as well as fields in subsubsection~\ref{LGT}.  The relation between the heavy field and the Dirac field is discussed in subsubsection~\ref{subsec:heavy-dirac}. The anti-particle d.o.f. in the Dirac field is discussed in subsubsection~\ref{ioapdof}. The comparison with the FW transformation and the NR reduction is exhibited in subsubsection~\ref{FWtransformation}. The discussion of the boost transformation is extended to the gauge interactions in subsection~\ref{GECC}, where the covariant derivatives and field strength are considered.

\subsection{Heavy Field and Nonlinear Transformation}\label{HEAVYFN}

\subsubsection{Nonlinear Heavy Field}
\label{LGT}

According to the $G/H = SO(3,1)/SO(3)$ coset description, let us introduce a two-component spinor field $N$, which transforms linearly under the $H =SO(3)$ group, while nonlinearly under the $G/H$ coset, which gives the Lorentz boost.  
The one-particle state is identified using the annihilation operator $|v,\vec k,\sigma\rangle = a^\sigma_{v,k} |\Omega\rangle$. 
Here $\sigma$ is the spin index, where the spin is quantized along the $z$ axis.
The two-component heavy field $N(x)$ can be written by the Fourier mode expansion
\begin{equation}\label{Nellmodeexp}
    N_{\ell}(x)=\int \frac{d^3\vec k}{(2\pi)^3\sqrt{2E}}\sum_{\sigma}u_{\ell}^{\sigma}(k)a_{v,k}^{\sigma}e^{-ikx}.
\end{equation}
Here $\ell$ is the $SO(3)$ index, $u_{\ell}^{\sigma}(k)$ is the wave function, and the energy is  $E = m+v\cdot k$. In this mode expansion, the residual momentum $k^{\mu}$ is the dynamical variable.
As a low-energy effective theory, the anti-particle d.o.f. is forbidden, so the negative frequency mode is absent.

Without  loss of generality, we choose the rest frame $v^{\mu}=(1,0,0,0)$. Additionally, note that due to the decomposition Eq.~\eqref{pequalmvk}, in the rest frame the energy and three-momentum of the heavy particle is
\begin{equation}
    p^{\mu}=mv^{\mu}+k^{\mu},\quad v^{\mu}=(1,0,0,0)\Longrightarrow\left\{
    \begin{array}{l}
          E=p^0=m+k^0,\\
          \\
         \vec p=\vec k.
    \end{array}
    \right.
\end{equation}
And due to the on-shell condition $E^2=m^2+\vec p^2$ we also have $E=\sqrt{m^2+\vec p^2}=\sqrt{m^2+\vec k^2}$. Therefore, the EOM of the heavy particle is 
\begin{equation}
    m+k^0=\sqrt{m^2+\vec k^2},
\end{equation}
in the momentum space, and according to the mode expansion Eq.~\eqref{Nellmodeexp}, the EOM of the heavy field is
\begin{equation}\label{eq:freeNeom}
\left(m+i\partial_t\right)N_{\ell}(x)=\sqrt{m^2-\nabla^2}N_{\ell}(x),
\end{equation}  
in the configuration space.

Under the $SO(3)$ group, the field $N$ transforms linearly as  
\begin{equation}
    U(R)^{-1} N_{\ell}(x)U(R)=D_{\ell\bar \ell}(R) N_{\bar \ell}(R^{-1}x).
\end{equation}
Note that,  $G=SO(3,1)\simeq SU(2)_L\times SU(2)_R$, and the unbroken subgroup is $H=SO(3)\simeq SU(2)_V$ with $D_{\ell\bar\ell}(R)\in H$. This rotation group $H$ is the diagonal subgroup generated by $J^i=\frac{1}{2}(A^i+B^i)$ given in Eq.~\eqref{eq:twosu2lrgen}.
In the following let us derive the nonlinear transformation under the Lorentz boost $G/H$. Following the same logic on defining the particle state from the rest one with the standard boost, we start from the heavy field at the rest, and then take the standard boost, as follow.

The heavy field $N(x)$ can be rewritten as the heavy field $N^0(x)$, where $N^0(x)$ contains the rest frame wave function.
Given that the boosts from the rest to momentum $\vec k$ for the left and right-handed Weyl spinors are
\begin{equation}
\left\{
\begin{array}{ccc}
       S_L&=&\frac{1}{\sqrt{2m(m+E)}}\left[(E+m)\mathbf{I}_{2\times2}-\vec\sigma\cdot\vec k\right],  \\
     \\
      S_R&=&\frac{1}{\sqrt{2m(m+E)}}\left[(E+m)\mathbf{I}_{2\times2}+\vec\sigma\cdot\vec k\right],
\end{array}  \right.
\end{equation}
where $\mathbf{I}_{2\times2}$ is the 2-dimensional identity matrix, thus the standard boost for the heavy field is
\begin{eqnarray}
    L_{\ell\bar\ell}(\vec k)=\frac{1}{2}\left(S_L + S_R \right)= \sqrt{\frac{m+E}{2m}}\delta_{\ell\bar\ell}.
\end{eqnarray}
The standard boost $L_{\ell\bar\ell}(\vec k)$ acting on the wave function at the rest $u^\sigma_\ell (v)$ gives
\bea
u_{\ell}^{\sigma}(k) = L_{\ell\bar\ell}(\vec k) u_{\bar\ell}^{\sigma}(v),
\eea
and we arrive the relation between $N(x)$ and $N^0(x)$
\begin{equation}
    N_{\ell}(x)=L_{\ell\bar\ell}(\vec\nabla)\underbrace{\int \frac{d^3\vec k}{(2\pi)^3\sqrt{2E}}\sum_{\sigma}u_{\bar\ell}^{\sigma}(v)a_{v,k}^{\sigma}e^{-ikx}}_{N^0_{\ell}(x)} 
    = L_{\ell\bar\ell}(\vec\nabla) N_\ell^0(x).
\end{equation}
In the rest frame, the wave function $u_{\ell}^{\sigma}(v)$ satisfies the equality between the rotation in the spin space and the rotation on the heavy field  
\begin{equation}\label{uaDmatrix}
    \sum_{\sigma}u_{\ell}^{\sigma}(v)D^{(s)}_{\sigma\sigma^{\prime}}(W[R])= D_{\ell\bar \ell}(W[R])u_{\bar \ell}^{\sigma}(v).
\end{equation}
This gives the solution of the wave function  
\begin{equation}
    u_{\ell}^{\sigma}(v)=\sqrt{2m}\delta_{\ell}^{\sigma}, \quad \Rightarrow \quad 
    u_{\ell}^{\sigma}(k)=\sqrt{m+E}\delta_{\ell}^{\sigma}.
\end{equation}

Let us consider the boost transformation of the heavy field $N^0(x)$, in which only the $a_{v,k}^\sigma$ transforms under the Lorentz boost with infinitesimal parameter $\frac{q^{\mu}}{m}\equiv(0,-\frac{q^i}{m})$. 
Given that under the boost $B(q)$ the particle state has Eq.~\eqref{boostofstate}
\begin{equation}
    U(B(q))^{-1}a_{v,k}^\sigma U(B(q))=\sum_{\sigma^{\prime}}D^{(s)}_{\sigma^{\prime}\sigma}(W[B(q),p]) a_{v,k'}^{\sigma'},
\end{equation}
with the Wigner rotation in Eq.~\eqref{Dsmatrix} up to the first order of $q/m$
\begin{equation}\label{eq:Dsmatrix1}
    D^{(\frac{1}{2})}(W[B(q),p])\equiv 1 +i \frac{\vec q}{m}\cdot \vec K, \quad\text{where}\quad
    K^i = \frac{\epsilon^{ijk}k^{j}\frac{\sigma^k}{2}}{2m+k^0}, 
\end{equation}
we obtain the boost transformation of the heavy field $N^0_{\ell}(x)$ due to Eq.~\eqref{uaDmatrix} and Eq.~\eqref{eq:Dsmatrix1}
\begin{equation}\label{restntildebost}
    U(B(q))^{-1} N^0_{\ell}(x)U(B(q)) = e^{i\vec q\cdot\vec x} D_{\ell\bar \ell}(W[B(q),i\vec\nabla]) N^0_{\bar \ell}(B(q)^{-1}x), 
\end{equation}
where the Wigner rotation correspondingly is
\begin{equation}\label{boostcord}
    \quad D(W[B(q),i\vec\nabla]) =  (1+i \frac{\vec q}{m}\cdot \vec{K}),\quad\text{where}\quad \vec{K}=\frac{i\frac{\vec \sigma}{2}\times\vec{\nabla}}{2m+i\partial_t}.
\end{equation}
As a result, the boost transformation of $N_{\ell}^0(x)$ in Eq.~\eqref{restntildebost} can be denoted as 
\begin{eqnarray}
    U(B(q))^{-1} N^0(x)U(B(q))&=&e^{i\vec q\cdot\vec x}\left(1+i\frac{\vec q}{m}\cdot \frac{i\frac{\vec \sigma}{2}\times\vec{\nabla}}{2m+i\partial_t}\right)N^0(B(q)^{-1}x)\nonumber
      \\
      &=&\left(1+i\frac{\vec q}{m}\cdot \vec K_x\right)N^0(B(q)^{-1}x), 
\end{eqnarray}
where the boost generator is 
\begin{equation}
   \vec K_x=m\vec x+\vec K=m\vec x+\frac{i\frac{\vec \sigma}{2}\times\vec{\nabla}}{2m+i\partial_t}.
\end{equation}
This result reproduces the so-called little group transformation of the field in Ref.~\cite{Heinonen:2012km}. However, in our work, it is derived directly from the calculation, rather than introduced as a postulate. More discussion about this boost generator and the phase factor are exhibited in appendix~\ref{app3}.

Besides, we need the  transformation property of the standard boost $L_{\ell\bar\ell}$. In the coordinate space, the standard boost takes the form
\bea
L_{\ell\bar\ell}(\vec\nabla)=\sqrt{\frac{2m+i\partial_t}{2m}}\delta_{\ell\bar\ell}.
\eea
Under the Lorentz boost $B(q)=\Lambda(v-\frac{q}{m},v)$ given by Eq.~\eqref{eq:Bqdef}, with $v^{\mu}=(1,0,0,0)$ and the infinitesimal parameter $\frac{q^{\mu}}{m}=(0,-\frac{\vec q}{m})$, we obtain the boost transformation of the derivatives up to the first order of $q/m$
\begin{equation}\label{bqofderiv}
    \left\{\begin{array}{rll}
         U(B(q))^{-1}i\partial_t~U(B(q))&= &i\partial_t-\frac{\vec q}{m}\cdot i\vec\nabla\ , \\
         \\
         U(B(q))^{-1}i\vec\nabla~U(B(q))&= &i\vec\nabla-\frac{\vec q}{m}i\partial_t\ , 
    \end{array}\right.
\end{equation}
and thus we obtain the boost transformation of the standard boost
\begin{eqnarray}\label{eq:bqlnabla}
     U(B(q))^{-1} L_{\ell\bar\ell}(\vec\nabla)U(B(q)) &=& \sqrt{\frac{2m+i\partial_t-\frac{\vec q}{m}\cdot i\vec\nabla}{2m}} \delta_{\ell\bar\ell}
         \nonumber\\
         &=& \sqrt{1-\frac{\vec q}{m}\cdot\frac{i\vec\nabla}{2m+i\partial_t}} \sqrt{\frac{2m+i\partial_t}{2m}}\delta_{\ell\bar\ell}
         \nonumber\\
         &= &\left( 1 - \frac{i\vec \nabla}{2(2m+ i\partial_t)} \cdot \frac{\vec q}{m}\right) L_{\ell\bar\ell}(\nabla).
\end{eqnarray}

Since $N_{\ell}(x)=L_{\ell\bar\ell}(\vec\nabla)N_{\bar\ell}^0(x)$, utilizing the transformations of $L_{\ell\bar\ell}(\vec\nabla)$ and $N^0_{\ell}(x)$ in Eq.~\eqref{eq:bqlnabla} and Eq.~\eqref{restntildebost}, the boost transformation of the heavy field $N_\ell(x)$ is  
\begin{eqnarray}\label{eq:nlbost}
    U(B(q))^{-1} N(x)U(B(q))&=& U(B(q))^{-1} L(\vec\nabla)U(B(q))  \  U(B(q))^{-1} N^0(x)U(B(q)) 
    \\
      &=& e^{i\vec q\cdot\vec x} \left( 1 - \frac{i\vec \nabla}{2(2m+i\partial_t)} \cdot \frac{\vec q}{m}+i\frac{\vec q}{m}\cdot \frac{i\frac{\vec \sigma}{2}\times\vec{\nabla}}{2m+i\partial_t} \right) N(x')\nonumber\\
    &=& e^{i\vec q\cdot\vec x} \left( 1 - \frac{i\vec \nabla}{2(m+ \sqrt{m^2-\nabla^2})} \cdot \frac{\vec q}{m}+i\frac{\vec q}{m}\cdot \frac{i\frac{\vec \sigma}{2}\times\vec{\nabla}}{m+\sqrt{m^2-\nabla^2}} \right) N(x'),\nonumber
\end{eqnarray}
where we denote $x'=B(q)^{-1}x$, and the EOM of $N(x)$ is used.
Using the vector calculus $(\vec\sigma\cdot\vec q)(\vec\sigma\cdot\vec \nabla)=\vec q\cdot\vec\nabla+i\vec\sigma\cdot(\vec q\times\vec \nabla)$, the above result can be rewritten 
\bea\label{eq:boostofnell}
    U(B(q))^{-1} N(x)U(B(q))
    &=&e^{i\vec q\cdot\vec x} \left( 1+\frac{-i\vec q\cdot\vec\nabla+\vec\sigma\cdot\vec q\times\vec\nabla}{2m(m+\sqrt{m^2-\nabla^2})} \right) N(x')
    \nonumber\\
    &=&e^{i\vec q\cdot\vec x} \left( 1+\frac{\vec\sigma\cdot\vec q}{2m} \frac{-i\vec \sigma\cdot \vec \nabla}{m+\sqrt{m^2-\nabla^2}}\right) N(x')\nonumber\\
    &=&e^{i\vec q\cdot\vec x}\left(1+\frac{\vec\sigma\cdot\vec q}{2m}f(\vec\nabla)\right)N(x'),
\eea
where we define
\begin{equation}\label{restfreefnabla}
    f(\vec\nabla)=\frac{-i\vec\sigma\cdot\vec\nabla}{2m+i\partial_t}.
\end{equation} 
The boost transformation in the last line of Eq.~\eqref{eq:boostofnell} can be denoted as 
\begin{equation}
  U(B(q))^{-1} N(x)U(B(q))=  e^{i\vec q\cdot\vec x}\left(1+i\frac{\vec q}{m}\cdot\vec {\mathcal K} \right)N(x')=\left(1+i\frac{\vec q}{m}\cdot\vec {\mathcal K}_x\right)N(x'),
\end{equation}
such that the boost generator of the heavy field $N(x)$ is
\begin{equation}\label{kprimegen}
\vec {\mathcal K}_x=m\vec x+ \vec{\mathcal K},\quad\text{where}\quad    \vec{\mathcal K}=-i\frac{\vec\sigma}{2}f(\vec \nabla). 
\end{equation}
This transformation is nonlinear, as evidenced by the boost transformation under two successive operations, where $\vec{\mathcal{K}}$ itself depends on the frame.

Expanding $f(\vec \nabla)$ with respect to $1/m$, we obtain
\begin{equation}\label{restfreenablaexact}
    f(\vec\nabla)=\frac{-i\vec\sigma\cdot\vec\nabla}{2m+i\partial_t}=\frac{-i\vec\sigma\cdot\vec \nabla}{m+\sqrt{m^2-\nabla^2}}=-\frac{i\vec\sigma\cdot\vec\nabla}{2m}-\frac{i\vec\sigma\cdot\vec\nabla\nabla^2}{8m^3}+...
\end{equation}
If we identify $f(\vec\nabla) N$ as the two-component spinor for anti-particle $\tilde{N}$~\footnote{Indeed, it is another combination of the chiral boost transformation
\bea
\tilde{N} \equiv f(\vec\nabla) N=\left(-S_L(\nabla) + S_R(\nabla)\right)N^0.
\eea
}, we obtain the relation
\bea
U(B(q))^{-1}N(x)U(B(q))=e^{i\vec q\cdot\vec x}\left(N(x') +\frac{\vec\sigma\cdot\vec q}{2m}\tilde{N}(x')\right).
\eea
In the following we will discuss how to derive this from the Dirac field in the top-down approach.


\subsubsection{Heavy Field From Dirac Field Projection}
\label{subsec:heavy-dirac}

As for NR field in the broken spacetime symmetry, the vector $v^{\mu}$ plays an essential role. Since every vector could be constructed out of spinors, we have
\begin{equation}\label{vzetadef}
     v^{\mu}=\bar\zeta\gamma^{\mu}\zeta=\bar{\tilde {\zeta}}\gamma^{\mu}\tilde \zeta,
\end{equation}
where $\zeta$ and $\tilde \zeta$ are the spinor projection representation of $v^{\mu}$. By definition, they have no dynamics and do not transform under the boost of momentum. Moreover, they satisfy
\begin{equation}
\left\{
    \begin{array}{l}
        \zeta^I\bar{\zeta}^{~I}=(1+v\!\!\!/)/2,
        \\
        \\v\!\!\!/\zeta^I=\zeta^I,
        \\
        \\
        \bar \zeta^{J}\zeta^I=\delta^{J I},
    \end{array}\right.
\quad
\left\{
    \begin{array}{l}
        \tilde{ \zeta}^I\bar{\tilde{ \zeta}}^{~I}=-(1-v\!\!\!/)/2,
        \\
        \\v\!\!\!/\tilde{ \zeta}^I=-\tilde{ \zeta}^I,
        \\
        \\
        \bar {\tilde{ \zeta}}^{J}\tilde{ \zeta}^I=-\delta^{JI}.
    \end{array}\right.
\end{equation}
where $I$ is the little group index of $v^{\mu}$. Besides, $\zeta$ and $\tilde{ \zeta}$ are orthogonal to each other
\begin{equation}
    \bar\zeta^I\tilde{ \zeta}^{J}=\bar{\tilde{ \zeta}}^I\zeta^{J}=0.
\end{equation}

Let us build the connection between the heavy field and the particle mode of the Dirac field. Extracting the phase factor $e^{-imvx}$, the positive energy modes of the Dirac spinor field can be written as
\begin{equation}\label{psiL}
    \Psi_A(x)= e^{-imvx} \int\frac{d^3\vec p}{(2\pi)^3}\frac{1}{\sqrt{2E}}\sum_{\sigma}a^{\sigma}_{\vec p}~\mathcal{U}_A^{\sigma}(p)e^{-ikx},
\end{equation}
where $A$ denotes the Lorentz index.
Due to the phase factor $e^{-imvx}$, the anti-particle mode $b_{\vec p}^{s\dagger}$ has an extremely high-frequency phase factor $e^{i(2mv+k)x}$ and is not relevant.
The NR two-component wave functions are the projection of the four-component wave function
\begin{equation}
\left\{
\begin{array}{lll}
     u_I^{\sigma}(k)&=&\bar\zeta^I\mathcal{U}^{\sigma}(p),\\
     \\
     \tilde u_I^{\sigma}(k)&=&\bar{\tilde{\zeta}}^I\mathcal{U}^{\sigma}(p),
\end{array}  \right.
\end{equation}
while the NR field is the projection of the Dirac field:
\begin{equation}\label{defntilden}
    \left\{\begin{array}{llrlr}
         N_{\ell}(x)&\equiv& N_I(x)&=&e^{imvx}\bar\zeta^I\Psi(x),  \\
         \\
         \tilde N_I(x)&\equiv& f(\nabla)N_I(x)&=&-e^{imvx}\bar{\tilde{\zeta}}^I\Psi(x).
    \end{array}\right.
\end{equation}
Here we identify the rotation index $\ell$ as the little group index $I$. In fact, the Lorentz group spontaneously breaks to the little group that keeps $v^{\mu}$ invariant, where the heavy field exactly transforms under this unbroken subgroup.
Performing the projection we obtains
\begin{equation}\label{uplowdiracfield}
\left\{
\begin{array}{lll}
   \zeta^IN_I(x)&=&\frac{1+v\!\!\!/}{2}e^{imvx}\Psi(x),
    \\
    \\
 \tilde{ \zeta}^I\tilde N_I(x)&=&\frac{1-v\!\!\!/}{2}e^{imvx}\Psi(x).
\end{array}\right.
\end{equation}

Having obtained the two-component spinor, the four-component Dirac spinor could be rewritten.
At momentum $k^{\mu}=0$, $p^{\mu}=mv^{\mu}$,  the Dirac wave function $\mathcal{U}_A^{\sigma}(v)$ is then projected to the NR wave function $u_I^{\sigma}(v)$
as
\begin{equation}
   \mathcal{U}_A^{\sigma}(v) =\zeta_A^{I}u_I^{\sigma}(v).
\end{equation}
This is because in the reference frame $v^{\mu}$, the  wave function $\mathcal{U}_A^{\sigma}(v)$ has only the particle state component. 
Then the Dirac wave function is boosted from the rest wave function 
\begin{equation}
    \mathcal{U}_A^{\sigma}(p)=L_{AB}(\vec k)\mathcal{U}_B^{\sigma}(v)=L_{AB}(\vec k)\zeta_B^{I}u_I^{\sigma}(v),
\end{equation}
where $L_{AB}(\vec k)$ is the four-component standard boost. Then in the coordinate space we have
\begin{equation}\label{psiNrelation}
     \Psi_A(x)= e^{-imvx}L_{AB}(\vec\nabla)\zeta_B^I N^0_I(x).
\end{equation}
On the other hand, due to Eq.~\eqref{uplowdiracfield} the Dirac spinor field is
\begin{eqnarray}\label{psinewNrelation}
      \Psi_A(x)=e^{-imvx}\left(\zeta_A^IN_I(x)+\tilde\zeta_A^I\tilde N_I(x)\right)
    =e^{-imvx}\left(\zeta_A^I+\tilde{ \zeta}_A^{I}f(\vec\nabla)\right)N_I(x).
\end{eqnarray}

From the expression of the Dirac field $\Psi(x)$ in Eq.~\eqref{psiNrelation} and Eq.~\eqref{psinewNrelation}, the two-component standard boost $L_{IJ}(\vec\nabla)$ and $f(\vec\nabla)$ derived in subsubsection~\ref{LGT} can be reproduced. For above two equations, multiplying $\bar\zeta^I$ on them gives
\begin{equation}
    \begin{array}{lrll}
        &\bar\zeta_A^I \left(e^{-imvx}L_{AB}(\vec\nabla)\zeta_B^J N^0_J(x)\right)&=&\bar\zeta_A^I\left[e^{-imvx}\left(\zeta_A^J+\tilde{ \zeta}_A^{J}f(\vec\nabla)\right)N_J(x)\right],  \\
         \\
         \Longrightarrow &\bar\zeta_A^I L_{AB}(\vec\nabla)\zeta_B^J N^0_J(x)&=&\delta_{IJ}N_J(x),
         \\
         \\
         \Longrightarrow&\bar\zeta_A^I L_{AB}(\vec\nabla)\zeta_B^J N^0_J(x)&=&L_{IJ}(\vec\nabla)N_J^0(x), 
    \end{array}
\end{equation}
thus we obtain the two-component boost transformation:
\begin{equation}\label{eq:twocomLIJ}
    L_{IJ}(\vec \nabla)=\bar\zeta _A^IL_{AB}(\vec\nabla)\zeta_B^J.
\end{equation}
Moreover, multiplying $\bar{\tilde{\zeta}}^I$ on the Eq.~\eqref{psiNrelation} and Eq.~\eqref{psinewNrelation} gives the identity
\begin{equation}
    \begin{array}{lrll}
        &\bar{\tilde{\zeta}}_A^I \left(e^{-imvx}L_{AB}(\vec\nabla)\zeta_B^J N^0_J(x)\right)&=&\bar{\tilde{\zeta}}_A^I\left[e^{-imvx}\left(\zeta_A^J+\tilde{ \zeta}_A^{J}f(\vec\nabla)\right)N_J(x)\right]  \\
         \\
         \Longrightarrow &\bar{\tilde{\zeta}}_A^I L_{AB}(\vec\nabla)\zeta_B^J N^0_J(x)&=&-\delta_{IJ}f(\vec\nabla)N_J(x)
         \\
         \\
         \Longrightarrow&\bar{\tilde{\zeta}}_A^I L_{AB}(\vec\nabla)\zeta_B^J N^0_J(x)&=&-f(\vec\nabla)L_{IJ}(\vec\nabla)N_J^0(x), 
    \end{array}
\end{equation}
which suggests the relation:
\begin{equation}\label{eq:twocomfLIJ}
     f(\vec\nabla) L_{IJ}(\vec \nabla) =-\bar{\tilde{\zeta}} _A^IL_{AB}(\vec\nabla)\zeta_B^J.
\end{equation}
As a result, the general expression of the $f(\vec\nabla)$ is obtained due to Eq.~\eqref{eq:twocomLIJ} and Eq.~\eqref{eq:twocomfLIJ}:
\begin{equation}\label{eq:generalfzeta}
    f(\vec\nabla)=-\bar{\tilde{\zeta}} _AL_{AB}(\vec\nabla)\zeta_B \left(\bar\zeta _{A'}L_{A'B'}(\vec\nabla)\zeta_{B'}\right)^{-1}.
\end{equation}

The nonlinear boost transformation of the heavy field $N(x)$, can also be derived from the linear transformation of the Dirac field $\Psi(x)$, utilizing the projection with $\zeta$ and $\tilde\zeta$.
Since the Dirac field transforms linearly under the Lorentz boost,
\begin{equation}\label{eq:diraclinearlambda}
U(\Lambda)^{-1}\Psi_A(x)U(\Lambda)=D_{AB}(\Lambda)\Psi_B(\Lambda^{-1}x),
\end{equation}
Multiplying $e^{imvx}\bar\zeta^I$ on the above equation, we find that the nonlinear boost transformation of the heavy field $N_I(x)$ is
\begin{eqnarray}\label{eq:generalboostnI}
    U(\Lambda)^{-1}N_{I}(x)U(\Lambda)&=&e^{imvx}\bar\zeta^I U(\Lambda)^{-1}\Psi(x)U(\Lambda)
    \nonumber\\
    &=&e^{imvx}\bar\zeta^ID(\Lambda)\Psi(\Lambda^{-1}x)
    \nonumber\\
     &=&e^{imvx}\bar\zeta^ID(\Lambda)\left[
    e^{-imv(\Lambda^{-1}x)}\left(\zeta^J+\tilde{ \zeta}^{J}f(\vec\nabla)\right)N_J(\Lambda^{-1}x)\right]
    \nonumber\\
   &=&e^{-im(\Lambda v-v)x}\bar\zeta^{I}D(\Lambda)
        \left(\zeta^{J}+\tilde{ \zeta}^{J}f(\vec\nabla)\right)N_{J}(\Lambda^{-1}x).
\end{eqnarray}
Similarly, multiplying $-e^{imvx}\bar{\tilde{\zeta}}^I$ on the linear transformation of the Dirac field in Eq.~\eqref{eq:diraclinearlambda}, we derive the boost transformation of $\tilde N_I(x)$
\begin{eqnarray}\label{eq:generalboostnI2}
    U(\Lambda)^{-1}\tilde{N}_{I}(x)U(\Lambda)&=&-e^{imvx}\bar{\tilde{\zeta}}^I U(\Lambda)^{-1}\Psi(x)U(\Lambda)
    \nonumber\\
   &=&-e^{-im(\Lambda v-v)x}\bar{\tilde{\zeta}}^ID(\Lambda)
        \left(\zeta^{J}+\tilde{ \zeta}^{J}f(\vec\nabla)\right)N_{J}(\Lambda^{-1}x).
\end{eqnarray}
Since the $f(\vec\nabla)$ is already determined by Eq.~\eqref{eq:generalfzeta}, the  corresponding transformation can be derived, and it is denoted as $ U(B(q))^{-1}f(\vec\nabla)U(B(q))\equiv f(\vec\nabla')$, we then derive the consistency condition:
\begin{equation}\label{eq:zetacc}
   f(\vec\nabla') e^{-im(\Lambda v-v)x}\bar\zeta^{I}D(\Lambda)
        \left(\zeta^{J}+\tilde{ \zeta}^{J}f(\vec\nabla)\right)N_{J}=-e^{-im(\Lambda v-v)x}\bar{\tilde{\zeta}}^ID(\Lambda)
        \left(\zeta^{J}+\tilde{ \zeta}^{J}f(\vec\nabla)\right)N_{J}.
\end{equation}
Utilizing this condition, we can check the validity of the expression of the $f(\vec\nabla)$.


In the following we take the Dirac representation, to derive the specific expression on the $N$ field under the Lorentz boost. 
Choosing the rest frame $v^{\mu}=(1,0,0,0)$, the Gamma matrix are
\begin{equation}
          \gamma^0=\left[ \begin{array}{cc}
           1&0\\
           0 &-1
      \end{array}\right],\quad \gamma^i=\left[ \begin{array}{cc}
           0&\sigma^i\\
           -\sigma^i &0
      \end{array}\right],\quad \gamma^5=\left[ \begin{array}{cc}
           0&1\\
           1 &0
      \end{array}\right],
      \end{equation} 
and the projection operators are
    \begin{equation}
        P_+=\frac{1+v\!\!\!/}{2}=\left[\begin{array}{cc}
             1& 0 \\
             0&0 
        \end{array}\right],\quad P_-=\frac{1-v\!\!\!/}{2}=\left[\begin{array}{cc}
             0& 0 \\
             0&1 
        \end{array}\right].
    \end{equation}
In this case, the reference spinors and the  wave functions are
\begin{equation}\label{fwave}
    \zeta_A^I=
         \left[\begin{array}{l}
              1  \\
              0 
         \end{array}\right]  , ~ \tilde{ \zeta}_A^I=\left[\begin{array}{l}
              0  \\
              1 
         \end{array}\right],~
    \mathcal{U}^{\sigma}(v)=\left[\begin{array}{c}
     u_{\ell}^{\sigma}(v)  \\
     0 
\end{array}\right],~ u_{\ell}^{\sigma}(v)=\sqrt{2m}\delta_{\ell}^{\sigma}.
\end{equation}
By definition, the standard boost $L_{AB}(\vec k)$ takes the rest to the momentum $\vec k$ is
\begin{equation}\label{standarddiracboost}
    L_{AB}(\vec k)=\frac{1}{\sqrt{2m(m+E)}}\left[\begin{array}{cc}
		(E+m)\mathbf{I}_{2\times2}&\vec{\sigma}\cdot\vec{k}\\\vec{\sigma}\cdot\vec{k}&(E+m)\mathbf{I}_{2\times2}
	\end{array}\right],
\end{equation}
Thus the two-component standard boost $L_{IJ}$ is again derived from Eq.~\eqref{eq:twocomLIJ} and $L_{AB}$ as
\begin{equation}
     L_{IJ}(\vec \nabla)=\bar\zeta _A^IL_{AB}(\vec\nabla)\zeta_B^J=\sqrt{\frac{2m+i\partial_t}{2m}}\delta_{IJ}, 
\end{equation}
consequently  $f(\vec\nabla)$ is derived from Eq.~\eqref{eq:generalfzeta} as
\begin{equation}
         f(\vec\nabla)=-\bar{\tilde{\zeta}} _AL_{AB}(\vec\nabla)\zeta_B \left(\bar\zeta _{A'}L_{A'B'}(\vec\nabla)\zeta_{B'}\right)^{-1} =\frac{-i\vec\sigma\cdot\vec\nabla}{2m+i\partial_t},
\end{equation}
and they are consistent with the results obtained in subsubsection~\ref{LGT}.

Now we are ready to perform the boost transformation of the heavy field $N(x)$, and examine the consistency condition. From the relation Eq.~(\ref{psinewNrelation}),  the Dirac field $\Psi(x)$ is expanded by the heavy field $N(x)$ as
\begin{equation}\label{expansionpsinrest}
    \Psi(x)=e^{-imvx}\left[\begin{array}{c}
          N(x)  \\
         f(\vec\nabla)  N(x) 
    \end{array}\right].
\end{equation}
 Taking $\Lambda = B(q)$  with infinitesimal parameter $\frac{q^{\mu}}{m}=(0,-\frac{\vec q}{m})$ in Eq.~\eqref{eq:diraclinearlambda}, the linear transformation matrix of $\Psi(x)$ up to the first order of $q/m$ is
\begin{equation}
    D_{AB}\left(B(q)\right)=\left[\begin{array}{cc}
          \mathbf{I}_{2\times2}&\frac{\vec \sigma \cdot \vec q}{2m} \\
          \frac{\vec \sigma \cdot \vec q}{2m}&\mathbf{I}_{2\times2}
          \end{array}\right].
\end{equation}
The Dirac field $\Psi(x)$ transforms linearly, while the heavy field $N(x)$ transforms nonlinearly under the Lorentz boost.  The nonlinear boost transformation of $N(x)$ in Eq.~\eqref{eq:generalboostnI} reduce to
\begin{equation}\label{nboostpd}
U(B(q))^{-1}N(x)U(B(q))=e^{i\vec q\cdot\vec x}\left(1+\frac{\vec\sigma\cdot\vec q}{2m}f(\vec\nabla)\right)N(B(q)^{-1}x),
\end{equation}
which is exactly the same as the result in Eq.~\eqref{eq:boostofnell} in the previous subsubsection. The nonlinear boost transformation of $\tilde N(x)$ in Eq.~\eqref{eq:generalboostnI2} become
\begin{equation}\label{bqfn}
    U(B(q))^{-1}\tilde N(x)U(B(q))=e^{i\vec q\cdot\vec x}\left(\frac{\vec \sigma\cdot\vec q}{2m}+f(\vec\nabla)\right)N(B(q)^{-1}x).
\end{equation}
Besides, from the transformations of derivatives in Eq.~\eqref{bqofderiv}, we obtain $f(\vec\nabla')$
\begin{equation}
    \begin{array}{rll}
       
          U(B(q))^{-1}f(\vec\nabla)U(B(q))\equiv f(\vec\nabla')&=&\frac{-i\vec\sigma\cdot\vec\nabla+\frac{\vec\sigma\cdot\vec qi\partial_t}{m}}{2m+i\partial_t-\frac{\vec q}{m}i\vec\nabla}.
    \end{array}
\end{equation}
Due to the definition $\tilde N(x)=f(\vec\nabla)N(x)$ in Eq.~\eqref{defntilden} and using the above three transformation
\begin{equation}
\left[U(B(q))^{-1}f(\vec\nabla)U(B(q))\right]\left[U(B(q))^{-1}N(x)U(B(q))\right]=\left[U(B(q))^{-1}\tilde N(x)U(B(q))\right],  
\end{equation}
we find the exact form of the consistency condition in Eq.~\eqref{eq:zetacc}
\begin{equation}\label{eq:zetacc1}
    f(\vec\nabla')e^{i\vec q\cdot\vec x}\left(1+\frac{\vec\sigma\cdot\vec q}{2m}f(\vec\nabla)\right)N=e^{i\vec q\cdot\vec x}\left(\frac{\vec \sigma\cdot\vec q}{2m}+f(\vec\nabla)\right)N.
\end{equation}
Note that the $f(\vec\nabla)$ is also dependent on the time derivative $\partial_t$ before using the EOM of the heavy field. The above equation yields
\begin{equation}
    \begin{array}{lrll}
       &\left(f(\vec\nabla)+\frac{ i\vec q\cdot\vec\nabla }{m(2m+i\partial_t)}f(\vec\nabla)+\frac{\vec\sigma\cdot\vec q~i\partial_t}{m(2m+i\partial_t)}\right)e^{i\vec q\cdot\vec x}\left(1+\frac{\vec\sigma\cdot\vec q}{2m}f(\vec\nabla)\right)N &=& e^{i\vec q\cdot\vec x}\left(\frac{\vec \sigma\cdot\vec q}{2m}+f(\vec\nabla)\right)N 
       \\
       \\
         \Longrightarrow& \left(\frac{-i\vec\sigma\cdot\vec\nabla+\vec\sigma\cdot\vec q}{2m+i\partial_t}+\frac{ i\vec q\cdot\vec\nabla }{m(2m+i\partial_t)}\frac{-i\vec\sigma\cdot\vec\nabla}{2m+i\partial_t}+\frac{\vec\sigma\cdot\vec q~i\partial_t}{m(2m+i\partial_t)}+\frac{-i\vec\sigma\cdot\vec\nabla}{2m+i\partial_t}\frac{\vec \sigma\cdot\vec q}{2m}\frac{-i\vec\sigma\cdot\vec\nabla}{2m+i\partial_t}\right)N&=&\left(\frac{\vec \sigma\cdot\vec q}{2m}+\frac{-i\vec\sigma\cdot\vec\nabla}{2m+i\partial_t}\right)N
         \\
         \\
         \Longrightarrow&\left(2\vec q\cdot\vec\nabla \vec\sigma\cdot\vec\nabla+(\vec\sigma\cdot\vec q~i\partial_t)(2m+i\partial_t)-(\vec\sigma\cdot\vec\nabla\vec \sigma\cdot\vec q\vec\sigma\cdot\vec\nabla)\right)N&=&0.
    \end{array}
\end{equation}
Utilizing the identity of the Pauli matrix $
\sigma^i\sigma^j\sigma^k=\sigma^i\delta^{jk}+\delta^{ij}\sigma^k-\delta^{ij}\sigma^j+i\epsilon^{ijk},
$
the consistency condition reduces to
\begin{equation}\label{eq:cceomfreen}
    \begin{array}{llll}
         &\left(2\vec q\cdot\vec\nabla \vec\sigma\cdot\vec\nabla+(\vec\sigma\cdot\vec q~i\partial_t)(2m+i\partial_t)-(2\vec q\cdot\vec\nabla\vec\sigma\cdot\vec\nabla-\vec\sigma\cdot\vec q~\nabla^2)\right)N&=&0 \\
         \\
        \Longrightarrow & i\partial_tN(x)=-\frac{\nabla^2}{2m+i\partial_t}N(x).
    \end{array}
\end{equation}
Due to the on-shell condition Eq.~\eqref{eq:freeNeom}
\begin{equation}
    (m+i\partial_t)^2N(x)=(m^2-\nabla^2)N(x),
\end{equation}
we find that the consistency condition Eq.~\eqref{eq:cceomfreen}  is exactly the EOM of the free heavy field $N(x)$.
Although this condition is always satisfied for the free field, it non-trivially constrains the expression of $f(\vec\nabla)\rightarrow f(D_{\mu})$ in the gauge interaction case, as we will discuss in subsection~\ref{GECC}.

The heavy field and the anti-particle d.o.f. can be combined as the Dirac field $\Psi(x)$ that transforms linearly under the Lorentz boost. Note that, the upper and lower components of the Dirac spinor field given in Eq.~\eqref{uplowdiracfield} are
\begin{equation}\label{eq:ntildenrest}
\left\{
\begin{array}{lllll}
   \zeta^IN_I(x)&=&\frac{1+v\!\!\!/}{2}e^{imvx}\Psi(x)&=&\left[\begin{array}{c}
        N(x)  \\
        0 
   \end{array}\right],
    \\
    \\
 \tilde{ \zeta}^I\tilde N_I(x)&=&\frac{1-v\!\!\!/}{2}e^{imvx}\Psi(x)&=&\left[\begin{array}{c}
        0 \\
        \tilde N(x) 
   \end{array}\right].
\end{array}\right.
\end{equation}
Since $\zeta$ and $\tilde\zeta$ are not dynamical and don't transform, thus $\zeta^IN_I$ and $\tilde\zeta^I \tilde N_I$ transform under the boost $B(q)$ similar to Eq.~\eqref{nboostpd} and Eq.~\eqref{bqfn} as
\begin{equation}
\left\{
\begin{array}{lllll}
   U(B(q))^{-1}\left[\begin{array}{c}
        N(x)  \\
        0 
   \end{array}\right] U(B(q))&=&\left[\begin{array}{c}
        e^{i\vec q\cdot\vec x}\left(1+\frac{\vec\sigma\cdot\vec q}{2m}f(\vec\nabla)\right)N(B(q)^{-1}x)  \\
        0 
   \end{array}\right],
    \\
    \\
  U(B(q))^{-1}\left[\begin{array}{c}
        0 \\
        \tilde N(x) 
   \end{array}\right] U(B(q))&=&\left[\begin{array}{c}
        0 \\
        
e^{i\vec q\cdot\vec x}\left(\frac{\vec \sigma\cdot\vec q}{2m}+f(\vec\nabla)\right)N(B(q)^{-1}x)
 
   \end{array}\right].
\end{array}\right.
\end{equation}
Utilizing these transformations and the following relation
\begin{equation}
    e^{-imvx}e^{i\vec q\cdot\vec x}=e^{-i(mv-q)x}=e^{-im(B(q)v)x}=e^{-imv(B(q)^{-1}x)},
\end{equation}
we can reproduce the linear transformation of the Dirac field $ \Psi(x)=e^{-imvx}\left[\begin{array}{c}
          N(x)  \\
         f(\vec\nabla)  N(x) 
    \end{array}\right]$ since:
\begin{equation}
\begin{array}{rll}
    e^{-imvx}U(B(q))^{-1}\left[\begin{array}{c}
         N(x)  \\
         f(\vec\nabla)  N(x) 
    \end{array}\right]U(B(q))&=&e^{-imv(B(q)^{-1}x)}\left[\begin{array}{cc}
          1&\frac{\vec \sigma \cdot \vec q}{2m} \\
          \frac{\vec \sigma \cdot \vec q}{2m}&1
          \end{array}\right]\left[\begin{array}{c}
          N(B(q)^{-1}x)  \\
         f(\vec\nabla) N(B(q)^{-1}x) 
    \end{array}\right],
    \\
    \\
    \Longrightarrow\quad U(B(q))^{-1}\Psi_A(x) U(B(q))&=&D_{AB}\left(B(q)\right)\Psi_B(B(q)^{-1}x).
    \end{array}
\end{equation}

\begin{table}
    \centering\scriptsize
    \begin{tabular}{|c|c|}
    \hline
          Relativistic &Heavy  \\
       \hline
       $
       \begin{array}{c}
            \textbf{State} 
            \\|\vec p,\sigma\rangle\equiv a_{\vec p}^{\sigma\dagger}|0\rangle  \\
            \\
            \\
            U(\Lambda  )^{-1}a_{\vec p}^{\sigma}U(\Lambda  )=D_{\sigma\sigma'}^{(s)}(W[\Lambda  ,p])a_{\vec p'}^{\sigma'}
       \end{array}
       $
          &
          $\begin{array}{c}
         \textbf{State}
         \\
              \begin{array}{llll}
                   |v,\vec k,\sigma\rangle&=& U(L(\vec k))&|v,0,\sigma\rangle    \\
                   |v,\vec k,\sigma\rangle\equiv a_{v,\vec k}^{\sigma\dagger}|\Omega\rangle &&&|v,0,\sigma\rangle\equiv a_{v,0}^{\sigma\dagger}|\Omega\rangle 
              \end{array} 
              \\
             \\
                    
                   a_{v,\vec k}^{\sigma\dagger}= U(L(\vec k))a_{v,0}^{\sigma\dagger}U(L(\vec k))^{-1}

         \end{array}$
         \\
         \hline 
      $\begin{array}{c}
         \textbf{Wave function}
         \\
         \mathcal{U}_A^{\sigma}(v) =\zeta_A^{I}u_I^{\sigma}(v)
         \\
         \mathcal{U}_A^{\sigma}(p)=L_{AB}(\vec k)\mathcal{U}_B^{\sigma}(v)
        
         \\
         \\
        \textbf{Constraint} 
                      \\
                      \sum_{\sigma}\mathcal{U}_{A}^{\sigma}(v)D^{(s)}_{\sigma\sigma^{\prime}}(W[\Lambda,p])
                      =D_{AB}(W[\Lambda,p])~\mathcal{U}_{B}^{\sigma}(v)
         \\
         \end{array}
         $& $\begin{array}{c}
             \textbf{Wave function}
             \\
             u_I^{\sigma}(v)=\sqrt{2m}~\delta_I^{\sigma}
             \\
             u_I^{\sigma}(k)=L_{IJ}(\vec k)u_J^{\sigma}(v)  \\
          
               \\

                      \textbf{Constraint} 
                      \\
                      \sum_{\sigma}u_{I}^{\sigma}(v)D^{(s)}_{\sigma\sigma^{\prime}}(W[\Lambda,p])
                      =D_{IJ}(W[\Lambda,p])~u_{J}^{\sigma}(v)

         \end{array}$
         \\
         \hline

         $\begin{array}{c}
         \textbf{Field at Rest}
         \\
             \Psi^0_A(x)= \int\frac{d^3\vec p}{(2\pi)^3}\frac{1}{\sqrt{2E}}\sum_{\sigma}a^{\sigma}_{\vec p}~\mathcal{U}_A^{\sigma}(v)e^{-ipx} \\
             \\
             \textbf{Boost Transformation}
             \\
            U(B(q))^{-1}\Psi^0_A(x)U(B(q))=D_{AB}(W[B(q),i\partial])\Psi^0_B(B(q)^{-1}x)  \\
            \\
            
            U(B(q))^{-1}L_{AB}(\partial)U(B(q))=L_{AB}(B(q)\partial)
            \\
            \\
        
            \textbf{Field}
        \\    
            \Psi_A(x)= \int\frac{d^3\vec p}{(2\pi)^3}\frac{1}{\sqrt{2E}}\sum_{\sigma}a^{\sigma}_{\vec p}~\mathcal{U}_A^{\sigma}(p)e^{-ipx}
            \\
            \\
            \textbf{Boost Transformation}
            \\
            U(B(q))^{-1}\Psi_A(x)U(B(q))=D_{AB}(\Lambda)\Psi_B(B(q)^{-1}x)
             \\
             \\
             \text{where}
             \\D_{AB}(B(q))=\left[L_{AA'}(B(q)\partial)\right]\left[D_{A'B'}(W[B(q),i\partial])\right]\left[L^{-1}_{B'B}(\partial)\right]
             \\
             \\
         \end{array}$&
         $\begin{array}{c}
              \textbf{Field at Rest}  \\
                N^0_{I}(x)=\int\frac{d^3k}{(2\pi)^3\sqrt{2E}}\sum_{\sigma}u_I^{\sigma}(v)a_{v,k}^{\sigma}e^{-ikx}\\
              \\
              \textbf{Boost Transformation}
              \\
            U(B(q))^{-1} N^0_{I}(x)U(B(q))
             
              =e^{i\vec q\cdot\vec x}D_{IJ}(W[B(q),i\vec\nabla]) N^0_{J}(B(q)^{-1}x)
              \\
    \\
     U(B(q))^{-1} L_{IJ}(\vec\nabla)U(B(q))=\left( 1 - \frac{i\vec \nabla}{2(2m+ i\partial_t)} \cdot \frac{\vec q}{m}\right) L_{IJ}(\nabla)
  \\
  \\
              \textbf{Field}
              \\
              N_I(x)=\int\frac{d^3k}{(2\pi)^3\sqrt{2E}}\sum_\sigma u_I^{\sigma}(k)a_{v,k}^{\sigma}e^{-ikx}
              \\
              \\
              \textbf{Boost Transformation}
              \\
              U(B(q))^{-1}N(x)U(B(q))=e^{i\vec q\cdot\vec x}\left(1+\frac{\vec\sigma\cdot\vec q}{2m}f(\vec\nabla)\right)N(B(q)^{-1}x),
              \\
              \\
              \text{where}
              \\
            U(B(q))^{-1}N_I(x)U(B(q))
               \\
               =\left[U(B(q))^{-1} L(\vec\nabla)_{IJ}U(B(q)) \right]\left[  U(B(q))^{-1} N^0_J(x)U(B(q))\right]
\end{array}$
          \\
         \hline
         \multicolumn{2}{|c|}{ \textbf{Connection}~\quad$\begin{array}{rcl}

         a_{\vec p}^{\sigma\dagger}&\leftrightarrow&a_{v,\vec k}^{\sigma\dagger}
         \\
         \\
              \mathcal{U}_A^{\sigma}(p)&=&L_{AB}(\vec k)\zeta_B^{I}u_I^{\sigma}(v)  \\
              \\
               \Psi_A(x) &=& e^{-imvx}L_{AB}(\vec\nabla)\zeta_B^I N^0_I(x)
                \\
                \\
                \Psi_A(x)&=&e^{-imvx}\left(\zeta_A^I+\tilde{ \zeta}_A^{I}f(\vec\nabla)\right)N_I(x)
         \end{array}$} 
         \\
         \hline
       
    \end{tabular}
    \caption{The constructive HPET versus  the Relativistic Theory.}\label{table1}
  
\end{table}

\subsection{Comparison with Traditional Methods}
\label{review on HPET}

Having obtained the heavy field from the nonlinear boost in the bottom-up way, in this subsection, for comparison we review two traditional treatments on obtaining the heavy field from the relativistic Dirac field in the top-down approach: the method of integrating out , and the FW transformation.

\subsubsection{Integrating out at Tree-level}\label{ioapdof}

The Dirac spinor field is written as
\begin{equation}\label{psiexpan}
      \Psi(x)=\int\frac{d^3p}{(2\pi)^3}\frac{1}{\sqrt{2E}}\sum_{\sigma}\left(a^{\sigma}_{\vec p}~\mathcal{U}^{\sigma}(p)e^{-ipx}+b_{\vec p}^{\sigma\dagger}~\mathcal{V}^{\sigma}(p)e^{ipx}\right).
  \end{equation} 
Here $a_{\vec p}^{s\dagger}$ ($b_{\vec p}^{s\dagger}$) and $\mathcal{U^{\sigma}}$ ($\mathcal{V}^{\sigma}$) are the particle (anti-particle) creation operators and the particle (anti-particle) wave functions, respectively. However, the anti-particle operator $b_{\vec p}^{s\dagger}$ is absent in this effective theory. 
For the relativistic spinor field $\Psi(x)$, it can be expanded by two kinds of modes labeled by their velocities $v^{\mu}$ as
\begin{equation}
\label{eq:projectv}
   \Psi(x) =\sum_{v}e^{-imv\cdot x}(Q_v(x)+B_v(x)),
\end{equation}
where the particle field $Q_v(x)$ and anti-particle field $B_v(x)$ are defined by
\begin{equation}
    Q_v(x)=e^{i\mathcal{P}x}P_+\Psi(x)=e^{imvx}\frac{1+v\!\!\!/}{2}\Psi(x),
\end{equation}
\begin{equation}\label{eq:bvdef}
     B_v(x)=e^{i\mathcal{P}x}P_-\Psi(x)=e^{imvx}\frac{1-v\!\!\!/}{2}\Psi(x),
 \end{equation}
where $\mathcal{P}$ is the label momentum operator and $P_{\pm}=\frac{1\pm v\!\!\!/}{2}$ are the projection operators. 
The phase factor $e^{imvx}$ ensures derivatives have definite power counting while $\frac{1\pm v\!\!\!/}{2}$ divides the d.o.f. in the Lorentz spinor space. 
%
Note that in the bottom-up construction of the HPET in subsubsection~\ref{LGT}, the heavy field is defined to describe excitations within a specific momentum region centered around a reference velocity $v^{\mu}$. This is in contrast to the traditionally defined relativistic Dirac field, which encompasses all possible momentum states. Consequently, the summation over $v^{\mu}$ in Eq.~\eqref{eq:projectv} formally represents the decomposition of the full relativistic theory into distinct, disjoint velocity sectors. In practical applications, however, one typically works within a single sector at a time, a principle known as the velocity superselection rule~\cite{Georgi:1990um}.
%
Above procedure is also known as the label formalism~\cite{Bauer:2000yr,Bauer:2001yt,Manohar:2006nz,Becher:2014oda}.
Consider the mode expansion 
\begin{equation}
    Q_v(x) = \int\frac{d^3k}{(2\pi)^3}\frac{1}{\sqrt{2E}}\sum_{\sigma}\left\{a^{\sigma}_{\vec p}~\frac{1+v\!\!\!/}{2}\mathcal{U}^{\sigma}(p)e^{-ikx}+b^{\sigma\dagger}_{\vec p}~\frac{1+v\!\!\!/}{2}\mathcal{V}^{\sigma}(p)e^{i(2mv+k)x}\right\} ,
\end{equation}
and
\begin{equation}
   B_v(x) = \int\frac{d^3k}{(2\pi)^3}\frac{1}{\sqrt{2E}}\sum_{\sigma}\left\{a^{\sigma}_{\vec p}~\frac{1-v\!\!\!/}{2}\mathcal{U}^{\sigma}(p)e^{-ikx}+b^{\sigma\dagger}_{\vec p}~\frac{1-v\!\!\!/}{2}\mathcal{V}^{\sigma}(p)e^{i(2mv+k)x}\right\} .
\end{equation}
Note that $k^{\mu}$ is the quantized momentum fluctuation, and every derivative $iD^{\mu}$ on the field corresponds to $k^{\mu}$ in the momentum space.  
The $B_v$ field should be integrated out since it contains the high-frequency modes which is not relevant to the effective theory.

Traditionally, in a chosen frame, the heavy component $B_v(x)$ projected by the frame velocity $v^{\mu}$ should be integrated out and this theory become manifestly non-relativistic. 
Consider the free Dirac field Lagrangian,
\begin{equation}
\mathcal{L}=\bar{\Psi}(x)(i\partial\!\!\!/-m)\Psi(x),
\end{equation}
using  $\Psi(x)=e^{-imvx}(Q_{v}(x)+B_{v}(x))$ to expand it as
\begin{eqnarray}
       \mathcal{L} = \bar{Q}_{v}iv\cdot \partial Q_{v}+\bar{Q}_{v}i\partial\!\!\!/_{\perp}B_{v}+\bar{B}_{v}i\partial\!\!\!/_{\perp}Q_{v}+\bar{B}_{v}(-iv\cdot \partial -2m)B_{v}, 
\end{eqnarray}
where we have used $v\!\!\!/Q_{v}=Q_{v}, v\!\!\!/B_{v}=-B_{v}$.
The derivatives are  projected,
$\partial_{\perp}^{\mu}=\partial^{\mu}-v^{\mu}v\cdot \partial$ is introduced such that $v\cdot \partial_{\perp}=0$. In the Lagrangian, the mass terms of field $Q_{v}$ is absent while $B_{v}$ has $2m$ mass term. When the interaction is small enough comparing to the mass $m$, $B_{v}$ is so heavy that it can be integrated out,  leaving only the light component $Q_{v}$.
At the tree-level, the EOM $\partial\mathcal{L}/\partial\bar{B}_{v}=0$ is utilized to perform such integrating out, and thus one obtain
\begin{equation}\label{eq:bvex}
   B_{v}=(iv\cdot \partial+2m)^{-1}i\partial\!\!\!/_{\perp}Q_{v}.
\end{equation}
Integrating out $B_{v}$ leads to the effective Lagrangian with only $Q_v$,
\begin{equation}
\mathcal{L}=\bar{Q}_{v}iv\cdot \partial Q_{v}+\bar{Q}_{v}i\partial\!\!\!/_{\perp}(iv\cdot \partial+2m)^{-1}i\partial\!\!\!/_{\perp}Q_{v}.
\end{equation}
This non-local propagator  can be expanded as a sequence of local operators suppressed by $m$, then the Lagrangian with the implicit Lorentz invariance are totally determined. More pedagogical derivation can be found in Ref.~\cite{Manohar:2000dt}, and functional methods are also developed in Ref.~\cite{Cohen:2019btp}.

Due to the underlying Lorentz symmetry and the independence of $v^{\mu}$, fields should transform properly under the  reparameterization~\cite{Luke:1992cs}. 
In practice, we set $\epsilon^{\mu}=\frac{q^{\mu}}{m}$ as an infinitesimal parameter with $m$ in the denominator to preserve power counting. The condition $v^2=1$ requires $(v+\frac{q}{m})^2=1$, or $v\cdot q=0$ to the leading order in $\frac{q^{\mu}}{m}$.
After the projection Eq.~($\ref{eq:projectv}$), $Q_v(x)$ and $B_v(x)$ transform under the reparameterization $v^{\mu}\rightarrow v^{\mu}+\frac{q^{\mu}}{m}$ as
\begin{equation}\label{QVRPI}
\left\{
\begin{array}{ll}
      Q_v&\rightarrow Q_{v+\frac{q}{m}}=e^{iqx}Q_v+e^{iqx}\frac{q\!\!\!/}{2m}(Q_v+B_v),\\
     \\
     B_v&\rightarrow B_{v+\frac{q}{m}}=e^{iqx}B_v-e^{iqx}\frac{q\!\!\!/}{2m}(Q_v+B_v).
\end{array}
  \right.
\end{equation}
Using the reparameterization of $Q_v$ requires the exact form of $B_v$, such as the tree-level integrating out
result in Eq.~\eqref{eq:bvex}. 
The integration should hold after the reparameterization in a general frame.  For  $v^{\mu}\rightarrow v^{\mu}+\frac{q^{\mu}}{m}$ up to $\frac{1}{m^3}$, LHS of ($\ref{eq:bvex}$) becomes
\begin{equation}
    B_v\rightarrow e^{iqx}B_v-e^{iqx}\frac{q\!\!\!/}{2m}\left(Q_v+(\frac{i\partial\!\!\!/_{\perp}}{2m}-\frac{iv\cdot \partial i\partial\!\!\!/_{\perp}}{4m^2})Q_v\right),
\end{equation}
while RHS of ($\ref{eq:bvex}$) becomes
\[(iv\cdot \partial+2m)^{-1}(i\partial\!\!\!/_{\perp})Q_v\rightarrow e^{iqx}B_v+e^{iqx}\left(-\frac{q\!\!\!/}{2m}+\frac{i\partial\!\!\!/_{\perp}q\!\!\!/}{4m^2} -\frac{v\!\!\!/iq\cdot \partial}{2m^2}\right.\]
\begin{equation}
   \left.-\frac{q\!\!\!/iv\cdot \partial}{4m^2}-\frac{iv\cdot \partial i\partial\!\!\!/_{\perp}q\!\!\!/}{8m^3}+\frac{i\partial\!\!\!\!/_{\perp}q\!\!\!/i\partial\!\!\!/_{\perp}}{8m^3}+\frac{iv\cdot \partial iv\cdot \partial q\!\!\!/}{8m^3}-\frac{iq\cdot \partial i\partial\!\!\!/_{\perp}}{4m^3}+\frac{iv\cdot \partial v\!\!\!/iq\cdot \partial}{4m^3}\right)Q_v.
\end{equation}
After tedious calculation, using the EOM 
$
    iv\cdot \partial Q_v=\frac{\partial^2_{\perp}}{2m}Q_v,
$
one finds that the above two equations are equal up to $\frac{1}{m^3}$.
However, there still exists a problem when finding the expression of $B_v$ in terms of $Q_v$. 
Traditionally, the free Dirac equation is used when performing reparameterization, considering only the kinematic effects. This works perfectly in the momentum space when dealing with the wave functions or amplitudes~\cite{Aoude:2020onz}, but is insufficient at the operator level, especially when the gauge fields are considered.  
%
Previous bottom-up methods have difficulty determining $B_v=B_v(Q_v)$ beyond kinematic contributions, such as the interactions with external fields.

In the rest frame, taking the Dirac representation, the $Q_v$ and $B_v$ are greatly simplified. In this case the relativistic field $\Psi(x)$ is separated into the upper and lower components as in Eq.~\eqref{eq:ntildenrest}
 \begin{equation}
      \Psi(x)=e^{-imt}\left\{\left[ \begin{array}{c}
           N(x)\\
           0 
      \end{array}\right]+\left[ \begin{array}{c}
           0\\
           \tilde N(x)
      \end{array}\right]\right\},
 \end{equation}
where
\begin{equation}\label{eq:restqvbvnchi}
    Q_{v}(x)=e^{imt}P_+\Psi(x)=\left[ \begin{array}{c}
           N(x)\\
           0 
      \end{array}\right],\quad B_{v}(x)=e^{imt}P_-\Psi(x)=\left[ \begin{array}{c}
           0\\
           \tilde N(x)
      \end{array}\right].
\end{equation} 
Here $N(x)$ and $\tilde N(x)$ are the two-component  heavy fields, and the $SU(2)$ index $I$ is suppressed for simplicity. Similarly, all four-component quantities in the general frame $v^{\mu}$ can be written in the rest frame. For example, 
 \begin{equation}
 \begin{array}{lll}
  \bar Q_{v}\gamma^0Q_{v}&=&N^{\dagger}(x)N(x),  \\
   \bar Q_{v}\gamma^iQ_{v}&=&0 , \\
       \bar Q_{v}\gamma^0\gamma^iQ_{v}&=&0 , \\
    \bar Q_{v}\gamma^i\gamma^jQ_{v}&=&-N^{\dagger}(x)\sigma^i\sigma^jN(x).
 \end{array}
 \end{equation}
As for the vector-like derivative $\partial^{\mu}$,  it can be divided into the parallel part and the perpendicular part as
 \begin{equation}
      \partial^{\mu}=v^{\mu}(v\cdot \partial)+\partial_{\perp}^{\mu}
       =(\partial_t,\vec 0)+(0,-\vec \nabla).
 \end{equation}
In the rest frame $v^{\mu}=(1,0,0,0)$,  one can verify that the operation becomes
\begin{equation}
    (iv\cdot \partial+2m)^{-1}(i\partial\!\!\!/_{\perp})\longrightarrow f(\vec\nabla)=\frac{-i\vec\sigma\cdot\vec\nabla}{2m+i\partial_t},
\end{equation}
for the $P_-$ component.  Then the EOM in Eq.~\eqref{eq:bvex} reduces to
\begin{equation}\label{treefd}
    \tilde N=f(\vec\nabla)N.
\end{equation}
 Therefore, the $f(\vec \nabla)$ also relates to the d.o.f. that are integrated out.

\subsubsection{The FW Transformation and NR Reduction}
\label{FWtransformation}

The nonlinear boost we obtained 
for the NR fields originates from the Lorentz symmetry in the bottom-up approach. However, this result coincides with the FW transformation~\cite{Foldy:1949wa, Foldy:1956kvk} and the NR reduction in a top-down view, providing a new interpretation for these conventional methods.

\paragraph{The FW Transformation}
The FW transformation is a unitary transformation $U_{\text{FW}}$ that block-diagonalizes the Dirac Hamiltonian, thereby decoupling the particle and anti-particle components. The transformed Hamiltonian is then expanded in powers of $1/m$, where $m$ is the heavy particle mass. In this representation, the low-energy sector is described by the upper two components of the four-component FW spinor, $\Psi_{\text{FW}}$,  which is related to the original Dirac spinor $\Psi$ by $\Psi_{\text{FW}} = U_{\text{FW}} \Psi$.

For the general cases with external electromagnetic fields, the FW transformation is derived through a series of successive unitary transformations (see~\cite{Eriksen:1958zz,Korner:1991kf,Balk:1993ev,Silenko:2003kg,Silenko:2007wi,Long:2010kt} for further discussions), while we use our generally constructed transformation in the following subsection instead. For the free field, with the matrix
\begin{equation}
     \alpha^i\equiv\left[\begin{array}{cc}
         0&\sigma^i  \\
         \sigma^i &0
    \end{array}\right],\quad\beta\equiv\left[\begin{array}{cc}
         \mathbf{I}_{2\times2}&0  \\
         0&-\mathbf{I}_{2\times2} 
    \end{array}\right],
\end{equation}
the Hamiltonian of the free Dirac field is 
\begin{eqnarray}
    H=\vec\alpha+\beta m=\left[\begin{array}{cc}
        m~\mathbf{I}_{2\times2} & \vec\sigma\cdot\vec k \\
       \vec\sigma\cdot\vec k   & -m~\mathbf{I}_{2\times2}
    \end{array}\right].
\end{eqnarray}
Then the FW transformation in Refs.~\cite{Foldy:1949wa, Foldy:1956kvk} 
\begin{equation}\label{ufw}
    U_{\text{FW}}=\frac{1}{\sqrt{2E(m+E)}}\left[\begin{array}{cc}
        (m+E)\mathbf{I}_{2\times2} & -i\vec\sigma\cdot\vec \nabla \\
        i\vec\sigma\cdot\vec \nabla & (m+E)\mathbf{I}_{2\times2}
    \end{array}\right],
\end{equation}
with $E\equiv\sqrt{m^2-\nabla^2}$ in the coordinate space, can diagonalize the free Dirac Hamiltonian
\begin{equation}
    H'=U_{\text{FW}}~H~U_{\text{FW}}^{\dagger}=\beta E=\left[\begin{array}{cc}
        E~\mathbf{I}_{2\times2} & 0\\
     0 & -E~\mathbf{I}_{2\times2}
    \end{array}\right].
\end{equation}

Below,  we derive the FW transformation Eq.~\eqref{ufw} through the connection between the heavy field and the Dirac field discussed in subsubsection~\ref{subsec:heavy-dirac}. Recall that in Eq.~\eqref{psiNrelation} we obtain the relation between the Dirac field $\Psi_A(x)$ and the heavy field $N^0_I(x)$, where $A$ is the Lorentz index and $I$ is the $SU(2)$ little group index,
\begin{equation}\label{diracfwexample}
     \Psi_A(x)= e^{-imvx}L_{AB}(\vec\nabla)\zeta_B^I N^0_I(x),
\end{equation}
where $\zeta$ is the reference spinor that is determined by the $v^{\mu}$. In the rest frame, we find that
\begin{equation}
  \zeta^I_BN_I^0(x)= \left[\begin{array}{c}
         N_I^0(x)  \\
         0 
    \end{array}\right]=\left[\begin{array}{c}
         \mathbf{I}_{2\times2}  \\
         0 
    \end{array}\right]N_I^0(x),
\end{equation}
while the standard boost $L_{AB}(\vec\nabla)$ derived in Eq.~\eqref{standarddiracboost} is
\begin{equation}
    L_{AB}(\vec \nabla)=\frac{1}{\sqrt{2m(m+E)}}\left[\begin{array}{cc}
		(E+m)\mathbf{I}_{2\times2}&\vec{\sigma}\cdot\vec{k}\\\vec{\sigma}\cdot\vec{k}&(E+m)\mathbf{I}_{2\times2}
	\end{array}\right],
\end{equation}
As a result, the Dirac field in Eq.~\eqref{diracfwexample} suppressing the Lorentz and little group indices reads
\begin{equation}\label{diracfw}
    \Psi(x)= e^{-imt}L(\vec\nabla)\left[\begin{array}{c}
         N^0(x)  \\
         0 
    \end{array}\right]=e^{-imt}\left[\begin{array}{c}
  	\sqrt{\frac{m+E}{2m}}\\\frac{-i\vec{\sigma}\cdot\vec{\nabla}}{\sqrt{2m(m+E)}}
  \end{array}\right]N^0(x),
\end{equation}
and then based on our procedure to derive the boost for the Dirac field $\Psi(x)$, we find that a map is suggested by Eq.~\eqref{diracfw}   
\begin{equation}\label{psinprime}
    \Psi(x)=e^{-imt}\sqrt{\frac{E}{m}}F(\vec\nabla)N^0(x),  \quad\text{where}\quad F(\vec \nabla)=\sqrt{\frac{m}{E}}L(\vec\nabla)\left[\begin{array}{c}
  		\mathbf{I}_{2\times2}\\0
  	\end{array}\right]=
  \left[\begin{array}{c}
  	\sqrt{\frac{m+E}{2E}}\\\frac{-i\vec{\sigma}\cdot\vec{\nabla}}{\sqrt{2E(m+E)}}
  \end{array}\right].
\end{equation}
On the other hand, defining the upper two-component NR field $N^{\prime}_I(x)$  as 
\begin{equation}\label{Nprime}
	N^{\prime}_I(x)\equiv\sqrt{\frac{E}{m}}N^0_I(x)=\int\frac{d^3k}{(2\pi)^3}\sum_{\sigma}\delta_I^{\sigma}a_{v, k}^{\sigma}e^{-ikx},
\end{equation}
such that $N_I^{\prime}(x)$ is merely a Fourier transformation of the creation-annihilation operator $a_{v,k}^{\sigma}$, with the same $SU(2)$ little group index $I=1,2$ as the field $N^0_I(x)$.
Then, we denote $\Psi_{\text{FW}}$ as
\begin{equation}\label{psifwnprime}
    \Psi_{\text{FW}}(x)\equiv e^{-imt}\left[\begin{array}{c}
         N_I'(x)  \\
         0 
    \end{array}\right]=e^{-imt}\left[\begin{array}{c}
         \mathbf{I}_{2\times2}  \\
         0 
    \end{array}\right]N_I'(x),
\end{equation}
where the phase factor $e^{-imt}$ remove the heavy mass from the energy. 
Combining Eq.~\eqref{psifwnprime} and Eq.~\eqref{psinprime}, we  reproduce the FW transformation
\begin{equation}
    \Psi=U_{\text{FW}}^{\dagger}\Psi_{\text{FW}},\quad\text{where}\quad U_{\text{FW}}^{\dagger}\left[\begin{array}{c}
  		\mathbf{I}_{2\times2}\\0
  	\end{array}\right]=F(\vec\nabla),
\end{equation}
and the $U_{\text{FW}}$  is the same as Eq.~\eqref{ufw}.

\paragraph{NR Reduction}
Starting from the Dirac field $\Psi(x)$, we can use the expansion in Eq.~(\ref{psinprime}) to maintain the NR d.o.f. such as $N'(x)$. This is called the NR reduction~\cite{Reinhard:1989zi,Epelbaum:2001fm,Cirelli:2013ufw, Girlanda:2010ya, Xiao:2018jot}. In this procedure, the relativistic field without the negative frequency part is
\begin{equation}
      \Psi(x)=\int\frac{d^3p}{(2\pi)^3}\frac{1}{\sqrt{2E}}\sum_{\sigma}a^{\sigma}_{\vec p}~\mathcal{U}^{\sigma}(p)e^{-ipx},
  \end{equation} 
while the NR field in this method is written by
\begin{equation}
	N^{\prime}_I(x)=\int\frac{d^3k}{(2\pi)^3}\delta_I^{\sigma}a_{v,k}^{\sigma}e^{-ikx}.
\end{equation}
Here we recognize that $a_{v,k}^{\sigma}=a_{\vec p}^{\sigma}$. Note that in the rest frame, $e^{-ipx}=e^{-imt}e^{-ikx}$. Thus we can expand the relativistic field in terms of the NR one as
\begin{equation}
	\Psi(x)=e^{-imt}\sqrt{\frac{m}{\sqrt{m^2-\nabla^2}}}L(-i\vec\nabla)\left[\begin{array}{c}
		N^{\prime}(x)\\0
	\end{array}\right]=e^{-imt}\left[\begin{array}{c}
		\sqrt{\frac{m+\sqrt{m^2-\nabla^2}}{2\sqrt{m^2-\nabla^2}}}N^{\prime}(x)\\\frac{-i\vec{\sigma}\cdot\vec{\nabla}}{\sqrt{2\sqrt{m^2-\nabla^2}(m+\sqrt{m^2-\nabla^2})}}N^{\prime}(x)
	\end{array}\right],
\end{equation}
where we have used $E=\sqrt{m^2-\nabla^2}$. Up to $\frac{1}{m^4}$, this equals
\begin{equation}\label{nrreduction}
	\Psi(x)=e^{-imt}\left[\left(\begin{array}{c}
		1\\0
	\end{array}\right)-\frac{i}{2m}\left(\begin{array}{c}
	0\\\vec{\sigma}\cdot\vec{\nabla}
\end{array}\right)+\frac{1}{8m^2}\left(\begin{array}{c}
\nabla^2\\0
\end{array}\right)-\frac{3i}{16m^3}\left(\begin{array}{c}
0\\\vec{\sigma}\cdot\vec{\nabla}\nabla^2
\end{array}\right)+\frac{11}{128m^4}\left(\begin{array}{c}
\nabla^4\\0
\end{array}\right)\right]N^{\prime}(x).
\end{equation}
Thus the NR reduction formula is obtained. Note that we have extracted the phase factor $e^{-imt}$ for the correct EOM of $N^{\prime}(x)$.

Both the FW transformation and the NR reduction suggest that for the two-component spinor $N^0(x)$ satisfying the free EOM $(m+i\partial_t)N^0(x)=\sqrt{m^2-\nabla^2}N^0(x)$, there exists the Lorentz boost leaving this EOM invariant, and the generator is written by 
  \begin{equation}\label{fwk}
  	\vec{K}=\frac{i{\frac{\vec\sigma}{2}}\times\vec{\nabla}}{m+\sqrt{m^2-\nabla^2}},
  \end{equation}
 which is the same as the boost generator  in Eq.~\eqref{boostcord}. The heavy field expansion for the relativistic Dirac field $\Psi(x)$ is
 \begin{equation}
     \Psi(x)=e^{-imt}\sqrt{\frac{m}{\sqrt{m^2-\nabla^2}}}\left[\begin{array}{c}
          L_{IJ}(\vec\nabla)N_J^{\prime}(x)  \\
          f(\vec\nabla)L_{IJ}(\vec\nabla)N_J^{\prime}(x) 
     \end{array}\right]=e^{-imt}\left[\begin{array}{c}
          N_I(x)  \\
          f(\vec\nabla)N_I(x)
     \end{array}\right],
 \end{equation}
with the same $f(\vec\nabla)$ and $L_{IJ}(\vec\nabla)$ given in subsection \ref{HEAVYFN}.  Furthermore, we note that the tree-level integrating out in Eq.~(\ref{treefd}) leads to the same result as the FW transformation and the NR reduction, due to the same role that $f(\vec\nabla)$ plays.

\subsection{Covariant Nonlinear Boost}\label{GECC}
Now we consider the gauge interaction. The nonlinear boost transformation is derivative-dependent. Due to gauge covariance, all derivatives $\partial^{\mu}$ should be replaced by covariant derivatives $D^{\mu}$ in a gauge theory, as $f(\vec\nabla)\longrightarrow f(\vec D)$, thus the nonlinear boost will depend on the gauge field strength.

\subsubsection{Consistence Condition for $f(D)$}

For the FW transformation, it is difficult to find an analytic transformation $F(\vec\nabla)$. For the NR reduction, there are more than one possibility to expand the relativistic field. In fact, only the kinematic information of the expansion is determined, while how to include the commutator $[D_{\mu},D_{\nu}]$ in Eq.~(\ref{restntildebost}) and (\ref{psinprime}) is not determined.  
As a result, neither the NR reduction nor the free field boost transformation or reparameterization works successfully when gauge fields are involved, since the requirement of covariant derivatives $D_{\mu}$ introduces the problem of covariant derivative commutators.

To address the problem associated with gauge fields and to find a general method constraining the Wilson coefficients of NR operators, we propose a general expansion in the field content,  to obtain the nonlinear boost and the $f(\frac{D^{\mu}}{m})$ order by order, where $1/m$ is the power counting we use.
A generally constructed $f(\frac{D_{\mu}}{m})$ is introduced to replace $f(\vec \nabla)$. The interpretation of $f(\frac{D_{\mu}}{m})$ is still related to anti-particle fields in Eq.~(\ref{x=fN}) and the boost generator in Eq.~(\ref{boostgen}), establishing connection between top-down and bottom-up methods.
This general expansion has a simple form at $v^{\mu}=(1,0,0,0)$ but can also be written in a covariant form with arbitrary $v^{\mu}$  based on the equivalence of reparameterization invariance and boost.

Based on the relation between the two-component free heavy field and the four-component free Dirac field in Eq.~\eqref{psinewNrelation}, specifically, similar to the rest frame result in Eq.~\eqref{expansionpsinrest}, in the gauge interaction case, we generally define
\begin{equation}\label{eq:tpd}
	\textbf{General Expansion:}\quad\Psi(x)= e^{-imt}\left[\begin{array}{c}
		 N(x)\\f(\frac{D_{\mu}}{m}) N(x)
	\end{array}\right],
\end{equation}
where $D_{\mu}=\partial_{\mu}+igA_{\mu}$ and field strength $\vec{E}$ and $\vec{B}$ are included in the function $f(\frac{D_{\mu}}{m})$. The factor $e^{-imt}$ is associated with the separation of the large momentum mentioned previously and is indispensable for obtaining the NR fields with the correct EOM. 
Once we determine the form of $f(\frac{D_{\mu}}{m})$, we obtain the combination of $N(x)$ and derivatives that transforms linearly under the Lorentz transformation. 
By definition, $\Psi(x)$ transforms as an irreducible representation of the Lorentz group, 
\begin{equation}
	U(\Lambda)^{-1}\Psi(x)U(\Lambda)=D(\Lambda)\Psi(\Lambda^{-1}x).
\end{equation}
The generators of the Lorentz transformation in the Dirac spinor representation is $S^{\mu\nu}=\frac{i}{4}[\gamma^{\mu},\gamma^{\nu}]$, and 
\begin{equation}
  J^i=\frac{1}{2}\epsilon^{ijk}S^{jk}=\frac{1}{2}\left[\begin{array}{cc}
 		\sigma^i&0\\
 		0&\sigma^i
 	\end{array}\right],\quad K^i=S^{0i}=\frac{i}{2}\left[\begin{array}{cc}
 		0&\sigma^i\\
 		\sigma^i&0
 	\end{array}\right],
\end{equation} 
are the rotation and boost generators, respectively.

To determine $f(\frac{D_{\mu}}{m})$, we use discrete symmetries, as well as the rotation and boost properties of the Dirac field $\Psi(x)$.
\begin{itemize}
    \item Parity and time reversal
    \begin{equation}
        \left\{
        \begin{array}{llr}
            P\Psi(t,\vec{x})P&=&\left[\begin{array}{cc}
		1&0\\0&-1
	\end{array}\right]\Psi(t,-\vec{x}),  \\
             \\T\Psi(t,\vec{x})T&=&\left[\begin{array}{cc}
		i\sigma^2&0\\0&i\sigma^2
	\end{array}\right]\Psi(-t,\vec{x}).
        \end{array}
        \right.
    \end{equation}
As a consequence of the above equations, we find that
\begin{equation}
\left\{
    \begin{array}{clr}
          P N(t,\vec{x})P&=& N(t,-\vec{x}),\\
          
          \\
          Pf(\frac{D_{\mu}}{m}) N(t,\vec{x})P&=&-f(\frac{D_{\mu}}{m}) N(t,-\vec{x}),  \\
         \\
            T N(t,\vec{x})T&=&i\sigma^2 N(t,-\vec{x}),\\
            \\
            Tf(\frac{D_{\mu}}{m}) N(t,\vec{x})T&=&i\sigma^2f(\frac{D_{\mu}}{m}) N(t,-\vec{x}),
    \end{array}\right.
\end{equation}
thus $f(\frac{D_{\mu}}{m})$ must be $P$-odd and $T$-even. Note that this result is independent of interactions in the theory.

    \item Rotation $\Lambda=R$
    \begin{equation}
        U(R)^{-1}\Psi(x) U(R)=\left(1-i\vec{J}\cdot\vec{\theta}\right)\Psi(R^{-1}x).
    \end{equation}
    This suggests that the two-component spinor $N(x)$ and arbitrary vectors $V^i$ such as $D^i$, $E^i$, and $B^i$ transform as
\begin{equation}\label{so3N}\left\{\begin{array}{lll}
   U(R)^{-1} N(x)U(R)&=&  \left(1-\frac{i}{2}\vec{\sigma}\cdot\vec{\theta} \right)N(R^{-1}x),\\
    \\
     U(R)^{-1}V^iU(R)&=& V^i-\epsilon^{ijk}V^j\theta^k.
\end{array}\right.
\end{equation}
Since every Pauli matrix $\sigma^i$ inside $f(\frac{D_{\mu}}{m})$ can be translated into the form of an inner product as $\vec{\sigma}\cdot\vec{V}$, and 
\begin{equation}
 U(R)^{-1}\vec{\sigma}\cdot\vec{V}  N(x)U(R)=\left(1 -\frac{i}{2}\vec{\sigma}\cdot\vec{\theta}\right)\vec{\sigma}\cdot\vec{V} N(R^{-1}x),    
\end{equation}
then we know that $f(\frac{D_{\mu}}{m}) N(x)$ transforms linearly under the rotation like $N(x)$:
\begin{equation}
     U(R)^{-1}f(\frac{D_{\mu}}{m}) N(x) U(R)= \left(1-\frac{i}{2}\vec{\sigma}\cdot\vec{\theta}\right)f(\frac{D_{\mu}}{m}) N(R^{-1}x).
\end{equation}

Therefore, the general expansion in Eq.~(\ref{eq:tpd}) is always a linear representation of rotation as long as we choose $N(x)$ as the linear representation of the $SO(3)$ group shown in Eq.~(\ref{so3N}), and $f(\frac{D_{\mu}}{m})$ is trivially constrained by the rotation symmetry.

    \item Boost $\Lambda=B(q)$
    
\begin{equation}
    U(B( q))^{-1}\Psi(x) U(B( q))=\left[\begin{array}{cc}
 		1&\frac{\vec{\sigma}\cdot\vec{q}}{2m}\\
 		\frac{\vec{\sigma}\cdot\vec{q}}{2m}&1
 	\end{array}\right]\Psi(B(q)^{-1}x).
\end{equation}
    This boost $B(q)$ is already defined as Eq.~(\ref{eq:Bqdef}), with small parameter $\frac{q^{\mu}}{m}=(0,-\frac{\vec q}{m})$.
    The RHS of the general expansion in Eq.~(\ref{eq:tpd}) should also transform properly, just like the LHS $\Psi(x)$ does, thus
\begin{equation}\label{twononlinear}
    \left\{
    \begin{array}{lll}
         U(B(q))^{-1} N(x)U(B(q))&=&e^{i\vec{q}\cdot\vec{x}}\left(1+\frac{\vec{\sigma}\cdot\vec{q}}{2m}f(\frac{D_{\mu}}{m})\right) N(B(q)^{-1}x),  \\
         \\
         U(B(q))^{-1}f(\frac{D_{\mu}}{m}) N(x)U(B(q))&=&e^{i\vec{q}\cdot\vec{x}}\left(\frac{\vec{\sigma}\cdot\vec{q}}{2m}+f(\frac{D_{\mu}}{m})\right)N(B(q)^{-1}x).
    \end{array} 
    \right.
\end{equation}

\end{itemize}

Under the boost transformation, $f(\frac{D_{\mu}}{m})$ takes the form
\begin{eqnarray}
    U(B(q))^{-1}f(\frac{D_{\mu}}{m}) U(B(q)) = f(\frac{B(q)_{\mu}^{\nu} D_{\nu}}{m}),
\end{eqnarray}
where from Eq.~\eqref{eq:Bqdef} we have
\begin{eqnarray}
    B(q)_{\mu}^{~\nu} D_{\nu} = D_{\mu}+\frac{1}{m}(v_{\mu}q\cdot D-q_{\mu}v\cdot D).
\end{eqnarray}
After combining the two transformations in Eq.~\eqref{twononlinear}, we find that
{\begin{small}
\begin{eqnarray}\label{ccconderivation}
\left[U(B(q))^{-1}f(\frac{D_{\mu}}{m}) U(B(q)) \right]\left[U(B(q))^{-1}N(x)U(B(q))\right]&=&f(\frac{B(q)_{\mu}^{\nu} D_{\nu}}{m})\, e^{i\vec{q}\cdot\vec{x}}\left(1+\frac{\vec{\sigma}\cdot\vec{q}}{2m}f(\frac{D_{\mu}}{m})\right) N(B(q)^{-1}x) \nonumber \\
  &=& 
e^{i\vec{q}\cdot\vec{x}}\left(\frac{\vec{\sigma}\cdot\vec{q}}{2m}+f(\frac{D_{\mu}}{m})\right)N(B(q)^{-1}x). 
\end{eqnarray}
\end{small}
}

Here we summarize the NR field $N(x)$ nonlinear transformation which will be used in the following
\begin{equation}\label{eq:nrboost}
    \textbf{Nonlinear boost:}\quad
U(B(q))^{-1} N(x)U(B(q))=e^{i\vec{q}\cdot\vec{x}}\left(1+\frac{\vec{\sigma}\cdot\vec{q}}{2m}f(\frac{D_{\mu}}{m})\right)N(B(q)^{-1}x),
\end{equation}
with the boost generator
\begin{equation}\label{boostgen}
    \vec{\mathcal{K}}=-i\frac{\vec\sigma}{2}f(\frac{D_{\mu}}{m}),
\end{equation}
and simultaneously, the $f(\frac{D_{\mu}}{m})$ must obey a constrain due to Eq.~\eqref{ccconderivation}:
\begin{equation}\label{eq:cccondition}
\begin{array}{l}
     \textbf{Consistency Condition:}\quad 
    f(\frac{B(q)_{\mu}^{\nu} D_{\nu}}{m})e^{i\vec q\cdot\vec x}\left(1+\frac{\vec\sigma\cdot\vec q}{2m}f(\frac{D_{\mu}}{m})\right)N=e^{i\vec q\cdot\vec x}\left(\frac{\vec \sigma\cdot\vec q}{2m}+f(\frac{D_{\mu}}{m})\right)N.
\end{array}
\end{equation}
The phase factors in this condition are essential since they change the derivatives in $f(\frac{D_{\mu}}{m})$. The $f(\frac{D_{\mu}}{m})$ is $P$-odd and $T$-even, constructed from building blocks in the given theory with the constraint in Eq.~(\ref{eq:cccondition}). Since this consistency condition is an operator equation acting on fields, the EOM of $N(x)$ will be used, which depends on the NR Lagrangian including interactions. Therefore, we will discuss the exact form of $f(\frac{D_{\mu}}{m})$ at the operator and Lagrangian level in section \ref{sec:b-HQET}. This consistency condition suggests the commutation relations of the boost generator, as it is discussed in appendix~\ref{app3}.

Once $N(x)$ and $f(\frac{D_{\mu}}{m})$ are determined, we have the following consequences.  Firstly, as a result of the general expansion, $f(\frac{D_{\mu}}{m})N(x)$ can be recognized as the  anti-particle field $\tilde N(x)$,
\begin{equation}\label{x=fN}
    \tilde N(x) \equiv f(\frac{D_{\mu}}{m})N(x),
\end{equation}
which is similar to the result from integrating out in Eq.~\eqref{treefd}. However, it should be emphasized that $f(\frac{D_{\mu}}{m})N(x)$ is not really obtained at the tree-level, instead it contains the most general form allowed by symmetry, beyond the tree-level.

In a general frame, instead of using the expansion (\ref{eq:tpd}) with $v^{\mu}=(1,0,0,0)$, we can also define the four-component NR spinor field $Q_v(x)$ as
\begin{eqnarray}\label{Qv}
    Q_v(x) &=& e^{imvx}\frac{1+v\!\!\!/}{2}\Psi(x), \\
    f_v(\frac{D_{\mu}}{m})Q_v(x) &=& e^{imvx}\frac{1-v\!\!\!/}{2}\Psi(x),\label{Bv}
\end{eqnarray}
thus the expansion is written by
\begin{equation}\label{covtpd}
\textbf{General Expansion:}\quad
    \Psi(x)=e^{-imvx}\left(Q_v(x)+f_v(\frac{D_{\mu}}{m})Q_v(x)\right).
\end{equation}
Away from the rest frame, $f_v(\frac{D_{\mu}}{m})$ is the $v$-dependent version of $f(\frac{D_{\mu}}{m})$, consisting of $v^{\mu}$, $D_{\mu}$ and $F^{\mu\nu}$. It is also $P$-odd and $T$-even and can be constructed order by order using building blocks. 
Recall that reparameterization and boost are equivalent, as discussed in subsection~\ref{coset}. Due to definitions (\ref{Qv}) and (\ref{Bv}), the reparameterization in Eq.~(\ref{QVRPI})
should be replaced by
\begin{equation}
\left\{
    \begin{array}{ccccc}\label{Qv+}
        Q_v&\longrightarrow& Q_{v+\frac{q}{m}}&=&e^{iqx}(1+\frac{q\!\!\!/}{2m})Q_v+e^{iqx}\frac{q\!\!\!/}{2m}f_v(\frac{D_{\mu}}{m})Q_v, \\
         \\
           f_v(\frac{D_{\mu}}{m})Q_v&\longrightarrow& f_{v+\frac{q}{m}}(\frac{D_{\mu}}{m})Q_{v+\frac{q}{m}}&=&-e^{iqx}\frac{q\!\!\!/}{2m}Q_v+e^{iqx}(1-\frac{q\!\!\!/}{2m})f_v(\frac{D_{\mu}}{m})Q_v.
    \end{array}\right.
\end{equation}
Therefore, we derive the general reparameterization  
\begin{equation}\label{eq:generalrpi}
\textbf{Reparameterization}\quad
\left\{
    \begin{array}{lll}
    v^{\mu}&\longrightarrow&v^{\mu}+\frac{q^{\mu}}{m},
           \\
         \\
          Q_v&\longrightarrow& Q_{v+\frac{q}{m}}=e^{iqx}(1+\frac{q\!\!\!/}{2m})Q_v+e^{iqx}\frac{q\!\!\!/}{2m}f_v(\frac{D_{\mu}}{m})Q_v,
    \end{array}\right.
\end{equation}
and the consistency condition 
\begin{equation}\label{covcc}
    \begin{array}{ll}
        &\textbf{Consistency Condition:}  \\
        \\
      &f_{v+\frac{q}{m}}(\frac{D_{\mu}}{m})e^{iqx}\left(1+\frac{q\!\!\!/}{2m}+\frac{q\!\!\!/}{2m}f_v(\frac{D_{\mu}}{m})\right)Q_v=e^{iqx}\left(-\frac{q\!\!\!/}{2m}+f_v(\frac{D_{\mu}}{m})-\frac{q\!\!\!/}{2m}f_v(\frac{D_{\mu}}{m})\right)Q_v.
    \end{array}
\end{equation}
Using Eq.~\eqref{eq:restqvbvnchi}, this condition reduces to the consistency condition (\ref{eq:cccondition}) in the rest frame.

\subsubsection{General Parametrization}\label{sec:generalparam}

To obtain the nonlinear boost for NR fields $N(x)$ at the operator level, we need the exact form of $f(\frac{D_{\mu}}{m})$. The following NR building blocks are identified:  the electric field $\vec{E}$, the magnetic field $\vec{B}$, the time derivative $D_{t}$ and spatial derivative $D_i$.
The electric-magnetic fields satisfy the CDC, 
 \begin{equation}\label{eq:EBCDC}
     E^i=\frac{i}{g}[D_t,D^i],\quad
         B^i=\frac{i}{2g}[D^j,D^k]\epsilon^{ijk},
 \end{equation}
according to $D^{\mu}=\partial^{\mu}+igA^{\mu}=(D_t, -\Vec{D})$, where $g$ is the gauge coupling constant. 
The $PT$ transformations of these building blocks take the form
\begin{eqnarray}
    \begin{array}{|c|c|c|c|c|c|}
    \hline
         &\boldsymbol{D}& D_t& \boldsymbol{B} &\boldsymbol{E}&\boldsymbol{\sigma}  \\
         \hline
         P & - &+ &+ &- &+ \\
         \hline
         T & + & - & - &+ &-
         \\
         \hline
    \end{array}\ .
\end{eqnarray}
Since the $f(\frac{D_{\mu}}{m})$ is $P$-odd and $T$-even, the combinations are chosen properly. 
\begin{itemize}
     \item Covariant derivative commutator (CDC): The relation in Eq.~\eqref{eq:EBCDC} shows that the CDC yields the electromagnetic field. Consequently, all derivatives acting on the fields are rendered symmetrized.

     \item Equation of motion (EOM): The EOMs for the electric and magnetic fields are
\begin{align}
    \boldsymbol{D}\cdot\boldsymbol{E}&=\rho,\nonumber\\
\boldsymbol{D}\cdot\boldsymbol{B}&=0,\nonumber\\
\boldsymbol{D}\times\boldsymbol{E}&=-D_t\boldsymbol{B},\nonumber\\
\boldsymbol{D}\times\boldsymbol{B}&=\boldsymbol{J}+D_t\boldsymbol{E},
\end{align}
where $\rho$ and $\boldsymbol{J}$ are the external sources. Therefore, we choose to eliminate the terms with $\boldsymbol{D}\cdot \boldsymbol{B}$ and $\boldsymbol{D}\times\boldsymbol{E}$. 
\end{itemize}
By contracting the $SO(3)$ indices of the  building blocks and eliminating the above redundancies, we construct the operator basis for $f(\frac{D_{\mu}}{m})$ up to order $1/m^3$ 
\begin{equation}\label{fform}
    \begin{array}{lll}
         f(\frac{D_{\mu}}{m})&=&c^{(1)}\frac{i\vec{\sigma}\cdot\vec{D}}{2m}+c^{(2)}_A\frac{\{D_t,\vec{\sigma}\cdot{\vec{D}}\}}{8m^2}+c^{(2)}_B\frac{ig\vec{\sigma}\cdot\vec{E}}{8m^2}\\
         \\
         & &+c_a^{(3)}\frac{ig\vec{B}\cdot\vec{D}}{8m^3}+c_b^{(3)}\frac{g\epsilon^{ijk}[D^i,B^j]\sigma^k}{8m^3}+c_b^{\prime(3)}\frac{g\epsilon^{ijk}\{D^i,B^j\}\sigma^k}{8m^3}+c_d^{(3)}\frac{\{(i\vec{D})^2,i\vec{\sigma}\cdot\vec{D}\}}{16m^3}\\
         \\
         &&+c_e^{(3)}\frac{g[D_t,\vec{\sigma}\cdot\Vec{E}]}{16m^3}+c_f^{(3)}\frac{g\{D_t,\vec{\sigma}\cdot\vec{E}\}}{16m^3}+c^{(3)}_g\frac{i\{D_t,\{D_t,\Vec{\sigma}\cdot\vec{D}\}\}}{16m^3},
    \end{array}
\end{equation}
with the parameters such as $c^{(1)}, c_A^{(2)},c_B^{(2)},\ldots,c_g^{(3)}$ to be determined by the consistency condition.

Notice that, the $f(\frac{D_{\mu}}{m})$ is constructed by $D_t, \vec D, \vec E$ and $\vec B$, whose transformation under the boost $B(q)$ are
\begin{equation}\label{dtdeboff}
\left\{\begin{array}{lll}
          D_t&\rightarrow &D_t-\frac{1}{m}\vec{q}\cdot\vec{D},\\
          \\
          \vec{D}&\rightarrow&\vec{D}-\frac{1}{m}\vec{q}~D_t,\\
          \\\Vec{E}&\rightarrow&\Vec{E}-\frac{1}{m}\Vec{q}\times\Vec{B},\\
          \\\Vec{B}&\rightarrow&\Vec{B}+\frac{1}{m}\Vec{q}\times\Vec{E}. 
\end{array}\right.
\end{equation}
Plugging the form of Eq.~\eqref{fform} into the consistency condition 
\begin{equation}\label{eq:fcccalcu}
    f(\frac{B(q)_{\mu}^{\nu} D_{\nu}}{m})e^{i\vec q\cdot\vec x}\left(1+\frac{\vec\sigma\cdot\vec q}{2m}f(\frac{D_{\mu}}{m})\right)N=e^{i\vec q\cdot\vec x}\left(\frac{\vec \sigma\cdot\vec q}{2m}+f(\frac{D_{\mu}}{m})\right)N,
\end{equation}
we can obtain the relations that constrain the parameters of $f(\frac{D_{\mu}}{m})$ in Eq.~\eqref{fform}. During this procedure, the term $iD_tN(x)$ need to be removed. Recall that in Eq.~\eqref{eq:freeNeom} the EOM of the heavy field can be expanded as
\begin{equation}
    i\partial_tN(x)=\left(-\frac{\nabla^2}{2m}-\frac{\nabla^4}{8m^3}+...\right)N(x).
\end{equation}
Since only the $1/m$ order term in the EOM is relevant in the calculation of the consistency condition Eq.~\eqref{eq:fcccalcu} in this work, the naive replacement $\partial_{\mu}\rightarrow D_{\mu}$ yields
\begin{equation}\label{eq:freeDtn}
    iD_tN(x)=-\frac{\vec D^2}{2m}N(x)+\mathcal{O}(1/m^3).
\end{equation}
Then we obtain that the parameters $c^{(1)}, c_A^{(2)}, c_B^{(2)}, c_a^{(3)}, c_b^{\prime(3)}, c_d^{(3)}$ are constrained by
\begin{equation}
    \begin{cases}
        c^{(1)}=-1,\\
        2c_2+c_2c_A^{(2)}-c_d^{(3)}=1,\\
        c_a^{(3)}=1,\\
        c_B^{(2)}+2c_b^{\prime(3)}=0,\\
        c_A^{(2)}-c_d^{(3)}=-1.
    \end{cases}
\end{equation}
Because of using the naive replacement of the free field EOM in Eq.~\eqref{eq:freeDtn}, there are some difference from the constraints of $f(\frac{D_{\mu}}{m})$ in the most general cases  at the order $1/m^3$ in the following section in Eq.~\eqref{eq:hqetcccondition}. However, since we only need the boost transformation up to $1/m^3$ , then $f(\frac{D_{\mu}}{m})$ up to $1/m^2$ is enough for our purpose.

Therefore, up to $1/m^2$ we obtain 
\begin{equation}
   f(\frac{D_{\mu}}{m})= -\frac{i\vec{\sigma}\cdot\vec{D}}{2m}+c^{(2)}_A\frac{\{D_t,\vec{\sigma}\cdot{\vec{D}}\}}{8m^2}+c^{(2)}_B\frac{ig\vec{\sigma}\cdot\vec{E}}{8m^2}+\mathcal{O}(\frac{1}{m^3}),
\end{equation}
where $c_A^{(2)}$ and $c_B^{(2)}$ are independent in this order.
As a result, the covariant nonlinear boost for the heavy field up to $1/m^3$ is
\begin{equation}
     U(B(\vec q))^{-1}NU(B(\vec q))=e^{i\vec q\cdot\vec x}\left(1-i\frac{\vec q\cdot\vec D}{4m^2}+\frac{\vec \sigma\cdot\vec q\times\vec D}{4m^2}+c^{(2)}_A\frac{\vec\sigma\cdot\vec q\{D_t,\vec{\sigma}\cdot{\vec{D}}\}}{16m^3}+c^{(2)}_B\frac{ig\vec\sigma\cdot\vec q~\vec{\sigma}\cdot\vec{E}}{16m^3}\right)N.
\end{equation}
If truncated up to $1/m^2$ order, after the redefinition of the NR field as $N^{\prime}(x)$ in Eq.~(\ref{Nprime}), the boost transformation would be the same as that in the literature~\cite{Heinonen:2012km}.
On the other hand, if considering $1/m^3$ order, additional terms including the gauge field, such as the $c_A^{(2)}$ and $c_B^{(2)}$ terms, appear, and thus, unlike the treatment in Ref.~\cite{Heinonen:2012km}, the most general nonlinear boost is needed.

\section{The Bottom-up Lagrangian}
\label{sec:b-HQET}

Having discussed the NR states and fields as well as their transformations, at the Lagrangian level (in a gauge theory), operators can be constructed with respect to the NR symmetry. All the NR building blocks can be written in either the rest frame or the general frame. In the rest frame $v^{\mu}=(1,0,0,0)$, the heavy field is a two-component spinor while the gauge field and its derivatives carry spatial indices. While in the general frame, these building blocks remain Lorentz covariant via the spurion $v^{\mu}$, as discussed in subsection~\ref{lagconstruction}. In this work, using these building blocks within the gauge theory, the Lagrangian would be constructed up to  $\mathcal{O}\left(\frac{1}{m^3}\right)$, while the systematic construction on the higher-order operators are discussed in a related work~\cite{nroperator}.

In the bottom-up approach, the constructed operators could not be independent due to the nonlinear Lorentz symmetry. The boost transformation obtained from the general expansion in Eq.~(\ref{eq:nrboost}) with the generally constructed $f(\frac{D_{\mu}}{m})$ reveals the non-trivial contribution from those gauge field strength terms at order $\frac{1}{m^3}$ which was missed in Ref.~\cite{Heinonen:2012km}, as in subsection~\ref{VM}. 
An alternative way is imposing the reparameterization invariance on the NR Lagrangian in the general frame, which turns out to be consistent with results in the boost transformation with $f(\frac{D_{\mu}}{m})$, as in subsection~\ref{sec8rpi}. 
Although our discussion is compatible with  Ref.~\cite{Berwein:2018fos}, detailed comparison with the slight differences is analyzed in subsection~\ref{bottomupcompare}.

\subsection{Lagrangian Construction}
\label{lagconstruction}

In the rest frame $v^{\mu}=(1,0,0,0)$, for a $U(1)$ gauge theory with gauge field $A^{\mu}$, the building blocks contain the heavy field $N(x)$ and its conjugation, the electromagnetic fields $\vec E$ and $\vec B$, and the derivatives $D_t$ and $\vec D$.  
In the general frame, the spurion $v^{\mu}$ is kept in the construction, contracting with other building blocks covariantly. The building blocks are the spurion, the projected field $Q_v$, the field strength $F^{\mu\nu}=\partial^{\mu}A^{\nu}-\partial^{\nu}A^{\mu}=-\frac{i}{g}\lbrack D^{\mu},D^{\nu}\rbrack$, $v\cdot D$ and  $D_{\perp}^{\mu}\equiv D^{\mu}-v^{\mu}(v\cdot D)=(0,-\vec D)$.
In fact, the results in the rest frame could be converted into the general frame directly: 
\begin{eqnarray}\label{eq:restgeneralframetab}
    \begin{array}{|c|c|c|c|c|c|c|}
    \hline
         \textrm{General frame}  & v^{\mu} & Q_v & v\cdot D& D_{\perp}^{\mu} & F^{\mu\nu} \\
              \hline
     \textrm{Rest frame}&  v^{\mu}=(1,\boldsymbol{0})& Q_v=\left[\begin{array}{l}
              N  \\
              0 
    \end{array}\right] & v\cdot D=D_t & D_{\perp}^{\mu}=\left(0,-D_i\right)& 
    \begin{array}{l}
      F^{0i}=-E^i
      \\
      F^{ij}=-\epsilon^{ijk}B^k
    \end{array} \\
    \hline
    \end{array} \ .
\end{eqnarray}
Both formalisms have their advantages. For the rest frame, the four-vector are divided into the time component scalar and the spatial vector. The little group of $v^{\mu}$ is exactly the spatial rotation group, satisfying the requirement that the bottom-up construction of fields and their boost transformations carry the $SU(2)$ indices.  As for the general frame, it is explicitly dependent on $v^{\mu}$, making the reparameterization of the spurion $v^{\mu}$ much easier.


In the HPET, the NR field $N(x)$ forms the fermion bilinear, the $SU(2)$ singlet $N(x)^{\dagger}N(x)$ or the $SU(2)$ triplet $N^{\dagger}(x)\vec\sigma N(x)$. Then with the $SU(2)$ triplets $\vec{E}$, $\vec{B}$,  $D_{i}$ and the singlet $D_t$, the fermion bilinear is contracted into the singlet using the invariant tensor $\delta^{ij}$ and $\epsilon^{ijk}$. Similar to the construction of $f(\frac{D_{\mu}}{m})$ in subsubsection~\ref{sec:generalparam}, the redundancies from the CDC and those related to the EOM for the electromagnetic fields ($\boldsymbol{D}\cdot\boldsymbol{B}$ and $\boldsymbol{D}\times\boldsymbol{E}$) are eliminated. Additionally,  the following procedure is also performed:
\begin{itemize}
   
    \item Integration by part (IBP): Given that total derivatives in the Lagrangian can be discarded, all  derivatives acting on $N^{\dagger}(x)$ can be converted into  derivatives acting on the other fields. Therefore, ensuring that no  derivatives act on $N^{\dagger}(x)$ removes the need for IBP.
    \item Equation of motion (EOM):  The term for the heavy field $iD_tN(x)$ is also removed, due to the on-shell condition $iD_tN(x)=\left(-\frac{\vec D^2}{2m}N(x)+...\right).$
    \item Schouten identity: Using the $SO(3)$ invariant tensors $\delta^{ij}$ and $\epsilon^{ijk}$, eliminating the redundancies such as $\epsilon^{ijk}\epsilon^{imn}=\delta^{jm}\delta^{kn}-\delta^{jn}\delta^{km}$, as well as $\sigma^i\sigma^j=\delta^{ij}+i\epsilon^{ijk}\sigma^k$.
\end{itemize} 
Moreover, since the electromagnetic interactions are invariant under the spatial and time reversal, all the operators are $P$-even and $T$-even. Detailed constructions of the independent bases can be found in Ref.~\cite{nroperator}. In this work, the bottom-up Lagrangian up to $1/m^3$ is enough for our purpose.


The HPET Lagrangian, in the rest frame up to $1/m^2$ order  is
\begin{equation}\label{m2HPET}
    \mathcal{L}=N^{\dagger}(x)\left(iD_t+c_2\frac{\vec D^2}{2m}+c_Fg\frac{\vec\sigma\cdot\vec B}{2m}+c_Dg\frac{[D_i,E^i]}{8m^2}+ic_Sg\frac{\epsilon^{ijk}\sigma^i\{D_j,E^k\}}{8m^2}\right)N(x).
\end{equation}
The leading-order term is $N^{\dagger}iD_tN$, depicting a static object when $m\rightarrow\infty$. Note that NRQED has the same Lagrangian as Eq.~(\ref{m2HPET}), but with a different leading-order term  $N^{\dagger}iD_tN+N^{\dagger}\frac{\vec D^2}{2m}N$ due to its different power counting, depicting the relative movement between the particle pair. Because of the hidden Lorentz symmetry, the Wilson coefficients in the HPET Lagrangian are related; $c_2, c_F, \ldots$ are not independent. For the free field $N(x)$, we have obtained the nonlinear boost transformation in Eq.~(\ref{nboostpd}), up to $1/m^2$ it reads
\begin{equation}
\begin{array}{lll}
     U(B(q))^{-1}N(x)U(B(q))  & = &e^{i\vec{q}\cdot\vec{x}}\left(1+\frac{\vec{\sigma}\cdot\vec{q}}{2m}f(\nabla)\right)N(\Lambda^{-1}x)\\
     \\
     & =&e^{i\vec q\cdot\vec x}\left(1-i\frac{\vec q\cdot\vec D}{4m^2}+\frac{\vec \sigma\cdot\vec q\times\vec D}{4m^2}\right)N(B(q)^{-1}x),
\end{array}       
\end{equation}
where we have used Eq.~\eqref{restfreefnabla}
\begin{equation}
    f(\vec\nabla)=\frac{-i\vec\sigma\cdot\vec \nabla}{2m+i\partial_t}=\left(-\frac{i\vec\sigma\cdot\vec\nabla}{2m}+\frac{i\vec\sigma\cdot\vec\nabla i\partial_t}{4m^2}+...\right),
\end{equation}
and naively replaced $\partial_{\mu}$ with $D_{\mu}$, without any electromagnetic field. As for this $f(\vec\nabla)$, we find that its consistency condition Eq.~\eqref{eq:cccondition} 
\begin{equation}
    f(\frac{B(q)_{\mu}^{\nu} D_{\nu}}{m})e^{i\vec q\cdot\vec x}\left(1+\frac{\vec\sigma\cdot\vec q}{2m}f(\frac{D_{\mu}}{m})\right)N=e^{i\vec q\cdot\vec x}\left(\frac{\vec \sigma\cdot\vec q}{2m}+f(\frac{D_{\mu}}{m})\right)N,
\end{equation}
gives the values 
\begin{equation}
    c_2=1,\quad c_F=1.
\end{equation}
Thus at the next-to-leading order, it recovers the Pauli equation 
\begin{equation}
    iD_tN(x)=\left(-\frac{1}{2m}\vec D^2-\frac{e}{2m}\vec\sigma\cdot\vec B\right)N(x),
\end{equation}
where the coupling constant is $g=e$ and the Landé $g$-factor is 2 for the electron. However, we stress that this is only the tree-level result. In fact, there will be an anomalous magnetic moment from the loop corrections, changing the value of $c_F$.

The building blocks of the Lagrangian in Eq.~\eqref{m2HPET} transform under the boost $B(q)$ as
\begin{equation}\label{nablaboost}
\left\{\begin{array}{lll}

    N(x)&\rightarrow &e^{i\vec{q}\cdot\vec{x}}\left(1+\frac{\vec{\sigma}\cdot\vec{q}}{2m}f(\nabla)\right)N(\Lambda^{-1}x),\\
    \\
          D_t&\rightarrow &D_t-\frac{1}{m}\vec{q}\cdot\vec{D},\\
          \\
          \vec{D}&\rightarrow&\vec{D}-\frac{1}{m}\vec{q}~D_t,\\
          \\\Vec{E}&\rightarrow&\Vec{E}-\frac{1}{m}\Vec{q}\times\Vec{B},\\
          \\\Vec{B}&\rightarrow&\Vec{B}+\frac{1}{m}\Vec{q}\times\Vec{E},
\end{array}\right.
\end{equation}
leading to the variation of the Lagrangian $\mathcal{L}\rightarrow \mathcal{L}+\delta \mathcal{L}$. Imposing invariance under boost $\delta \mathcal{L}=0$ up to $1/m^2$, we obtain the non-trivial relations among the Wilson coefficients in the Lagrangian
\begin{equation}
    c_2=1,\quad c_S=2c_F-1.
\end{equation}
These relations are independent of the radiative corrections at the loop level. However, up to $1/m^3$, the boost transformation of $N(x)$ becomes more complicated, as a consequence of the field strength dependence.
The naive replacement $f(\vec\nabla)\rightarrow f(\vec D)$ in the boost fails at higher orders, and we need to construct the most general $f(\frac{D_{\mu}}{m})$ including the electromagnetic fields order by order as in subsection~\ref{GECC}, to realize the Lorentz symmetry nonlinearly.

\subsection{Covariant Boost Transformation}
\label{VM}

The HPET Lagrangian up to $1/m^3$, in the rest frame, can be written as 
\begin{equation}\label{HQETnrlag}
    \begin{array}{lll}
      \mathcal{L}&=&N^{\dagger}\left\{iD_t+c_2\frac{ \vec{D}^2}{2m}+c_Fg\frac{\vec{\sigma}\cdot\vec{B}}{2m}+c_Dg\frac{[D_i,E^i]}{8m^2}+ic_Sg\frac{\epsilon^{ijk}\sigma^i\{D_j,E^k\}}{8m^2}+c_4\frac{\vec{D}^4}{8m^3}\right.\\
      \\
        & & +c_{W1}g\frac{\{\vec{D}^2,\vec{\sigma}\cdot\vec{B}\}}{8m^3}-c_{W2}g\frac{D^i\vec{\sigma}\cdot\vec{B}D^i}{4m^3}+c_{p^{\prime}p}g\frac{\{\vec{\sigma}\cdot\vec{D},\vec{B}\cdot\vec{D}\}}{8m^3}\\
        \\
        &&\left.+ic_Mg\frac{\{D_i,\epsilon^{ijk}[D_j,B^k]\}}{8m^3}+c_{A1}g^2\frac{B^2-E^2}{8m^3}-c_{A2}g^2\frac{E^2}{16m^3}+\mathcal{O}(1/m^4)\right\}N,
    \end{array}
\end{equation}
following the construction procedure in Ref.~\cite{nroperator}, 
which agrees with result in Refs.~\cite{Manohar:1997qy,Gunawardana:2017zix}. 

For the generally constructed $f(\frac{D_{\mu}}{m})$ in Eq.~\eqref{fform}, utilizing the consistency condition 
\begin{equation}
   f(\frac{B(q)_{\mu}^{\nu} D_{\nu}}{m})e^{i\vec q\cdot\vec x}\left(1+\frac{\vec\sigma\cdot\vec q}{2m}f(\frac{D_{\mu}}{m})\right)N=e^{i\vec q\cdot\vec x}\left(\frac{\vec \sigma\cdot\vec q}{2m}+f(\frac{D_{\mu}}{m})\right)N,
\end{equation}
and using the EOM of $N(x)$ from the Lagrangian in Eq.~\eqref{HQETnrlag},
\begin{equation}\label{EOMofN}
    iD_t N(x)=\left(-c_2\frac{\vec D^2}{2m}-c_Fg\frac{\vec\sigma\cdot\vec B}{2m}+...\right)N(x),
\end{equation}
we obtain that the parameters $c^{(1)}, c_A^{(2)}, c_B^{(2)}, c_a^{(3)}, c_b^{\prime(3)}, c_d^{(3)}$ are constrained by
\begin{equation}\label{eq:hqetcccondition}
    \begin{cases}
        c^{(1)}=-1,\\
        -2c_2-c_2c_A^{(2)}+c_d^{(3)}=-1,\\
        -c_a^{(3)}-2c_F-c_Fc_A^{(2)}=-1,\\
        -c_B^{(2)}-2c_F-c_Fc_A^{(2)}-2c_b^{\prime(3)}=0,\\
        -c_A^{(2)}+c_d^{(3)}=1.
    \end{cases}
\end{equation}
As a result, up to $1/m^3$ we obtain 
\begin{equation}\label{bupboost}
   f(\frac{D_{\mu}}{m})= -\frac{i\vec{\sigma}\cdot\vec{D}}{2m}+c^{(2)}_A\frac{\{D_t,\vec{\sigma}\cdot{\vec{D}}\}}{8m^2}+c^{(2)}_B\frac{ig\vec{\sigma}\cdot\vec{E}}{8m^2}+\mathcal{O}(\frac{1}{m^3}),
\end{equation}
where $c_A^{(2)}$ and $c_B^{(2)}$ are independent up to $1/m^3$, while they are constrained by Eq.~(\ref{eq:hqetcccondition}) if higher order contributions are considered.
Consequently, the nonlinear boost for $N(x)$ up to $1/m^3$ is
\begin{equation}\label{nbfd}
     U(B(\vec q))^{-1}NU(B(\vec q))=e^{i\vec q\cdot\vec x}\left(1-i\frac{\vec q\cdot\vec D}{4m^2}+\frac{\vec \sigma\cdot\vec q\times\vec D}{4m^2}+c^{(2)}_A\frac{\vec\sigma\cdot\vec q\{D_t,\vec{\sigma}\cdot{\vec{D}}\}}{16m^3}+c^{(2)}_B\frac{ig\vec\sigma\cdot\vec q~\vec{\sigma}\cdot\vec{E}}{16m^3}\right)N.
\end{equation}

Now we use the covariant boost transformation in Eq.~(\ref{nbfd}) to obtain the relations among the Wilson coefficients in the rest frame Lagrangian Eq.~\eqref{HQETnrlag} up to $1/m^3$. From the symmetry analysis that combines the NR d.o.f. into the Lorentz covariant ones, we know that under boost these building blocks transform as
\begin{equation}\label{hqetboost}
\left\{\begin{array}{lll}

    N(x)&\rightarrow &e^{i\vec{q}\cdot\vec{x}}\left(1+\frac{\vec{\sigma}\cdot\vec{q}}{2m}f(\frac{D_{\mu}}{m})\right)N(\Lambda^{-1}x),\\
    \\
          D_t&\rightarrow &D_t-\frac{1}{m}\vec{q}\cdot\vec{D},\\
          \\
          \vec{D}&\rightarrow&\vec{D}-\frac{1}{m}\vec{q}D_t,\\
          \\
          \Vec{E}&\rightarrow&\Vec{E}-\frac{1}{m}\Vec{q}\times\Vec{B},\\
          \\
          \Vec{B}&\rightarrow&\Vec{B}+\frac{1}{m}\Vec{q}\times\Vec{E},
\end{array}\right.
\end{equation}
where we use the result in Eq.~\eqref{bupboost}
\begin{equation}
   f(\frac{D_{\mu}}{m})= -\frac{i\vec{\sigma}\cdot\vec{D}}{2m}+c^{(2)}_A\frac{\{D_t,\vec{\sigma}\cdot{\vec{D}}\}}{8m^2}+c^{(2)}_B\frac{ig\vec{\sigma}\cdot\vec{E}}{8m^2}+\mathcal{O}(\frac{1}{m^3}).
\end{equation}
Under the above transformations, the variation of the Lagrangian (\ref{HQETnrlag}) is
\begin{equation}\label{deltaL1}
    \begin{array}{lll}
         \delta\mathcal{L}&=&N^{\dagger}\left\{(c_2-1)\frac{i\Vec{q}\cdot\Vec{D}}{m}+(1-2c_2)\frac{\{D_t,\Vec{q}\cdot\Vec{D}\}}{4m^2}+\frac{g}{4m^2}\Vec{\sigma}\cdot\vec{q}\times\Vec{E}(-1+2c_F-c_S)+\frac{\{i\Vec{q}\cdot\Vec{D},\Vec{D}^2\}}{8m^3}(2c_4-c_2)\right.\\
         \\
         &&+\frac{ig}{8m^3}\{\Vec{\sigma}\cdot\Vec{B},\Vec{q}\cdot\Vec{D}\}(-c_F-c_2+c_S+2c_{W1}-2c_{W2})+\frac{g}{8m^3}\epsilon^{ijk}q^i[D^j,B^k](-c_D+2c_M+c_F)\\
         \\
        && +\frac{ig}{8m^3}\Vec{\sigma}\cdot\Vec{q}\Vec{D}\cdot\Vec{B}(-2c_S+2c_{p^{\prime}p}+2c_F)+\frac{ig}{8m^3}\{\Vec{\sigma}\cdot\Vec{D},\Vec{q}\cdot\Vec{B}\}(c_{p^{\prime}p}-c_F+c_2)\\
        \\
        &&\left.+\frac{c_A^{(2)}\{\{D_t,\Vec{q}\cdot\Vec{D}\},iD_t\}}{16m^3}+\frac{g[D_t,\Vec{q}\cdot\Vec{E}]}{16m^3}(-2c_D-c_B^{(2)})+\frac{g\{iD_t,\Vec{\sigma}\cdot\Vec{q}\times\vec
E\}}{16m^3}(-2c_S-c_B^{(2)}-c_A^{(2)})\right\}N\\
\\
&&+\mathcal{O}(1/m^4),
         
    \end{array}
\end{equation}
after using the EOM, we have
\begin{equation}
    N^{\dagger}\left(i\frac{\{iD_t,\Vec{q}\cdot\Vec{D}\}}{4m^2}\right)N= N^{\dagger}\left(-c_2\frac{\{i\Vec{q}\cdot\Vec{D},\Vec{D}^2\}}{8m^3}-gc_F\frac{\{i\Vec{q}\cdot\Vec{D},\Vec{\sigma}\cdot\Vec{B}\}}{8m^3}\right)N+\mathcal{O}(1/m^4),
\end{equation}
and the variation becomes
\begin{equation}\label{deltaL}
    \begin{array}{lll}
         \delta\mathcal{L}&=&N^{\dagger}\left\{(c_2-1)\frac{i\Vec{q}\cdot\Vec{D}}{m}+(2-2c_2)\frac{\{D_t,\Vec{q}\cdot\Vec{D}\}}{4m^2}+\frac{g}{4m^2}\Vec{\sigma}\cdot\vec{q}\times\Vec{E}(2c_F-c_S-1)+\frac{\{i\Vec{q}\cdot\Vec{D},\Vec{D}^2\}}{8m^3}(2c_4-2c_2)\right.\\
         \\
         & &+\frac{ig}{8m^3}\{\Vec{\sigma}\cdot\Vec{B},\Vec{q}\cdot\Vec{D}\}(2c_{W1}-2c_{W2}-2c_F-c_2+c_S)+\frac{g}{8m^3}\epsilon^{ijk}q^i[D^j,B^k](-c_D+2c_M+c_F)\\
         \\
         &&+\frac{ig}{8m^3}\Vec{\sigma}\cdot\Vec{q}\Vec{D}\cdot\Vec{B}(-2c_S+2c_{p^{\prime}p}+2c_F)+\frac{ig}{8m^3}\{\Vec{\sigma}\cdot\Vec{D},\Vec{q}\cdot\Vec{B}\}(c_{p^{\prime}p}-c_F+c_2)\\
         \\
         &&\left.
          +\frac{c_A^{(2)}\{\{D_t,\Vec{q}\cdot\Vec{D}\},iD_t\}}{16m^3}+\frac{g[D_t,\Vec{q}\cdot\Vec{E}]}{16m^3}(-2c_D-c_B^{(2)})+\frac{g\{iD_t,\Vec{\sigma}\cdot\Vec{q}\times\vec
E\}}{16m^3}(-2c_S-c_B^{(2)}-c_A^{(2)})\right\}N\\
\\
&&+\mathcal{O}(1/m^4).
\end{array}\end{equation}
Symmetry demands that the variation of the Lagrangian vanishes under the boost, $\delta \mathcal{L}=0$ in each order, so that one can solve this equation order by order and obtain the relations among the Wilson coefficients as
\begin{equation}\label{wcr}
    \begin{cases}
        c_2=1,\\
        c_4=1,\\
        c_S=2c_F-1,\\
        2c_M=c_D-c_F,\\
        c_{W2}=c_{W1}-1,\\
        c_{p^{\prime}p}=c_F-1.
    \end{cases}
\end{equation}
Note that at the $\mathcal{O}(1/m^3)$ there are terms containing $D_t$, which can be converted to $\mathcal{O}(1/m^4)$ via the EOM of $iD_t N(x)$ from Eq.~\eqref{EOMofN}, so the following terms should belong to the $\mathcal{O}(1/m^4)$
\begin{equation}\label{deltal3}
\begin{array}{lll}
       \delta\mathcal{L}& \supset &N\left\{\frac{c_A^{(2)}\{\{D_t,\Vec{q}\cdot\Vec{D}\},iD_t\}}{16m^3}+\frac{g[D_t,\Vec{q}\cdot\Vec{E}]}{16m^3}(-2c_D-c_B^{(2)})\right.
       \\
       \\&&\left.+\frac{g\{iD_t,\Vec{\sigma}\cdot\Vec{q}\times\vec
E\}}{16m^3}(-2c_S-c_B^{(2)}-c_A^{(2)})\right\}N.
\end{array}
\end{equation}
Therefore, up to $1/m^3$, these coefficients  do not have to vanish. For example, this does not necessarily lead to $c_A^{(2)}=0$.
One expect, at order $1/m^4$ and higher, extra non-trivial constraints on $c_D$ and $c_S$ from $c_A^{(2)}$ and $c_B^{(2)}$ could arise, and then the gauge field dependent terms in $f(\frac{D_{\mu}}{m})$ in the covariant boost become relevant.

\subsection{Reparameterization Invariance}
\label{sec8rpi}

The boost transformation in the previous subsection should be equivalent to the reparameterization discussed in subsection~\ref{GECC}. Now we examine the variation of the HPET Lagrangian in the general frame, and validate the relations between the Wilson coefficients in this subsection. The general reparameterization is given in Eq.~\eqref{eq:generalrpi} and is dependent on $f_v(\frac{D_{\mu}}{m})$, satisfying the consistency condition. 


From the Lagrangian Eq.~\eqref{HQETnrlag} in the rest frame, the Lagrangian in the general frame up to $1/m^3$ is obtained using Eq.~\eqref{eq:restgeneralframetab}
\begin{equation}\label{Lv}
    \begin{array}{lll}
        \mathcal{L}&=&\Bar{Q}_v\left\{iv\cdot D-c_2\frac{D^2_{\perp}}{2m}-c_Fg\frac{\sigma_{\mu\nu}F^{\mu\nu}}{4m}-c_Dg\frac{v^{\alpha}[D_{\perp}^{\beta}F_{\alpha\beta}]}{8m^2}+ic_Sg\frac{v_{\lambda}\sigma_{\alpha\beta}\{D_{\perp}^{\alpha},F^{\lambda\beta}\}}{8m^2}+c_4\frac{D_{\perp}^4}{8m^3}\right.\\
         \\
         &&+c_{W1}g\frac{\{D_{\perp}^2,\sigma_{\mu\nu}F^{\mu\nu}\}}{16m^3}-c_{W2}g\frac{D_{\perp}^{\lambda}\sigma_{\mu\nu}F^{\mu\nu}D_{\perp \lambda}}{8m^3}+c_{p^{\prime}p}g\frac{\sigma^{\alpha\beta}(D_{\perp}^{\lambda}F_{\lambda\alpha}D_{\perp\beta}+D_{\perp \beta}F_{\lambda\alpha}D_{\perp}^{\lambda}+D_{\perp}^{\lambda}F_{\alpha\beta}D_{\perp\lambda})}{8m^3}\\
         \\
         && \left.-ic_Mg\frac{(D_{\perp\alpha}[D_{\perp\beta}F^{\alpha\beta}]+[D_{\perp\beta}F^{\alpha\beta}]D_{\perp\alpha})}{8m^3}+c_{A1}g^2\frac{F_{\mu\nu}F^{\mu\nu}}{16m^3}+c_{A2}g^2\frac{F_{\mu\alpha}F^{\mu\beta}v^{\alpha}v_{\beta}}{16m^3}+\mathcal{O}(1/m^4)\right\}Q_v,
    \end{array}
\end{equation}
which is the same as that in Ref.~\cite{Manohar:1997qy}. 
Following the similar procedure, $f_v(\frac{D_{\mu}}{m})$ up to $1/m^3$ is obtained by converting Eq.~\eqref{fform} to the general frame 
\begin{equation}
    \begin{array}{lll}
f_v(\frac{D_{\mu}}{m})&=&-c^{(1)}\frac{iD\!\!\!\!/_{\perp}}{2m}-c^{(2)}_A\frac{\{v\cdot D,D\!\!\!\!/_{\perp}\}}{8m^2}+c^{(2)}_B\frac{g\sigma_{\mu\nu}v^{\mu}v_{\alpha}F^{\alpha\nu}}{8m^2}\\
\\
         & &-c_a^{(3)}\frac{ig\gamma_5\epsilon^{\mu\nu\rho\sigma}F_{\mu\nu}D_{\perp\rho}v_{\sigma}}{16m^3}+c_b^{(3)}\frac{ig\sigma_{\mu\nu}v^{\mu}[D_{\perp\alpha},F^{\nu\alpha}]}{8m^3}+c_b^{\prime(3)}\frac{ig\sigma_{\mu\nu}v^{\mu}\{D_{\perp\alpha},F^{\nu\alpha}\}}{8m^3}-c_d^{(3)}\frac{\{D^2_{\perp},iD\!\!\!\!/_{\perp}\}}{16m^3}\\
         \\
         &&-c_e^{(3)}\frac{ig[v\cdot D,\sigma_{\mu\nu}v^{\mu}v_{\alpha}F^{\alpha\nu}]}{16m^3}-c_f^{(3)}\frac{ig\{v\cdot D,\sigma_{\mu\nu}v^{\mu}v_{\alpha}F^{\alpha\nu}\}}{16m^3}-c^{(3)}_g\frac{i\{v\cdot D,\{v\cdot D,D\!\!\!\!/_{\perp}\}\}}{16m^3},
    \end{array}
\end{equation}
while the constraints on the parameters calculated from the covariant consistency condition (\ref{covcc}) 
\begin{equation}
  f_{v+\frac{q}{m}}(\frac{D_{\mu}}{m})e^{iqx}\left(1+\frac{q\!\!\!/}{2m}+\frac{q\!\!\!/}{2m}f_v(\frac{D_{\mu}}{m})\right)Q_v=e^{iqx}\left(-\frac{q\!\!\!/}{2m}+f_v(\frac{D_{\mu}}{m})-\frac{q\!\!\!/}{2m}f_v(\frac{D_{\mu}}{m})\right)Q_v,
\end{equation}
are exactly the same as the results in Eq.~(\ref{eq:hqetcccondition})
\begin{equation}
    \begin{cases}
        c^{(1)}=-1,\\
        -2c_2-c_2c_A^{(2)}+c_d^{(3)}=-1,\\
        -c_a^{(3)}-2c_F-c_Fc_A^{(2)}=-1,\\
        -c_B^{(2)}-2c_F-c_Fc_A^{(2)}-2c_b^{\prime(3)}=0,\\
        -c_A^{(2)}+c_d^{(3)}=1.
    \end{cases}
\end{equation}
This result is expected due to the equivalence of the boost and the reparameterization invariance, then the constraints from the consistency condition Eq.~(\ref{eq:cccondition}) and the covariant consistency condition Eq.~(\ref{covcc}) should coincide, which are derived from the Lorentz boost of the relativistic fields and from the reparameterization invariance, respectively. 
Besides, this also shows that the constraints from the symmetry are unique no matter what building blocks we use, the explicit NR ones or the covariant ones.

In the covariant construction, under the general reparameterization in Eq.~\eqref{eq:generalrpi}
\begin{equation}\label{rpi1}
     \begin{cases}
         v^{\mu}\rightarrow v^{\mu}+\frac{q^{\mu}}{m},\\
         \\
         Q_v\rightarrow Q_{v+\frac{q}{m}}=e^{iqx}(1+\frac{q\!\!\!/}{2m})Q_v+e^{iqx}\frac{q\!\!\!/}{2m}f_v(\frac{D_{\mu}}{m})Q_v,
         \\
         \\
         D^{\mu}\rightarrow D^{\mu},
         \\
         \\
         F^{\mu\nu}\rightarrow F^{\mu\nu},
         
     \end{cases}
 \end{equation}
the variation of the Lagrangian in Eq.~(\ref{Lv}) should vanish, $\delta \mathcal{L}=0$. The variation of the Lagrangian (\ref{Lv}) under the transformation (\ref{rpi1}) is
\begin{equation}
    \begin{array}{lll}
        \delta\mathcal{L}&=&\Bar{Q}_v\left\{\frac{iq\cdot D_{\perp}}{m}(1-c_2)+\frac{\{v\cdot D,q\cdot D_{\perp}\}}{4m^2}(2c_2-1)-\frac{gv_{\lambda}\sigma_{\alpha\beta}q^{\alpha}F^{\lambda\beta}}{4m^2}(1-2c_F+c_S)\right.\\
        \\
         & &-\frac{\{iq\cdot D_{\perp},D_{\perp}^2\}}{8m^3}(-2c_4+c_2)+\frac{ig\{\sigma_{\mu\nu}F^{\mu\nu},q\cdot D_{\perp}\}}{8m^3}(c_F+c_2-c_S-2c_{W1}+2c_{W2})\\
         \\
         &&-\frac{gq^{\alpha}[D_{\perp}^{\beta}F_{\alpha\beta}]}{8m^3}(c_D-2c_M-c_F)+\frac{ig(2q_{\lambda}\sigma_{\alpha\beta}\{D_{\perp}^{\alpha},F^{\lambda\beta}\}-\{\sigma_{\mu\nu}F^{\mu\nu},q\cdot D_{\perp}\})}{32m^3}(2c_S-2c_{p^{\prime}p}-2c_F)\\
         \\
         &&+\frac{ig(-2\sigma^{\alpha\beta}q_{\beta}\{D_{\perp}^{\lambda},F_{\lambda\alpha}\}-\{\sigma_{\mu\nu}F^{\mu\nu},q\cdot D_{\perp}\})}{16m^3}(-c_{p^{\prime}p}+c_F-c_2) +\frac{c_A^{(2)}\{\{v\cdot D,q\cdot D_{\perp}\},iv\cdot D\}}{16m^3}\\
         \\
         &&\left.+\frac{gv^{\alpha}q^{\beta}[v\cdot D,F_{\alpha\beta}]}{16m^3}(2c_D+c_B^{(2)})-\frac{gv_{\lambda}\sigma_{\alpha\beta}q^{\alpha}\{v\cdot D,F^{\lambda\beta}\}}{16m^3}(2c_S+c_B^{(2)}+c_A^{(2)})\right\}Q_v+\mathcal{O}(1/m^4),
    \end{array}
\end{equation}
and it reduces to Eq.~(\ref{deltaL1}) when choosing $v^{\mu}=(1,0,0,0)$ and $q^{\mu}=(0,-\Vec{q})$. The extra minus sign in the transformation parameters $q$ appears when connecting the boost and the reparameterization invariance, just as discussed in the coset construction in subsection \ref{coset}, and the above results explicitly show their equivalence $\delta v=-\delta_{\Lambda'}$. 

Finally, using the EOM, the relations among the Wilson coefficients are the same as in Eq.~(\ref{wcr})
\begin{equation}
    \begin{cases}
        c_2=1,\\
        c_4=1,\\
        c_S=2c_F-1,\\
        2c_M=c_D-c_F,\\
        c_{W2}=c_{W1}-1,\\
        c_{p^{\prime}p}=c_F-1,
    \end{cases}
\end{equation}
as well as the additional terms in the Eq.~\eqref{deltal3} depending on $c_A^{(2)}$ and $c_B^{(2)}$.

\subsection{Comparison with the Boost Commutator}
\label{bottomupcompare}

The consistency condition we derived turns out to be equivalent to the constraints from the boost commutator in Eq.~\eqref{eq:comuk}. This boost commutator is also discussed in Ref.~\cite{Berwein:2018fos}. However, we note that the above boost transformation contains less independent parameters than that in Ref.~\cite{Berwein:2018fos}. The reason is as follows. What we construct is the singlet $f(\frac{D_{\mu}}{m})$ 
while they construct the vector $\vec{\mathcal{K}}_x$, and the difference can come from the order of the products $\Vec{\sigma}f(\frac{D_{\mu}}{m})$ and $f(\frac{D_{\mu}}{m})\Vec{\sigma}$ in Eq.~(\ref{boostgen}). To obey the linear transformation of the Dirac spinor field, only $\Vec{\sigma}f(\frac{D_{\mu}}{m})$ is kept in this work.


 %

According to Ref.~\cite{Berwein:2018fos},  the general form of boost can be parametrized as
\begin{equation}
    \begin{array}{lll}
        \vec {\mathcal{K}}_x &=& m\Vec{x}-\frac{k_D}{2m}\Vec{D}-\frac{ik_{DS}}{4m}\vec{D}\times\Vec{\sigma}+\frac{gk_E}{8m^2}\Vec{E}+\frac{ik_{D0}}{8m^2}\{D_t,\Vec{D}\}+\frac{igk_{ES}}{8m^2}\Vec{E}\times\Vec{\sigma}-\frac{k_{DS0}}{8m^2}\{D_t,\Vec{D}\times\Vec{\sigma}\} \\
        \\
         & &-\frac{k_{D3}}{8m^3}\{\Vec{D}^2,\Vec{D}\}-\frac{ik_{D3S}}{32m^3}\{\Vec{D}^2,\Vec{D}\times\Vec{\sigma}\}+\frac{igk_{B1}}{16m^3}(\Vec{D}\times\Vec{B}-\Vec{B}\times\Vec{D})+\frac{igk_{B2}}{16m^3}(\Vec{D}\times\vec{B}+\Vec{B}\times\Vec{D})\\
         \\
         &&
         +\frac{gk_{BS1}}{16m^3}[\Vec{D},\Vec{\sigma}\cdot\Vec{B}]+\frac{gk_{BS2}}{16m^3}\{\Vec{D},\Vec{\sigma}\cdot\Vec{B}\}
+\frac{gk_{BS3}}{16m^3}[\Vec{\sigma}\cdot\Vec{D},\Vec{B}]+\frac{gk_{BS4}}{16m^3}\{\Vec{\sigma}\cdot\Vec{D},\Vec{B}\}+\frac{gk_{BS5}}{16m^3}\Vec{\sigma}\Vec{B}\cdot\Vec{D} \\

         \\
          &&+\frac{k_{D00}}{16m^3}\{D_t,\{D_t,\Vec{D}\}\}+\frac{igk_{E01}}{16m^3}[D_t,\Vec{E}] +\frac{igk_{E02}}{16m^3}\{D_t,\Vec{E}\} +\frac{ik_{DS00}}{16m^3}\{D_t,\{D_t,\Vec{D}\times\Vec{\sigma}\}\}
          \\
          \\&&
         +\frac{gk_{ES01}}{16m^3}[D_t,\Vec{E}\times\Vec{\sigma}] +\frac{gk_{ES02}}{16m^3}\{D_t,\Vec{E}\times\Vec{\sigma}\},
    \end{array}
\end{equation}
and the boost generators satisfy the commutation rule
\begin{equation}
    \left[\left(1+i\frac{\Vec{q}_2}{m}\cdot\vec {\mathcal{K}}_x\right),\left(1+i\frac{\Vec{q}_1}{m}\cdot\vec {\mathcal{K}}_x\right)\right]N=i\frac{\Vec{q}_2\times\Vec{q}_1}{m^2}\cdot\frac{\Vec{\sigma}}{2}N.
\end{equation}
The LHS of the commutator together with the EOM becomes 
\begin{equation}
    \begin{array}{ll}
      
         &i\frac{\Vec{q}_2\times\Vec{q}_1}{m^2}\cdot\left\{\frac{k_{DS}}{2}\Vec{\sigma}+\frac{1}{4m^2}(k_{D3S}+k_{DS0}-c_2k_{DS}+c_2k_{DS0})\Vec{D}^2\Vec{\sigma }\right.\\
         \\
         &+\frac{1}{16m^2}(k_{DS}^2-k_{D3S}-2k_{DS0})\{\vec{D},\Vec{\sigma}\cdot\Vec{D}\}
         \\
         \\
         &-\frac{1}{16m^2}(4k_E-4k_D^2+k_{DS}^2+4k_{B1}+4k_{DS}c_F-4k_{DS0}c_F)e\Vec{B}\\
         \\
         &
    -\frac{i}{8m^2}(k_{ES}-k_Dk_{DS}-k_{BS4}+k_{BS5}+2k_{DS}c_F-2k_{DS0}c_F)e\Vec{B}\times\Vec{\sigma}\left.\right\}N,
    \end{array}
\end{equation}
which leads to
\begin{equation}\label{constraink}
    \textbf{Results from Commutator:}\begin{cases}
        k_{DS}=1,\\
         k_{D3S}+k_{DS0}-c_2+c_2k_{DS0}=0,\\
         1-k_{D3S}-2k_{DS0}=0,\\
         4k_E-4k_D^2+1+4k_{B1}+4c_F-4k_{DS0}c_F=0,\\
         k_{ES}-k_D-k_{BS4}+k_{BS5}+2c_F-2k_{DS0}c_F=0.
    \end{cases}
\end{equation}

In this work,  for the boost from the general expansion, it can be written as 
\begin{equation}\label{HQETboostgen}
    \begin{array}{lll}
        \vec{\mathcal{K}}_x &=&-i\frac{\vec{\sigma}}{2}f(\frac{D_{\mu}}{m})+m\Vec{x}
        \\
        \\&& =m\Vec{x}+\frac{c^{(1)}}{4m}\Vec{D}+\frac{ic^{(1)}}{4m}\vec{D}\times\Vec{\sigma}+\frac{gc_B^{(2)}}{16m^2}\Vec{E}-\frac{ic_A^{(2)}}{16m^2}\{D_t,\Vec{D}\}+\frac{igc_B^{(2)}}{16m^2}\Vec{E}\times\Vec{\sigma}+\frac{c_A^{(2)}}{16m^2}\{D_t,\Vec{D}\times\Vec{\sigma}\} \\
        \\
         & &+\frac{gc_a^{(3)}}{16m^3}\Vec{\sigma}\Vec{B}\cdot\Vec{D}+\frac{igc_b^{(3)}}{16m^3}(\Vec{D}\times\vec{B}+\Vec{B}\times\Vec{D})-\frac{gc_b^{(3)}}{16m^3}[\Vec{\sigma}\cdot\Vec{D},\Vec{B}]+\frac{gc_b^{(3)}}{16m^3}[\Vec{D},\Vec{\sigma}\cdot\Vec{B}]\\
         \\
         &&+\frac{igc_b^{\prime(3)}}{16m^3}(\Vec{D}\times\Vec{B}-\Vec{B}\times\Vec{D})-\frac{gc_b^{\prime(3)}}{16m^3}\{\Vec{\sigma}\cdot\Vec{D},\Vec{B}\}+\frac{gc_b^{\prime(3)}}{16m^3}\{\Vec{D},\Vec{\sigma}\cdot\Vec{B}\}-\frac{c_d^{(3)}}{32m^3}\{\Vec{D}^2,\Vec{D}\}\\
         \\
          &&-\frac{ic_d^{(3)}}{32m^3}\{\Vec{D}^2,\Vec{D}\times\Vec{\sigma}\}-\frac{igc_e^{(3)}}{32m^3}[D_t,\Vec{E}] +\frac{gc_e^{(3)}}{32m^3}[D_t,\Vec{E}\times\Vec{\sigma}]-\frac{igc_f^{(3)}}{32m^3}\{D_t,\Vec{E}\} +\frac{gc_f^{(3)}}{32m^3}\{D_t,\Vec{E}\times\Vec{\sigma}\}\\
          \\
          &&-\frac{igc_e^{(3)}}{32m^3}[D_t,\Vec{E}] +\frac{gc_e^{(3)}}{32m^3}[D_t,\Vec{E}\times\Vec{\sigma}]-\frac{igc_f^{(3)}}{32m^3}\{D_t,\Vec{E}\} +\frac{gc_f^{(3)}}{32m^3}\{D_t,\Vec{E}\times\Vec{\sigma}\}\\
          \\
         && +\frac{c_g^{(3)}}{32m^3}\{D_t,\{D_t,\Vec{D}\}\}+\frac{ic_g^{(3)}}{32m^3}\{D_t,\{D_t,\Vec{D}\times\Vec{\sigma}\}\},
    \end{array}
\end{equation}
with the constraints
\begin{equation}\label{eq:hqetcccondition1}
   \textbf{Results from Consistency Condition} \begin{cases}
        c^{(1)}=-1,\\
        -2c_2-c_2c_A^{(2)}+c_d^{(3)}=-1,\\
        -c_a^{(3)}-2c_F-c_Fc_A^{(2)}=-1,\\
        -c_B^{(2)}-2c_F-c_Fc_A^{(2)}-2c_b^{\prime(3)}=0,\\
        -c_A^{(2)}+c_d^{(3)}=1.
    \end{cases}
\end{equation}

Compared to the boost constrained by the commutation relation in Ref.~\cite{Berwein:2018fos}, the boost in Eq.~(\ref{HQETboostgen}) is constrained by the consistency condition (\ref{eq:hqetcccondition}). 
Note that these parameters in~\cite{Berwein:2018fos} can be related to the parameters in the boost in Eq.~(\ref{HQETboostgen}) by

\textbf{Conversion of Parameters:}
\[k_D=-\frac{c^{(1)}}{2},\quad k_{DS}=-c^{(1)},\quad k_E=\frac{c_B^{(2)}}{2},\quad k_{D0}=-\frac{c_A^{(2)}}{2},\]
\[k_{ES}=\frac{c_B^{(2)}}{2},\quad k_{DS0}=-\frac{c_A^{(2)}}{2},\quad k_{D3}=\frac{c_d^{(3)}}{4},\quad k_{D3S}=c_d^{(3)},\]
\[k_{B1}=c_b^{\prime(3)},\quad k_{B2}=c_{b}^{(3)},\quad k_{BS1}=c_b^{(3)},\quad k_{BS2}=c_b^{\prime(3)},\]
\[k_{BS3}=-c_b^{(3)},\quad k_{BS4}=-c_b^{\prime(3)},\quad k_{BS5}=\frac{c_a^{(3)}}{2}\]
\[k_{D00}=\frac{c_g^{(3)}}{2},\quad k_{E01}=-\frac{c_e^{(3)}}{2},\quad k_{E02}=-\frac{c_f^{(3)}}{2},\]
\begin{equation}\label{kwithourk}
    k_{DS00}=\frac{c_g^{(3)}}{2},\quad k_{ES01}=\frac{c_e^{(3)}}{2},\quad k_{ES02}=\frac{c_f^{(3)}}{2}.
\end{equation}
Note that in above numbers of independent parameters are different. Together with Eq.~(\ref{constraink}) and (\ref{kwithourk}), we find exactly the same constraint as in Eq.~(\ref{eq:hqetcccondition1}). 

\section{The Top-down Lagrangian}
\label{ahqet}

Having written down the HPET Lagrangian in the bottom-up approach, we construct the general form of $f(\frac{D_{\mu}}{m})$ and use the consistency condition to constrain it. Then we derive the covariant nonlinear boost transformation of the NR field $N(x)$ based on $f(\frac{D_{\mu}}{m})$, and obtain the relation among the Wilson coefficients of the HPET operators. On the other hand, in the top-down approach not only the relation among the Wilson coefficients, but also the form of each Wilson coefficient should be obtained by matching the UV Lagrangian to the HPET one. Although the Wilson coefficients are UV-dependent, if the most general UV Lagrangian is written down, the obtained Wilson coefficients should reproduce the one in the bottom-up approach. The only concern would be the integrating out should be beyond the tree-level one, which can be realized by replacing the tree-level form with the one containing the $f(\frac{D_{\mu}}{m})$.  



In the top-down approach, the QED Lagrangian is extended to the most general $P$ and $T$ even relativistic Lagrangian up to $1/m^3$  
\begin{equation}\label{rhqet}
    \begin{array}{lll}
         \mathcal{L}&=&\bar{\Psi}\left\{(iD\!\!\!\!/-m)-a_1g\frac{\sigma_{\mu\nu}F^{\mu\nu}}{m}-a_2g\frac{\gamma_{\mu}[\partial_{\nu}F^{\mu\nu}]}{m^2}\right.\\
         \\
         & &\left.+a_3g^2\frac{F_{\mu\nu}F^{\mu\nu}}{m^3}+a_4ig^2\frac{\gamma^5F_{\mu\nu}\tilde{F}^{\mu\nu}}{m^3}-a_5g\frac{\sigma_{\mu\nu}[\partial^2F^{\mu\nu}]}{m^3}\right\}\Psi+\mathcal{O}(1/m^4),
    \end{array}
\end{equation}
where $\sigma^{\mu\nu}=\frac{i}{2}[\gamma^{\mu},\gamma^{\nu}]$, $F^{\mu\nu}=\frac{-i}{g}[D^{\mu},D^{\nu}]$ and $\tilde{F}^{\mu\nu}=\frac{1}{2}\epsilon^{\mu\nu\rho\sigma}F_{\rho\sigma}$, and the gauge field is seen as the external field. There are five effective operators and $a_i$ are free parameters, where $i=1,2,3,4,5$. By considering the most general relativistic Lagrangian including the effective operators without redundancy up to certain order, we are allowed to obtain the complete information and the correct constraints on the NR Wilson coefficients up to the same order.

Given the most general relativistic Lagrangian, the tree-level integrating out of the anti-particle modes $B_v$ defined in Eq.~\eqref{eq:bvdef} gives 
\bea\label{eq:vbtree}
B_v =  \left[ \frac{iD\!\!\!\!/_{\perp}}{2m} +\frac{\{v\cdot D,D\!\!\!\!/_{\perp}\}}{8m^2}-(1+8a_1)\frac{g\sigma_{\mu\nu}v^{\mu}v_{\alpha}F^{\alpha\nu}}{8m^2}\right] Q_v.
\eea
In the rest frame $v^{\mu}=(1,0,0,0)$, above equation can be rewritten by the heavy field $N$ and $\tilde N$ using Eq.~\eqref{eq:restqvbvnchi}, and we obtain that
\begin{equation}
    \tilde N=\left(-\frac{i\vec{\sigma}\cdot\vec{D}}{2m}-\frac{\{D_t,\vec{\sigma}\cdot{\vec{D}}\}}{8m^2}-(1+8a_1)\frac{ig\vec{\sigma}\cdot\vec{E}}{8m^2}\right)N.
\end{equation}
The tree-level integrating out  of the QED Lagrangian has $a_1=0$, and it gives
\begin{equation}
    \tilde N=\left(-\frac{i\vec{\sigma}\cdot\vec{D}}{2m}-\frac{\{D_t,\vec{\sigma}\cdot{\vec{D}}\}}{8m^2}-\frac{ig\vec{\sigma}\cdot\vec{E}}{8m^2}\right)N.
\end{equation}
To incorporate the most general result beyond the tree-level, the tree-level EOM  are extended to the following parametrization
\bea\label{eq:vbvint}
f_v(\frac{D_{\mu}}{m}) Q_v = \left[ \frac{iD\!\!\!\!/_{\perp}}{2m} - c^{(2)}_A\frac{\{v\cdot D,D\!\!\!\!/_{\perp}\}}{8m^2}+c^{(2)}_B\frac{g\sigma_{\mu\nu}v^{\mu}v_{\alpha}F^{\alpha\nu}}{8m^2}\right] Q_v.
\eea
In the rest frame, the above equation reduces to
\begin{equation}
     f(\frac{D_{\mu}}{m})N=\left(-\frac{i\vec{\sigma}\cdot\vec{D}}{2m}+c^{(2)}_A\frac{\{D_t,\vec{\sigma}\cdot{\vec{D}}\}}{8m^2}+c^{(2)}_B\frac{ig\vec{\sigma}\cdot\vec{E}}{8m^2}\right)N.
\end{equation}
The parameters  $c_A^{(2)}$ and $c_B^{(2)}$ are constrained by the consistency condition. At the leading-order, they give the same results, while beyond the leading-order, integrating out is a special solution of the general expansion, as we explain in the subsection \ref{HQETcompare}.

Finally, let us perform (beyond) tree-level integrating out. 
After determining the anti-particle d.o.f.,  the Dirac field $\Psi(x)$ is expanded by the heavy field $N(x)$ according to
\begin{equation}
    \Psi(x)=e^{-imt}\left[\begin{array}{c}
         N(x)  \\
         f(\frac{D_{\mu}}{m})N(x) 
    \end{array}\right].
\end{equation}
The relativistic Lagrangian in Eq.~\eqref{rhqet} matches to the NR Lagrangian Eq.~(\ref{HQETnrlag}),
\begin{equation}
    \begin{array}{lll}
      \mathcal{L}&=&N^{\dagger}\left\{iD_t+c_2\frac{ \vec{D}^2}{2m}+c_Fg\frac{\vec{\sigma}\cdot\vec{B}}{2m}+c_Dg\frac{[D_i,E^i]}{8m^2}+ic_Sg\frac{\epsilon^{ijk}\sigma^i\{D_j,E^k\}}{8m^2}+c_4\frac{\vec{D}^4}{8m^3}\right.\\
      \\
        & & +c_{W1}g\frac{\{\vec{D}^2,\vec{\sigma}\cdot\vec{B}\}}{8m^3}-c_{W2}g\frac{D^i\vec{\sigma}\cdot\vec{B}D^i}{4m^3}+c_{p^{\prime}p}g\frac{\{\vec{\sigma}\cdot\vec{D},\vec{B}\cdot\vec{D}\}}{8m^3}\\
        \\
        &&\left.+ic_Mg\frac{\{D_i,\epsilon^{ijk}[D_j,B^k]\}}{8m^3}+c_{A1}g^2\frac{B^2-E^2}{8m^3}-c_{A2}g^2\frac{E^2}{16m^3}+\mathcal{O}(1/m^4)\right\}N.
    \end{array}
\end{equation}
The results using the integrated-out method up to $1/m^3$ is showed in subsection~\ref{RFIO}, while the results given by the general expansion up to $1/m^3$ is discussed in subsection~\ref{RFGE}. Detailed comparison between these two matching are exhibited in subsection~\ref{HQETcompare}.

\subsection{Integrating out}
\label{RFIO}

In this subsection we explicitly derive the Wilson coefficients by  tree-level integrating out the relativistic Lagrangian Eq.~(\ref{rhqet}).  For $v^{\mu}=(1,0,0,0)$, the Dirac spinor field is expanded as
 \begin{equation}
 	\Psi(x)=e^{-imt}\left[\begin{array}{c}
 		N(x)\\ \tilde N(x)
 	\end{array}\right],
 \end{equation}
 and then the relativistic Lagrangian (\ref{rhqet}) can be expressed as
\begin{equation}
	\mathcal{L}=N^{\dagger}(iD_t+O_A)N+N^{\dagger}(i\vec{\sigma}\cdot\vec{D}+O_B)\tilde N+\tilde N^{\dagger}(i\vec{\sigma}\cdot\vec{D}+O_B^{\dagger})N+\tilde N^{\dagger}(2m+iD_t+O_C)\tilde N,
\end{equation}
where for simplicity we define the variables
\begin{equation}
    \begin{cases}
        O_A\equiv\frac{2a_1}{m}g\vec{\sigma}\cdot\vec{B}+\frac{a_2}{m^2}g[D_i,E^i]+\frac{2a_3}{m^3}g^2(B^2-E^2)+\frac{2a_5}{m^3}g\{\vec{\sigma}\cdot\vec{B},\vec D^2\}-\frac{4a_5}{m^3}gD^i\vec{\sigma}\cdot\vec{B}D^i,\\
        O_B\equiv-\frac{2a_1}{m}ig\vec{\sigma}\cdot\vec{E}+\frac{a_2}{m^2}[D_t,\vec{\sigma}\cdot\vec{E}]+\frac{a_2}{m^2}g\epsilon^{ijk}\sigma^{i}[D^j,B^k],\\
        O_C\equiv-\frac{2a_1}{m}g\vec{\sigma}\cdot\vec{B}+\frac{a_2}{m^2}g[D_i,E^i].
    \end{cases}
\end{equation}
Using the EOM of $\tilde N^{\dagger}$, one can find that the heavy component is
\begin{equation}
	\tilde N=-(2m+iD_t+O_C)^{-1}(i\vec{\sigma}\cdot\vec{D}+O_B^{\dagger})N,
\end{equation}
with the explicit expression
\begin{equation}\label{eq:eomchi}
    \begin{array}{lll}
      \tilde N&=&\left\{-\frac{i\vec{\sigma}\cdot\vec{D}}{2m}-\frac{\{D_t,\vec{\sigma}\cdot{\vec{D}}\}}{8m^2}-(1+8a_1)\frac{ig\vec{\sigma}\cdot\vec{E}}{8m^2}+(-4a_1)\frac{ig\vec{B}\cdot\vec{D}}{8m^3}\right.\\
      \\
        & & +(-2a_1-4a_2)\frac{g\epsilon^{ijk}[D^i,B^j]\sigma^k}{8m^3}+(2a_1)\frac{g\epsilon^{ijk}\{D^i,B^j\}\sigma^k}{8m^3}+0\times\frac{\{(i\vec{D})^2,i\vec{\sigma}\cdot\vec{D}\}}{16m^3}\\
        \\
        &&\left.(4a_1+8a_2+1)\frac{g[D_t,\vec{\sigma}\cdot\Vec{E}]}{16m^3}-(4a_1+1)\frac{g\{D_t,\vec{\sigma}\cdot\vec{E}\}}{16m^3}+0\times \frac{i\{D_t,\{\Vec{\sigma}\cdot\vec{D},D_t\}\}}{16m^3}+\mathcal{O}(1/m^4)\right\}N.
    \end{array}
\end{equation}
Thus, after integrating out $\tilde N$, the Lagrangian after symmetry breaking with only the small components $N$ reads
\begin{equation}
    \begin{array}{lll}
    \mathcal{L}&=&N^{\dagger}(iD_t+O_A)N-N^{\dagger}(i\vec{\sigma}\cdot\vec{D}+O_B)(2m+iD_t+O_C)^{-1}(i\vec{\sigma}\cdot\vec{D}+O_B^{\dagger})N\\
    \\
        &=&N^{\dagger}\left\{iD_t+\frac{ \vec D^2}{2m}+(1+4a_1)g\frac{\vec{\sigma}\cdot\vec{B}}{2m}+(1+8a_1+8a_2)g\frac{[D_i,E^i]}{8m^2}\right.\\
        \\
         && +(1+8a_1)ig\frac{\epsilon^{ijk}\sigma^i\{D_j,E^k\}}{8m^2}+c_2\frac{\vec D^4}{8m^3}+(\frac{c_F+c_2}{2}+4a_2+16a_5)g\frac{\{\vec D^2,\vec{\sigma}\cdot\vec{B}\}}{8m^3}\\
         \\
         &&-(2a_1+4a_2+16a_5)g\frac{D^i\vec{\sigma}\cdot\vec{B}D^i}{4m^3}+4a_1g\frac{\{\vec{\sigma}\cdot\vec{D},\vec{B}\cdot\vec{D}\}}{8m^3}	+(2a_1+4a_2)ig\frac{\{D^i,\epsilon^{ijk}[D^j,B^k]\}}{8m^3}\\
         \\
         &&\left.+(c_F-4a_1+16a_3)g^2\frac{B^2-E^2}{8m^3}-(-2c_F+32a_1^2+24a_1+16a_2+2)g^2\frac{E^2}{16m^3}+\mathcal{O}(1/m^4)\right\}N.
    \end{array}
\end{equation}
 Matching it to the NR Lagrangian Eq.~\eqref{HQETnrlag}, we find the same relations of the Wilson coefficients  as those in the bottom-up approach in Eq.~\eqref{wcr},
\begin{equation}
    \begin{cases}
        c_2=1,\\
        c_4=1,\\
        c_S=2c_F-1,\\
        2c_M=c_D-c_F,\\
        c_{W2}=c_{W1}-1,\\
        c_{p^{\prime}p}=c_F-1,
    \end{cases}
\end{equation}
except for one more relation
\begin{equation}\label{additionwlc}
   c_{A2} =2c_{F}^2-4c_F+2c_D.
\end{equation}
Although we perform the tree-level integrating out in this part, we have started from the most general relativistic Lagrangian symmetry allowed in Eq.$~$(\ref{rhqet}). The results are discussed further in subsection~\ref{HQETcompare}.

\subsection{General Expansion}
\label{RFGE}

Instead of integrating out, starting from the relativistic Lagrangian Eq.~(\ref{rhqet}),  we can expand it based on the general expansion, 
\begin{equation}\label{gexp}
	\Psi(x)=e^{-imt}\left[\begin{array}{c}
		N(x)\\f(\frac{D_{\mu}}{m})N(x)
	\end{array}\right],
\end{equation}
with the same $f(\frac{D_{\mu}}{m})$ as Eq.~\eqref{fform} 
\begin{equation}\label{fform2}
    \begin{array}{lll}
         f(\frac{D_{\mu}}{m})&=&c^{(1)}\frac{i\vec{\sigma}\cdot\vec{D}}{2m}+c^{(2)}_A\frac{\{D_t,\vec{\sigma}\cdot{\vec{D}}\}}{8m^2}+c^{(2)}_B\frac{ig\vec{\sigma}\cdot\vec{E}}{8m^2}\\
         \\
         & &+c_a^{(3)}\frac{ig\vec{B}\cdot\vec{D}}{8m^3}+c_b^{(3)}\frac{g\epsilon^{ijk}[D^i,B^j]\sigma^k}{8m^3}+c_b^{\prime(3)}\frac{g\epsilon^{ijk}\{D^i,B^j\}\sigma^k}{8m^3}+c_d^{(3)}\frac{\{(i\vec{D})^2,i\vec{\sigma}\cdot\vec{D}\}}{16m^3}\\
         \\
         &&+c_e^{(3)}\frac{g[D_t,\vec{\sigma}\cdot\Vec{E}]}{16m^3}+c_f^{(3)}\frac{g\{D_t,\vec{\sigma}\cdot\vec{E}\}}{16m^3}+c^{(3)}_g\frac{i\{D_t,\{D_t,\Vec{\sigma}\cdot\vec{D}\}\}}{16m^3}.
    \end{array}
\end{equation}
Exactly the same as the bottom-up approach, the parameters in $f(\frac{D_{\mu}}{m})$ are constrained by the consistency condition Eq.~\eqref{eq:cccondition}
\begin{equation}
    \left(-e^{-i\vec{q}\cdot\vec{x}}f\left(\frac{B(q)_{\mu}^{\nu} D_{\nu}}{m}\right)e^{i\vec{q}\cdot\vec{x}}\frac{\vec{\sigma}\cdot\vec{q}}{2m}f(\frac{D_{\mu}}{m})+\frac{\vec{\sigma}\cdot\vec{q}}{2m}\right)N=	\left(e^{-i\vec{q}\cdot\vec{x}}f\left(\frac{B(q)_{\mu}^{\nu} D_{\nu}}{m}\right)e^{i\vec{q}\cdot\vec{x}}-f(\frac{D_{\mu}}{m})\right)N,
\end{equation}
and the EOM in Eq.~\eqref{EOMofN}
\begin{equation}
    iD_t N(x)=\left(-c_2\frac{\vec D^2}{2m}-c_Fg\frac{\vec\sigma\cdot\vec B}{2m}+...\right)N(x),
\end{equation}
such that these parameters in $f(\frac{D_{\mu}}{m})$ have the relations
\begin{equation}
    \begin{cases}
        c^{(1)}=-1,\\
        -2c_2-c_2c_A^{(2)}+c_d^{(3)}=-1,\\
        -c_a^{(3)}-2c_F-c_Fc_A^{(2)}=-1,\\
        -c_B^{(2)}-2c_F-c_Fc_A^{(2)}-2c_b^{\prime(3)}=0,\\
        -c_A^{(2)}+c_d^{(3)}=1,
    \end{cases}
\end{equation}
where $c_2, c_F$ are the undetermined Wilson coefficients in the NR Lagrangian Eq.~\eqref{HQETnrlag}.
Then we match the relativistic Lagrangian Eq.~\eqref{rhqet} to the NR Lagrangian Eq.~\eqref{HQETnrlag}, with the Wilson coefficients determined as
\begin{equation}
    \begin{array}{lll}
         \mathcal{L}&=&N^{\dagger}\left\{iD_t+\frac{\vec D^2}{2m}+(1+4a_1)g\frac{\vec{\sigma}\cdot\vec{B}}{2m}+(1+8a_1+8a_2)g\frac{[D_i,E^i]}{8m^2}\right.\\
         \\
         & &+(1+8a_1)ig\frac{\epsilon^{ijk}\sigma^i\{D_j,E^k\}}{8m^2}+c_2\frac{\vec D^4}{8m^3}+(\frac{c_2+1}{2}+2a_1+4a_2+16a_5)g\frac{\{\vec D^2,\vec{\sigma}\cdot\vec{B}\}}{8m^3}\\
         \\
         &&-(2a_1+4a_2+16a_5)g\frac{D^i\vec{\sigma}\cdot\vec{B}D^i}{4m^3}+4a_1g\frac{\{\vec{\sigma}\cdot\vec{D},\vec{B}\cdot\vec{D}\}}{8m^3}	+(2a_1+4a_2)ig\frac{\{D^i,\epsilon^{ijk}[D^j,B^k]\}}{8m^3}\\
         \\
         &&+(c_F-4a_1+16a_3)g^2\frac{B^2-E^2}{8m^3}-\left(-2(c_F-4a_1+16a_3)-\frac{1}{2}(c_A^{(2)}+c_B^{(2)})^2\right.\\
         \\
         &&
         \left.\left.-2(c_A^{(2)}+c_B^{(2)})-8a_1(c_A^{(2)}+c_B^{(2)})+16a_2+32a_3\right)g^2\frac{E^2}{16m^3}+\mathcal{O}(1/m^4)\right\}N,
    \end{array}
\end{equation}
which are also compatible with the one-loop level results in Refs.~\cite{Kinoshita:1995mt,Manohar:1997qy}. We eventually reproduce the relations Eq.~\eqref{wcr}
\begin{equation}\label{wilson relation }
    \begin{cases}
        c_2=1,\\
        c_4=1,\\
        c_S=2c_F-1,\\
        2c_M=c_D-c_F,\\
        c_{W2}=c_{W1}-1,\\
        c_{p^{\prime}p}=c_F-1.
    \end{cases}
\end{equation}
However, for $c_{A2}$, the  Wilson coefficient of the operator $\frac{g^2}{16m^3}N^{\dagger}E^2N$ defined in the NR Lagrangian Eq.~\eqref{HQETnrlag},  there are non-trivial contributions of $c^{(2)}_A\frac{\{D_t,\vec{\sigma}\cdot{\vec{D}}\}}{8m^2}$ and $c^{(2)}_B\frac{ig\vec{\sigma}\cdot\vec{E}}{8m^2}$ in the $f(\frac{D_{\mu}}{m})$.

\subsection{Matching and Comparison}
\label{HQETcompare}
 In this subsection, we compare the matching results from the integrating out and the general expansion. In the integrating out, Eq.~(\ref{eq:eomchi}) expresses the anti-particle component $\tilde N(x)$ in terms of the particle component $N(x)$, and can be understood as 
 \begin{equation}
     \tilde N(x)=f(\frac{D_{\mu}}{m})N(x).
 \end{equation}
 Comparing to the general construction $f(\frac{D_{\mu}}{m})$, integrating out is one special solution of the general expansion:
\[c^{(1)}=-1, \hspace{0.1in}c_A^{(2)}=-1, \hspace{0.1in}c_B^{(2)}=-1-8a_1,\]
\[c_a^{(3)}=-4a_1, \hspace{0.1in} c_b^{(3)}=-2a_1-4a_2, \hspace{0.1in} c_b^{\prime(3)}=2a_1,\hspace{0.1in} \]
\begin{equation}\label{onesetcoeff}
    c_d^{(3)}=0,\hspace{0.1in} c_e^{(3)}=-1-4a_1-8a_2,\hspace{0.1in} c_f^{(3)}=-1-4a_1,\hspace{0.1in}c_g^{(3)}=0,
\end{equation}
satisfying the consistency condition (\ref{eq:cccondition}) and (\ref{eq:hqetcccondition}). These result are also exhibited in Tab.~\ref{hqetcomparelist}.

\begin{table}\small
    \centering
    \begin{tabular}{|c|c|c|c|c|c|}
    \hline
    \multicolumn{6}{|c|}{Parameters of $f(\frac{D_{\mu}}{m})$}
         \\
         \hline
        $f(\frac{D_{\mu}}{m})$ &\scriptsize{General }&\scriptsize{Integrating }&  $f(\frac{D_{\mu}}{m})$&\scriptsize{General }&\scriptsize{Integrating }\\
         &\scriptsize{ Expansion}&\scriptsize{Out}&&\scriptsize{Expansion}&\scriptsize{Out}\\
         \hline
         $c^{(1)}\frac{i\vec{\sigma}\cdot\vec{D}}{2m}$& $c^{(1)}=-1$&$c^{(1)}=-1$&$c^{(2)}_A\frac{\{D_t,\vec{\sigma}\cdot{\vec{D}}\}}{8m^2}$&$c^{(2)}_A$&$c^{(2)}_A=-1$\\
         \hline
         $c^{(2)}_B\frac{ig\vec{\sigma}\cdot\vec{E}}{8m^2}$&$c^{(2)}_B$&$c^{(2)}_B=-1-8a_1$&$c_a^{(3)}\frac{ig\vec{B}\cdot\vec{D}}{8m^3}$&$c_a^{(3)}$&$c_a^{(3)}=-4a_1$\\
         \hline
         $c_b^{(3)}\frac{g\epsilon^{ijk}[D^i,B^j]\sigma^k}{8m^3}$&$c_b^{(3)}$&$c_b^{(3)}=-2a_1-4a_2$&$c_b^{\prime(3)}\frac{g\epsilon^{ijk}\{D^i,B^j\}\sigma^k}{8m^3}$&$c_b^{\prime(3)}$&$c_b^{\prime(3)}=2a_1$\\
         \hline
         $c_d^{(3)}\frac{\{(i\vec{D}^2),i\vec{\sigma}\cdot\vec{D}\}}{16m^3}$&$c_d^{(3)}$&$c_d^{(3)}=0$&$c_e^{(3)}\frac{g[D_t,\vec{\sigma}\cdot\Vec{E}]}{16m^3}$&$c_e^{(3)}$&$c_e^{(3)}=-1-4a_1-8a_2$\\
         \hline
         $c_f^{(3)}\frac{g\{D_t,\vec{\sigma}\cdot\vec{E}\}}{16m^3}$&$c_f^{(3)}$&$c_f^{(3)}=-1-4a_1$&$c^{(3)}_g\frac{i\{D_t,\{D_t,\Vec{\sigma}\cdot\vec{D}\}\}}{16m^3}$&$c^{(3)}_g$&$c^{(3)}_g=0$\\
         \hline
         \multicolumn{6}{|c|}{Consistency Condition}
         \\
         \hline
         \multicolumn{3}{|c|}{ General Expansion}&\multicolumn{3}{c|}{Integrating Out}\\
         \hline
           \multicolumn{3}{|c|}{$c^{(1)}=-1$}&\multicolumn{3}{c|}{$c^{(1)}=-1$}\\
         \hline
           \multicolumn{3}{|c|}{$-2c_2-c_2c_A^{(2)}+c_d^{(3)}=-1$}&\multicolumn{3}{c|}{$c_2=1,\quad-2+1+0=-1$}\\
         \hline
           \multicolumn{3}{|c|}{$	-c_a^{(3)}-2c_F-c_Fc_A^{(2)}=-1$}&\multicolumn{3}{c|}{$c_F=1+4a_1,\quad4a_1-2(1+4a_1)+(1+4a_1)=-1$}\\
         \hline
           \multicolumn{3}{|c|}{$-c_B^{(2)}-2c_F-c_Fc_A^{(2)}-2c_b^{\prime(3)}=0$}&\multicolumn{3}{c|}{$1+8a_1-2(1+4a_1)+(1+4a_1)-4a_1=0$}\\
         \hline
           \multicolumn{3}{|c|}{$-c_A^{(2)}+c_d^{(3)}=1$}&\multicolumn{3}{c|}{$1+0=1$}\\
         \hline
    \end{tabular}
    \caption{The parameters of $f(\frac{D_{\mu}}{m})$, and the constraints from the consistency condition.}\label{hqetcomparelist}
    \label{tab:f_comparison}
\end{table}
\begin{table}
    \centering
    \begin{tabular}{|c|c|c|}
    \hline
        Wilson coefficient & General Expansion & Integrating  Out\\
        \hline
         $c_2$&1 &1\\
          \hline
         $c_F$&$1+4a_1$ &$1+4a_1$\\
          \hline
         $c_D$& $1+8a_1+8a_2$&$1+8a_1+8a_2$\\
          \hline
         $c_S$& $1+8a_1$&$1+8a_1$\\
          \hline
         $c_4$&1&1\\
         \hline
         $c_{W1}$&$1+2a_1+4a_2+16a_5$ &$1+2a_1+4a_2+16a_5$\\
          \hline
         $c_{W2}$&$2a_1+4a_2+16a_5$ &$2a_1+4a_2+16a_5$ \\
          \hline
         $c_{p^{\prime}p}$& $4a_1$&$4a_1$\\
          \hline
         $c_M$& $2a_1+4a_2$& $2a_1+4a_2$\\
          \hline
         $c_{A1}$&$1+16a_3$ &$1+16a_3$\\
          \hline
         $c_{A2}$& 
         \textcolor{blue}{$-2-\frac{1}{2}(c_A^{(2)}+c_B^{(2)})^2-(2+8a_1)(c_A^{(2)}+c_B^{(2)})+16a_2$}&$16a_1+16a_2+32a_1^2$\\
         \hline  
    \end{tabular}
    \caption{The Wilson coefficients of the NR operators. }
    \label{Wilson coefficients of NR operators}
\end{table}

The Wilson coefficients of the HPET basis up to $1/m^3$ derived from the integrating out and the general expansion are listed in Tab.~\ref{Wilson coefficients of NR operators}. These Wilson coefficients have relations Eq.~\eqref{wcr}, and reduce to those tree-level matching results in Ref.~\cite{Manohar:1997qy} when $a_i=0, i=1,2,3,4,5$, with the non-perturbative relations Eq.~(\ref{wcr}) disappear.

Non-trivial contributions of the general expansion show up in the Wilson coefficient $c_{A2}$, 
which reduce to the result from the tree-level integrating out when the set of parameters in Eq.~(\ref{onesetcoeff}) is adopted. Specifically, the Wilson coefficient $c_{A2}$ in the general expansion is
\begin{equation}
    c_{A2}=-2-\frac{1}{2}(c_A^{(2)}+c_B^{(2)})^2-(2+8a_1)(c_A^{(2)}+c_B^{(2)})+16a_2.
\end{equation}
However, in the tree-level integrating out, due to Eq.~\eqref{onesetcoeff} we have
\begin{equation}\label{abvalue}
    c_A^{(2)}=-1,\quad c_B^{(2)}=-1-8a_1,
\end{equation}
then in the tree-level integrating out, the Wilson coefficient $c_{A2}$ reduces to 
\begin{eqnarray}\label{ca2reduce}
    c_{A2}&=&-2-\frac{1}{2}(c_A^{(2)}+c_B^{(2)})^2-(2+8a_1)(c_A^{(2)}+c_B^{(2)})+16a_2\nonumber\\
    &=&16a_1+16a_2+32a_1^2.
\end{eqnarray}
At the tree-level, the parameters $c_A^{(2)}=-1, c_B^{(2)}=-1-8a_1$ have determined values, by integrating out anti-particle components as in Eq.~\eqref{eq:eomchi}. But in the general expansion with $f(\frac{D_{\mu}}{m})$, the parameters $c_A^{(2)}$ and $c_B^{(2)}$ are independent even with respect to the $a_1$ in the UV Lagrangian. Although the results in the integrating out are included in the results from the general expansion through the reduction Eq.~\eqref{ca2reduce}, there are more information contained by the general expansion: the $c_{A2}$ is the Wilson coefficient of   $\frac{g^2}{16m^3}N^{\dagger}E^2N$, depending on the $c^{(2)}_A\frac{\{D_t,\vec{\sigma}\cdot{\vec{D}}\}}{8m^2}$ and $c^{(2)}_B\frac{ig\vec{\sigma}\cdot\vec{E}}{8m^2}$ in the $f(\frac{D_{\mu}}{m})$, and this deviation allowed by Lorentz symmetry, can originates from the loop correction. 
%

Notice that in the tree-level integrating out, apart from the 6 relations in Eq.~\eqref{wcr}, one additional relation exists: $c_{A2}$ can be written as 
\begin{equation}
   c_{A2} =2c_{F}^2-4c_F+2c_D,
\end{equation}
which are not found in our general expansion and in the bottom-up approach. In the tree-level integrating out, there are only 4 free parameters in the UV Lagrangian $a_1, a_2, a_3, a_5$, since $a_4$ $ig^2\frac{1}{m^3}$ $\bar\Psi\gamma^5F_{\mu\nu}\tilde{F}^{\mu\nu}\Psi$
 contributes at order $1/m^4$ and higher. Thus, up to $1/m^3$, among the 11 Wilson coefficients in the NR Lagrangian listed in Tab.~\ref{Wilson coefficients of NR operators}, there are naturally only 4 independent Wilson coefficients and $11-4=6+1$ relations in the tree-level integrating out. 
However, the bottom-up approach is free from the UV-dependence, the Lorentz symmetry only requires 6 relations Eq.~\eqref{wcr}, allowing $c_{A2}$ to be independent of $c_{F}$ and $c_D$ up to $1/m^3$. Unsurprisingly, our general expansion has the consistent results with our calculation in the bottom-up approach: it yields 6 relations in Eq.~\eqref{wcr} and 5 independent Wilson coefficients out of the 11 Wilson coefficients in Tab.~\ref{Wilson coefficients of NR operators}. These discussions are summarized in Tab.~\ref{wcrcomparing}.
\begin{table}
    \centering\small
    \begin{tabular}{|c|c|c|c|}
    \hline
        &$\begin{array}{c}
            \text{Tree-level} 
            \\\text{Integrating out} 
        \end{array}$& $\begin{array}{c}
            \text{General} 
            \\\text{Expansion} 
        \end{array}$& $\begin{array}{c}
            \text{Bottom-up} 
            \\\text{approach}
            \\
            \text{(boost/RP)}
        \end{array}$  \\
         \hline
         &&&\\
         UV Parameters & $a_1, a_2, a_3, a_5$&$a_1, a_2, a_3, a_5$& 
         $\times$
         \\
           &&&\\
         \hline
    
         $
    \begin{array}{l}
          c_2=1,\\
        c_4=1,\\
        c_S=2c_F-1,\\
        2c_M=c_D-c_F,\\
        c_{W2}=c_{W1}-1,\\
        c_{p^{\prime}p}=c_F-1.
    \end{array}
$&$\surd$&$\surd$&$\surd$
\\

\hline
  &&&\\
$
   c_{A2} =2c_{F}^2-4c_F+2c_D$&$\surd$&$\times$&$\times$
   \\
     &&&\\
     \hline
     $\begin{array}{c}
         \text{Independent}  \\
          c_F, c_D, c_{W1}, c_{A1}
     \end{array}$&$\surd$&$\surd$&$\surd$
     \\
     \hline
     $\begin{array}{c}
         \text{Independent}  \\
          c_{A2}
     \end{array}$&$\times$&$\surd$&$\surd$
     \\
     \hline
    \end{tabular}
    \caption{There are only 6 relations of the Wilson coefficients in the general expansion and in the bottom-up approach.}
    \label{wcrcomparing}
\end{table}

\section{Conclusion}

\begin{figure}
    \centering
\includegraphics[height=15cm, width=15cm]{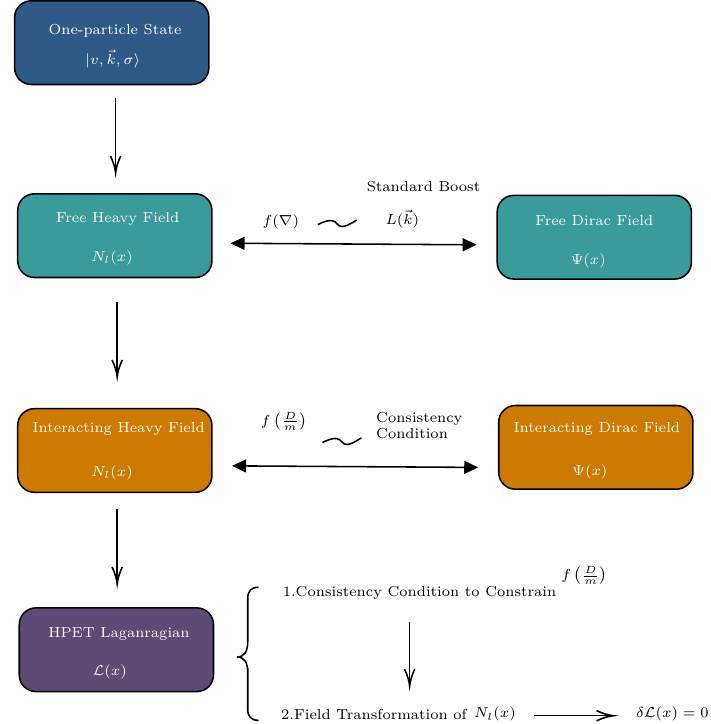}
    \caption{The workflow of the constructive HPET.}
    \label{figflow}
\end{figure}

Through the nonlinear realization of the spontaneously broken Poincar\'{e} symmetry, we present a  step-by-step construction of the Heavy Particle Effective Theory, where the workflow is exhibited in Fig.~\ref{figflow}. We begin by constructing the heavy one-particle state in the context of spontaneously broken Poincar\'{e} symmetry. With the translational symmetry remaining unbroken, the homogeneous Lorentz group is broken in such a way that only the rotational symmetry is preserved.  Using the coset construction of the Poincar\'{e} group, the boost transformation of the heavy one-particle state is obtained.  In this framework, the shift symmetry originating from the spacetime symmetry breaking pattern, leads to the reparameterization invariance.

We then establish the nonlinear boost transformation for the heavy field $N(x)$ by the properly determined non-relativistic wave function. The central element in the boost transformation is the $ f(\frac{D_{\mu}}{m})$. It appears as a set of parameterized non-relativistic operators, which are constructed from the  derivatives and external fields. The parameters in $ f(\frac{D_{\mu}}{m})$ are constrained by the consistency conditions derived from the linear transformation of the relativistic Dirac field under the Lorentz boost. This $ f(\frac{D_{\mu}}{m})$ plays two crucial roles: 
\begin{itemize}
    \item It governs the boost transformation of the non-relativistic field $N(x)$ through the  nonlinear boost generator $\vec{\mathcal{K}}_x$,
    \begin{equation}
      \vec{\mathcal{K}}_x=\vec{\mathcal{K}}+m\vec{x},\quad\text{where}\quad  \vec{\mathcal{K}}=-i\frac{\vec\sigma}{2}f(\frac{D_{\mu}}{m}).
    \end{equation}

    \item It encodes the anti-particle information via 
    \begin{equation}
        \tilde N(x)=f(\frac{D_{\mu}}{m})N(x).
    \end{equation}
    
\end{itemize}
 Compared to the previous work, the $f(\frac{D_{\mu}}{m})$ in this work is generally  parameterized and constrained by the consistency condition from the Lorentz symmetry. It is extended beyond the tree-level results.

After that, the effective Lagrangian is constructed through both the bottom-up and the top-down approach:
\begin{itemize}
    \item In the bottom-up approach, we start from the non-relativistic Lagrangian and impose the invariance under the  boost transformation, which generates non-trivial relations among the Wilson coefficients. 

    \item In the top-down approach, we utilize the general expansion including $f(\frac{D_{\mu}}{m})$ to represent the relativistic Dirac field in terms of the non-relativistic field, allowing the  $SL(2,C)$-invariant relativistic Lagrangian to be systematically expanded into the  $SU(2)$-invariant non-relativistic form with the Wilson coefficients directly determined. 
\end{itemize}
Both approaches yield compatible results, with several significant findings: 
\begin{itemize}

\item Up to $\mathcal{O}(1/m^3)$, we derive the relations among the Wilson coefficients, given in Eq.~\eqref{wcr}.

\item At $\mathcal{O}(1/m^3)$, both methods reveal the non-trivial contributions from the gauge field-dependent term in constraining the Wilson coefficient, which are absent in the little group transformation~\cite{Heinonen:2012km} in the bottom-up approach.

\item Up to $\mathcal{O}(1/m^3)$, in the bottom-up approach, our constructed boost generator satisfies the Lorentz boost commutation relations, with fewer free parameters compared to the previous work~\cite{Berwein:2018fos}. 

\item Up to $\mathcal{O}(1/m^3)$, in the top-down approach, the tree-level anti-particle equation of motion represents a special solution of our consistency condition. Besides, at $\mathcal{O}(1/m^3)$, the tree-level integrating out yields an additional relation,  Eq.~\eqref{additionwlc}, which is absent in either the bottom-up or the top-down approach using $f(\frac{D_{\mu}}{m})$. This indicates that our general construction corrects the tree-level matching.
\end{itemize}

Our constructive procedure is fundamentally based on the symmetry considerations and remains independent of specific ultraviolet completions. The framework can be systematically extended to higher orders in the power counting, providing a powerful tool for constructing and matching various non-relativistic effective theories. Potential applications include the heavy quark effective theory, nucleon-nucleon interactions~\cite{Girlanda:2010ya,Filandri:2023qio}, three-nucleon contact operators~\cite{Girlanda:2011fh,Nasoni:2023adf}, and other systems where the separation of relativistic and non-relativistic degrees of freedom is essential. Furthermore, the methodology of utilizing the little group index of $v^{\mu}$ is directly applicable to Heavy Black Hole Effective Theory~\cite{Damgaard:2019lfh} and to the on-shell scattering amplitude constructions~\cite{Aoude:2020onz} for the gravitational wave in the binary black hole systems. 

\section*{Acknowledgements}
We would like to thank Yu-Ming Wang for valuable discussion. This work is supported by the National Science Foundation of China under Grants No. 12347105, No. 12375099 and No. 12047503, and the National Key Research and Development Program of China Grant No. 2020YFC2201501, No. 2021YFA0718304.

\appendix

\begin{appendix}

\section{Little Group Transformation}
\label{ap1}

The standard boost $\Lambda(w,v)$ that transforms $mv$ to $mw$ is defined as 
\begin{equation}
    \Lambda(w,v)\equiv\text{exp}[-iJ_{\alpha\beta}w^{\alpha}v^{\beta}\theta],
\end{equation}
where $J_{\alpha\beta}$ is the generator of the Lorentz group and $\theta\equiv\frac{\sinh^{-1}\left(\sqrt{(w\cdot v)^2-1}\right)}{ \sqrt{(w\cdot v)^2-1}}$ for $w^2=v^2=1$. 
For the spinor representation $J^{\rho\sigma}=\frac{i}{4}(\gamma^\rho\gamma^\sigma-\gamma^\sigma\gamma^\rho)$, one obtains
\begin{align}\label{eq:StandardBoostInSpinorRep}
\Lambda_{\frac{1}{2}}(w,v) &\equiv \frac{1+w \mkern -9.5 mu / v \mkern -9.5 mu /}{\sqrt{2(1+w\cdot v)}}.
\end{align}
Specially, if $v^\mu=(1,0,0,0)$ and $p/m=w$, then we find
\bea
\Lambda_{\frac{1}{2}}(\frac{p}{m},v)&=&\begin{pmatrix}
    S_L&0\\
    0&S_R
\end{pmatrix}
=\sqrt{\frac{m}{2(m+E)}}
\begin{pmatrix}
    \mathbf{I}_{2\times2}+\frac{p\cdot\sigma}{m}&0\\
    0&\mathbf{I}_{2\times2}+\frac{p\cdot\Bar{\sigma}}{m}
\end{pmatrix}\nonumber\\
&=&\sqrt{\frac{1}{2m(m+E)}}
\begin{pmatrix}
    (m+E)\mathbf{I}_{2\times2}-\Vec{p}\cdot \Vec{\sigma}&0\\
    0&(m+E)\mathbf{I}_{2\times2}+\Vec{p}\cdot \Vec{\sigma}
\end{pmatrix}.
\eea
For the vector representation $(J^{\rho\sigma})^{\mu}_{~\nu}=i(g^{\rho\mu}\delta^{\sigma}_{~\nu}-g^{\sigma\mu}\delta^{\rho}_{~\nu})$, one obtains
\begin{align}\label{eq:StandardBoostInVectorRep}
\Lambda_{~\ \nu}^{\mu}(w,v) &\equiv g_{\nu}^{\mu}+(w^{\mu}v_{\nu}-v^{\mu}w_{\nu})-\frac{w^{\mu}w_{\nu}+v^{\mu}v_{\nu}-w\cdot v(w^{\mu}v_{\nu}+v^{\mu}w_{\nu})}{1+w\cdot v}\nonumber\\
 & =1+wv-vw-\frac{ww+vv-w\cdot v(wv+vw)}{1+w\cdot v}.
\end{align}
Notice that the indices are left inexplicit for simplicity, for example,  $wv$ refers to $w^{\mu}v_{\nu}$. This choice of standard boost satisfying  $\Lambda(w,v)v=w$ as in Ref.\cite{Luke:1992cs}. Recall that for the conventional transformation $\text{exp}[-\frac{i}{2}J_{\alpha\beta}\omega^{\alpha\beta}]$ with parameter $\omega^{\alpha\beta}$ in the special relativity, a common Lorentz boost is then written by 
\begin{equation}\label{usualboost}
    S(\eta)\equiv\text{exp}[-iJ_{0i}\omega^{0i}]=\text{exp}[-i\boldsymbol{K}\cdot\boldsymbol{\eta}],
\end{equation}
where $K^i\equiv J^{0i}$ is the boost generator and $\eta^i\equiv\omega^{i0}$ is the boost parameter. Therefore, the standard boost $\Lambda(w,v)$ we choose is exactly the boost Eq.~(\ref{usualboost}) with the rapidity 
\begin{equation}
    \eta^i=w^i\theta=\frac{w^i}{|\boldsymbol{w}|}\sinh^{-1}|\boldsymbol{w}|,
\end{equation}
for $v^{\mu}=(1,0,0,0)$.

For a infinitely small transformation, a general Lorentz boost transformation (in vector and spinor representation) can be parametrized by the standard boost with three parameters as
\begin{equation}\label{boostbq}
    B(q)\equiv \Lambda(v-q,v)=1+vq-qv+\mathcal{O}(q^{2}),~\ B_{\frac{1}{2}}(q)=1-\frac{q \mkern -9.5 mu / v \mkern -9.5 mu /}{2},
\end{equation}
with $v^2=(v-q)^2=1$ and thus we obtain
\begin{equation}\label{eq:bqw}
B(q)w=w+v(w\cdot q)-q(w\cdot v)= w+t,
\end{equation}
with $t\equiv v(w\cdot q)-q(w\cdot v)$. The induced little group transformation is then
\begin{equation}
    W[B,w]\equiv \Lambda(Bw,v)^{-1}B(q)\Lambda(w,v).
\end{equation}
To derive the explicit form of $W(B,w)$,  firstly we 
define a variable $a\equiv wv-vw$, one can check that 
\begin{align}
    a^{2} & =-(ww+vv)+w\cdot v(wv+vw),\nonumber\\
    a^{3} & =((w\cdot v)^{2}-1)a,\nonumber\\
    a^{4} & =((w\cdot v)^{2}-1)a^{2},
\end{align}
and thus obtain a compact expression 
\begin{equation}
    \Lambda(w,v)=1+a+\frac{a^{2}}{1+w\cdot v}.
\end{equation}
The validity of this boost transformation $\Lambda(w,v)$ is examined, as it's easy to see that for the metric $g_{\mu\nu}=(1,-1,-1,-1)$, the following equation holds
\begin{equation}
g^{-1}\Lambda^{T}g\Lambda=(1-a+\frac{a^{2}}{1+w\cdot v})(1+a+\frac{a^{2}}{1+w\cdot v})=1.
\end{equation}
Thus given in the above $\Lambda^T$ we can perform the replacement $a$ as $a\rightarrow a+b$ with $b\equiv tv-vt$, and $w \rightarrow Bw= w + t$ in Eq.~\eqref{eq:bqw}, and thus obtain 
\begin{align}
\Lambda(Bw,v)^{-1} = \Lambda(w+t,v)^{-1}= & 1-a-b+\frac{(a+b)^{2}}{1+w\cdot v}-\frac{a^{2}(t\cdot v)}{(1+w\cdot v)^{2}}.
\end{align}

The little group transformation can be written as the product of two parts
\begin{equation}\label{2termW}
W[B,w]=\left(\Lambda(w+t,v)^{-1}\Lambda(w,v)\right)\left(\Lambda(w,v)^{-1}B\Lambda(w,v)\right).
\end{equation}
We will calculate each part separately. 
For the first part, 
%
with the following replacements
\begin{align}
    ab&=wv(v\cdot t)-vv(w\cdot t)-wt+vt(w\cdot v),\nonumber\\
    ba&=tv(w\cdot v)-tw-vv(w\cdot t)+vw(v\cdot t),\nonumber\\
    ab-ba&=a(v\cdot t)-(wt-tw)-b(w\cdot v),\nonumber\\
    aba & =(wv-vw)(v\cdot t)(w\cdot v)=(v\cdot t)(w\cdot v)a,
\end{align}
we obtain the following expression for the first part
\begin{align}\label{firsttermW}
\Lambda(w+t,v)^{-1}\Lambda(w,v) & =(1-a-b+\frac{(a+b)^{2}}{1+w\cdot v}-\frac{a^{2}(t\cdot v)}{(1+w\cdot v)^{2}})(1+a+\frac{a^{2}}{1+w\cdot v})\nonumber\\
 & =1+(-b+\frac{ab+ba}{1+w\cdot v}-\frac{a^{2}(t\cdot v)}{(1+w\cdot v)^{2}})(1+a+\frac{a^{2}}{1+w\cdot v})\nonumber\\
 & =1-b+\frac{ab-ba}{1+w\cdot v}-\frac{a^{2}(t\cdot v)(w\cdot v)}{(1+w\cdot v)^{2}}+\frac{aba}{1+w\cdot v}-\frac{a(t\cdot v)(w\cdot v-1)}{(1+w\cdot v)}\nonumber\\
 &+\frac{aba^{2}}{(1+w\cdot v)^{2}}.
\end{align}
While for the second part in Eq.~(\ref{2termW}) it reads
\begin{align}
\Lambda(w,v)^{-1}\Lambda(v-q,v)\Lambda(w,v)= & (1+a+\frac{a^{2}}{1+w\cdot v})(1+vq-qv)(1-a+\frac{a^{2}}{1+w\cdot v})\nonumber\\
= & 1-d+(e+d(w\cdot v))-\frac{a^{2}d+da^{2}}{1+w\cdot v},
\end{align}
where we have used $q=\frac{v(q\cdot w)-t}{w\cdot v}$, $d\equiv qv-vq,e\equiv wq-qw$. With the following replacements
\begin{equation}
    \begin{array}{rll}
         -ad&=&wv(v\cdot q)-vv(w\cdot q)-wq+vq(w\cdot v),  \\
         -da&=&qv(w\cdot v)-qw-vv(w\cdot q)+vw(v\cdot q),\\
         -(ad-da)&=&a(v\cdot q)-e-d(w\cdot v),\\
         ada & =&(wv-vw)(v\cdot q)(w\cdot v)=(v\cdot q)(w\cdot v)a=0,\\
         -a^{2}d & =&vq((w\cdot v)^{2}-1)+wv(q\cdot w)-w\cdot v(q\cdot w)vv,\\
         -da^{2} & =&-vw(q\cdot w)+(w\cdot v)(q\cdot w)vv+qv(1-(w\cdot v)^{2}),\\
         a^{2}d+da^{2}&=&-  (vq-qv)((w\cdot v)^{2}-1)+(w\cdot q)(wv-vw),\
    \end{array}
\end{equation}
we eventually obtain the result for the second part
\begin{align}\label{secondtermW}
 & \Lambda(w,v)^{-1}\Lambda(v-q,v)\Lambda(w,v)\nonumber\\
= & 1-d-(e+d(w\cdot v))-\frac{a^{2}d+da^{2}}{1+w\cdot v}\nonumber\\
= & 1+(vq-qv)(1+w\cdot v)-(wq-qw)+\frac{(vq-qv)((w\cdot v)^{2}-1)-(w\cdot q)(wv-vw)}{1+w\cdot v}\nonumber\\
= & 1+2(vq-qv)(w\cdot v)-(wq-qw)-\frac{(w\cdot q)(wv-vw)}{1+w\cdot v}.
\end{align}
Combining Eq.~(\ref{firsttermW}) and (\ref{secondtermW}), expanding to the leading order of boost parameter $q$ and noticing that
$q\cdot w=v\cdot t$, the final result of $W[B,w]$ is 
\begin{align}
W[B,w] &=\left(\Lambda(w+t,v)^{-1}\Lambda(w,v)\right)\left(\Lambda(w,v)^{-1}B\Lambda(w,v)\right)\nonumber\\
 &= 1+b-(wq-qw)+\frac{a(v\cdot t)-c-b(w\cdot v)}{1+w\cdot v}\nonumber\\
& =1-\frac{(qv-vq)w\cdot v}{1+w\cdot v}+\frac{(qw-wq)}{(1+w\cdot v)}\nonumber\\
& =1+\frac{(qw_{\perp}-w_{\perp}q)}{(1+w\cdot v)},
\end{align}
where we have used the definition $w_{\perp}=w-v(w\cdot v)$. Notice that, the vector representation of the Lorentz generator is $(J^{\rho\sigma})^{\mu}_{~\nu}=i(g^{\rho\mu}\delta^{\sigma}_{~\nu}-g^{\sigma\mu}\delta^{\rho}_{~\nu})$, therefore, in the general representation it is
\begin{equation}
    W[B,w]=1-\frac{i}{2}\frac{(q_{\alpha}w_{\perp\beta}-w_{\perp\alpha}q_{\beta})}{(1+w\cdot v)}J^{\alpha\beta}.
\end{equation}

\section{Goldstone Theorem for Boost Breaking}\label{ap2}

Since the boost symmetry of the physical vacuum breaks, $ K^i|\Omega\rangle\neq0$, the boost reads
\begin{equation}
     K^i=\int d^3\vec x \mathcal{J}^{0,i}(x),
\end{equation}
where $\mathcal{J}^{\mu,i}(x)$ is the Noether current of the $i$ transformation.  Thus for some operator $\mathcal{O}$, 
\begin{equation}
\begin{array}{lll}
    \langle \Omega|[K^i,\mathcal{O}]|\Omega\rangle&=&\langle \Omega|\int d^3\vec x \mathcal{J}^{0i}(x)\mathcal{O}|\Omega\rangle-\langle \Omega|\int d^3\vec x \mathcal{O}\mathcal{J}^{0i}(x)|\Omega\rangle,\\
    \\
    &=&\int \frac{d^3\vec k}{(2\pi)^3} \sum_{n}\int d^3\vec x\left\{ 
    \langle \Omega|\mathcal{J}^{0i}(x)|n,\vec k\rangle\langle n,\vec k|\mathcal{O}|\Omega\rangle\right.
    \\
    \\
    &&
    \left.-\langle \Omega|\mathcal{O}|n,\vec k\rangle\langle n,\vec k|\mathcal{J}^{0i}(x)|\Omega\rangle\right\},
    \end{array}
\end{equation}
here $|\Omega\rangle=|v,0,0\rangle$, $\mathbf{1}=\int \frac{d^3\vec k}{(2\pi)^3} \sum_{n}|n,\vec k\rangle\langle n,\vec k|$ serves as the expansion of intermediate excitation states, $n$ denotes the undetermined quantum number. Using $\mathcal{J}^{0i}(x)=e^{i\hat kx}\mathcal{J}^{0i}(0)e^{-i\hat kx}$, we find
\begin{equation}\label{GBeq1}
    \begin{array}{lll}
         \langle \Omega|[K^i,\mathcal{O}]|\Omega\rangle&=&\int \frac{d^3\vec k}{(2\pi)^3} \sum_{n}\int d^3\vec x\left\{ 
    \langle \Omega|\mathcal{J}^{0i}(0)|n,\vec k\rangle\langle n,\vec k|\mathcal{O}|\Omega\rangle e^{-ikx}\right.
     \\
    \\
    &&
    \left.-\langle \Omega|\mathcal{O}|n,\vec k\rangle\langle n,\vec k|\mathcal{J}^{0i}(0)|\Omega\rangle e^{ikx}\right\},\\
    \\
    &=&\int d^3\vec k\delta^3(\vec k)\sum_{n}\left\{ 
    \langle \Omega|\mathcal{J}^{0i}(0)|n,\vec k\rangle\langle n,\vec k|\mathcal{O}|\Omega\rangle e^{-ik^0t}\right.
     \\
    \\
    &&
    \left.-\langle \Omega|\mathcal{O}|n,\vec k\rangle\langle n,\vec k|\mathcal{J}^{0i}(0)|\Omega\rangle e^{ik^0t}\right\}\neq 0.
    \end{array}
\end{equation}
Besides,  a conserved current $\partial_{\mu}\mathcal{J}^{\mu i}=0$, then 
\begin{equation}
         0=\int d^3\vec x[\partial_{\mu}\mathcal{J}^{\mu i}(x),\mathcal{O}]= \frac{d}{dt}[K^i,\mathcal{O}], 
\end{equation}
and therefore,
\begin{equation}\label{GBeq2}
    \begin{array}{lll}
        \langle \Omega| \frac{d}{dt}[K^i,\mathcal{O}]|\Omega\rangle& =&-i\int d^3\vec k\delta^3({\vec k})k^0\sum_{n}\left\{ 
    \langle \Omega|\mathcal{J}^{0i}(0)|n,\vec k\rangle\langle n,\vec k|\mathcal{O}|\Omega\rangle e^{-ik^0t}\right.
     \\
    \\
    &&
    \left.+\langle \Omega|\mathcal{O}|n,\vec k\rangle\langle n,\vec k|\mathcal{J}^{0i}(0)|\Omega\rangle e^{ik^0t}\right\}=0.
    \end{array}
\end{equation}
In order that Eq.~(\ref{GBeq1}) and Eq.~(\ref{GBeq2}) hold, there must be state $|n,\vec k\rangle$ such that
\begin{equation}\label{gbeq}
\begin{cases}
     \langle \Omega|\mathcal{O}|n,\vec k\rangle\langle n,\vec k|\mathcal{J}^{0i}|\Omega\rangle\neq 0,\\
     \\
    k^0=0,\quad\text{when}  \quad\vec k=0.
     
\end{cases}
\end{equation}

\section{Boost Transformation and Generators}\label{app3}

We discuss the origin of the phase factor in the transformation of heavy field.
Similar to the Eq.~\eqref{relativisticphilambda}, a general Lorentz transformation for relativistic field $\Psi(x)$ is 
\begin{eqnarray}
    \Psi^{\prime}(x')=U(\Lambda)^{-1}\Psi(x)U(\Lambda)=D(\Lambda)\Psi(\Lambda^{-1}x),
\end{eqnarray}
where the transformation of field index $\{\alpha\}$ is
\begin{eqnarray}
    \delta_0\Psi\equiv\Psi^{\prime}(x')-\Psi(x')=\left(D(\Lambda)-1\right)\Psi(x'),
\end{eqnarray}
such that in this variation $\delta_0\Psi$ the coordinate $x'$ (or $x$) could be suppressed. However, this is not the entire transformation including the coordinate. The missing part is 
\begin{equation}
\begin{array}{lll}
     \delta_x\Psi&=&\Psi^{\prime}(x^{\prime})-\Psi(x)-\delta_0\Psi
     =\Psi(x+\delta x)-\Psi(x)
     \\
    & =&\delta x^{\mu}\partial_{\mu}\Psi(x),
\end{array}
\end{equation}
where $\delta x\equiv x'-x$. Here, the coordinates $x$ transform as
\begin{eqnarray}
    \delta x^{\mu}=-\frac{i}{2}\omega_{\rho\sigma}(J^{\rho\sigma})^{\mu}_{~\nu}x^{\nu},
\end{eqnarray}
where $\omega_{\rho\sigma}$ is the parameter, and $(J^{\rho\sigma})^{\mu}_{~\nu}=i(g^{\rho\mu}\delta^{\sigma}_{~\nu}-g^{\sigma\mu}\delta^{\rho}_{~\nu})$ is the Lorentz generator in the vector representation. Under the boost with parameter $\omega_{0i}=\eta^i$, we have
\begin{eqnarray}
\delta_x\Psi=\frac{1}{2}\omega_{\rho\sigma}\left(x^{\sigma}\partial^{\rho}-x^{\rho}\partial^{\sigma}\right)\Psi(x)
      =i\vec\eta\cdot\left(\vec x~i\partial_t+t~i\vec\nabla\right)\Psi(x).
\end{eqnarray}

To derive the transformation of $N^0_{\ell}(x)$, notice that the transformation of the annihilation operator under the Lorentz transformation $\Lambda  $ is
\begin{equation}
    U(\Lambda  )^{-1}a_{v, k}^{\sigma}U(\Lambda  )=D_{\sigma\sigma'}^{(s)}(W[\Lambda  ,p])a_{v,k'}^{\sigma'},
\end{equation}
where $k'$ is obtained by Eq.~\eqref{klorentz}, 
\begin{equation}\label{klorentz1}
    k'=\Lambda  ^{-1} k+m\Lambda  ^{-1} v-mv.
\end{equation}
Besides, the wave function $u_{\ell}^{\sigma}(v)$  translates the change in the spin index $\sigma$ into the field index $\ell$.  Notice that $W[R,p]=R$. Therefore, the corresponding wave function $u_{\ell}^{\sigma}(v)$ should have the following transformation property under the  Lorentz transformation $\Lambda  $ as
\begin{equation}
    \sum_{\sigma}u_{\ell}^{\sigma}(v)D^{(s)}_{\sigma\sigma^{\prime}}(W[\Lambda  ,p])= D_{\ell\bar \ell}(W[\Lambda  ,p])u_{\bar \ell}^{\sigma}(v).
\end{equation}
As a result, $N^0_{\ell}(x)$ transforms under the Lorentz transformation $\Lambda  $ as
\begin{equation}
    \begin{array}{lll}
        U(\Lambda  )^{-1}N^0_{\ell}(x)U(\Lambda  ) &=&\int\frac{d^3\vec k}{(2\pi)^3\sqrt{2E}}\sum_{\sigma}u_{\ell}^{\sigma}(v)U(\Lambda  )^{-1}a_{v,k}^{\sigma}U(\Lambda  )e^{-ikx}  \\
        \\
         &=& \int\frac{d^3\vec k}{(2\pi)^3\sqrt{2E}}\sum_{\sigma}u_{\ell}^{\sigma}(v)D_{\sigma\sigma'}^{(s)}(W[\Lambda  ,p])a_{v,k'}^{\sigma'}e^{-ikx} 
         \\
         \\
         &=&\int\frac{d^3\vec k}{(2\pi)^3\sqrt{2E}}D_{\ell\bar \ell}(W[\Lambda  ,p])u_{\bar \ell}^{\sigma}(v)a_{v,k'}^{\sigma'}e^{-ikx} 
         \\
         \\
         &=&e^{-im(\Lambda   v-v)x}\int\frac{d^3\vec k'}{(2\pi)^3\sqrt{2E'}}D_{\ell\bar \ell}(W[\Lambda  ,p])u_{\bar \ell}^{\sigma}(v)a_{v,k'}^{\sigma'}e^{-ik'(\Lambda  ^{-1}x)} 
        \\
        \\
        &=&e^{-im(\Lambda   v-v)x}D_{\ell\bar \ell}(W[\Lambda  ,i\vec\nabla]) N^0_{\bar \ell}(\Lambda  ^{-1}x).
    \end{array}
\end{equation}
In the penultimate line, we have used the transformation property of $k^{\mu}$ Eq.~\eqref{klorentz1} and
\begin{equation}
    \begin{array}{lll}
         e^{-ikx}&=&e^{-i(\Lambda  ^{-1}k)(\Lambda  ^{-1}x)}  \\
         &=&e^{-i(k'-m\Lambda  ^{-1}v+mv)(\Lambda  ^{-1}x)} \\
         &=&e^{-ik'(\Lambda  ^{-1}x)}\times e^{-im(\Lambda   v-v)x}.
    \end{array}
\end{equation}
Therefore,  the field $N^0_{\ell}(x)$ transforms under the Lorentz transformation  $\Lambda  $ as
\begin{equation}\label{lgtoffield}
U(\Lambda)^{-1} N^0_{\ell}(x)U(\Lambda)=e^{-im(\Lambda v-v)x}D_{\ell\bar \ell}(W[\Lambda,i\vec\nabla]) N^0_{\bar \ell}(\Lambda^{-1}x).
\end{equation}
In the rest frame with $v^{\mu}=(1,0,0,0)$, the boost transformation of heavy field $ N^0_{\ell}$ under the boost $B(q)$ with infinitesimal parameter $\frac{q^{\mu}}{m}\equiv(0,-\frac{q^i}{m})$ is   
\begin{equation}
    U(B(q))^{-1} N^0_{\ell}(x)U(B(q)) = e^{i\vec q\cdot\vec x} D_{\ell\bar \ell}(W[B(q),i\vec\nabla]) N^0_{\bar \ell}(B(q)^{-1}x), 
\end{equation}
where we use $ e^{-im (B(q)v-v)x}=e^{iqx}=e^{i\vec q\cdot\vec x}$, and the Wigner rotation correspondingly is directly calculated  to the first order of $q/m$ as
\begin{equation}\label{DWbqapp}
    \quad D(W[B(q),i\vec\nabla]) =  1+i \frac{\vec q}{m}\cdot \frac{i\frac{\vec \sigma}{2}\times\vec{\nabla}}{2m+i\partial_t}.
\end{equation}
Due to the above two equation, we write down
\begin{equation}
     U(B(q))^{-1} N^0(x)U(B(q))=\left(1+\frac{\vec q}{m}\cdot \vec K_x\right)N^0(x')=e^{i\vec q\cdot\vec x}\left(1+\frac{\vec q}{m}\cdot\vec K\right)N^0(x'),
\end{equation}
with $x'=B(q)^{-1}x$, then due to Eq.~\eqref{DWbqapp} the boost generator of $N^0_{\ell}(x)$ reads
\begin{equation}\label{eq:kgenapp}
    \vec K_x=\vec K+m\vec x,\quad\text{where}\quad   \vec{K}=\frac{i\frac{\vec \sigma}{2}\times\vec{\nabla}}{2m+i\partial_t}.
\end{equation}

Note that the time translation operator for the relativistic field $\Psi(x)$ is $ \hat P_0$  while for the heavy field $N^0(x)$ it is $\hat k_0= \hat P_0-m$ discussed in subsection~\ref{hilbertspace}. Comparing to  $\delta_x\Psi$, the coordinate variation of heavy field $\delta_xN^0$ has a compensation for $i\partial_t$  
\begin{eqnarray}
    E=i\partial_t\longrightarrow E=i\partial_t+m.
\end{eqnarray}
The interpretation is that the total energy $E=p_0$ for the heavy particle is the small residual kinematic energy $k_0$ plus the mass $m$. And then we have
\begin{equation}
    \delta_xN^0=i\vec\eta\cdot m\vec x~N^0(x)+i\vec\eta\cdot\left(\vec x~i\partial_t+t~i\vec\nabla\right)N^0(x).
\end{equation}
Therefore, the complete boost generator for the heavy field $N^0(x)$ is
\begin{equation}
    \vec K+m\vec x+\underbrace{\left(\vec x~i\partial_t+t~i\vec\nabla\right)}_{\text{universal part}},
\end{equation}
as in Eq.~(\ref{eq:kgenapp}). Here $\vec K$ is the spin part. The universal part $\vec x~i\partial_t+t~i\vec\nabla$ works for the coordinates of both heavy and relativistic field, and it is suppressed for simplicity. We focus only on the generator $\vec K$ as well as $\vec K_x$
\begin{equation}\label{kxboost}
    \vec K_x=\vec K+m\vec x=\frac{i\frac{\vec \sigma}{2}\times\vec{\nabla}}{2m+i\partial_t}+m\vec x.
\end{equation}
Here we divide the boost generator into two parts: the spin part $\Vec{K}$ and the phase part $m\Vec{x}$.  It should be noted that the phase part dominates when $m\rightarrow \infty$, then $\Vec{K}_x$ becomes the pure non-relativistic (Galilean) boost generator $m\Vec{x}$ in the traditional quantum mechanics. The phase $\exp(im\Vec{x}\cdot \Vec{v})$ shifts the momentum by a fixed value $m\Vec{v}$, which acts as the boost.

The nonlinear boost transformation defines the boost generator, as follow 
\begin{equation}   U(B(q))^{-1}N(x)U(B(q))=e^{i\Vec{q}\cdot\vec{x}}\left(1+i\frac{\Vec{q}}{m}\cdot\Vec{\mathcal{K}}\right)N(B(q)^{-1}x)=\left(1+i\frac{\Vec{q}}{m}\cdot\Vec{\mathcal{K}}_x\right)N(B(q)^{-1}x),
\end{equation}
where the boost generator of the spin part is identified as
\begin{equation}
    \vec{ \mathcal{K}}=-i\frac{\vec\sigma}{2}f(\frac{D_{\mu}}{m}),
\end{equation}
while the boost generator including the phase factor is 
\begin{equation}
     \Vec{\mathcal{K}}_x=-i\frac{\vec{\sigma}}{2}f(\frac{D_{\mu}}{m})+m\Vec{x}.
\end{equation}
We find that, the boost generator $\vec {\mathcal{K}}_x$ satisfying the consistency condition (\ref{eq:cccondition}) also satisfies the commutation relation of the Lorentz boost as well, 
\begin{equation}\label{eq:comuk}
    \left[\left(1+i\frac{\Vec{q}_2}{m}\cdot\Vec{\mathcal{K}}_x\right),\ \left(1+i\frac{\Vec{q}_1}{m}\cdot\Vec{\mathcal{K}}_x\right)\right]N(x)=i\frac{\Vec{q}_2\times\Vec{q}_1}{m^2}\cdot\vec J~N(x),
\end{equation}
where $\vec J=\frac{\Vec{\sigma}}{2}$ on the RHS is the rotation generator of NR fields and $\frac{\Vec{q}_1}{m}$, $\frac{\Vec{q}_2}{m}$ are the infinitesimal parameters. Given that the consistency condition is 
\begin{eqnarray}
 &&f(\frac{\Lambda_{1\mu}^{\nu} D_{\nu}}{m})e^{i\vec q_1\cdot\vec x}\left(1+\frac{\vec\sigma\cdot\vec q_1}{2m}f(\frac{D_{\mu}}{m})\right)N=e^{i\vec q_1\cdot\vec x}\left(\frac{\vec \sigma\cdot\vec q_1}{2m}+f(\frac{D_{\mu}}{m})\right)N
 \\
&\Rightarrow&\frac{\Vec{\sigma}\cdot\Vec{q}_1}{2m}N = \left(-f(\frac{D_{\mu}}{m})+e^{-i\vec{q}_1\cdot\vec{x}}f(\frac{\Lambda_{1\mu}^{\nu} D_{\nu}}{m})e^{i\vec{q}_1\cdot\vec{x}}+e^{-i\vec{q}_1\cdot\vec{x}}f(\frac{\Lambda_{1\mu}^{\nu} D_{\nu}}{m})e^{i\vec{q}_1\cdot\vec{x}}\frac{\vec{\sigma}\cdot\vec{q}_1}{2m}f(\frac{D_{\mu}}{m})\right)\nonumber N, 
\end{eqnarray}
and the successive actions give
\begin{eqnarray}
&&\frac{\Vec{\sigma}\cdot\Vec{q}_2}{2m} \frac{\Vec{\sigma}\cdot\Vec{q}_1}{2m}N \\
&=& \left(-\frac{\Vec{\sigma}\cdot\Vec{q}_2}{2m}f(\frac{D_{\mu}}{m})+\frac{\Vec{\sigma}\cdot\Vec{q}_2}{2m}e^{-i\vec{q}_1\cdot\vec{x}}f(\frac{\Lambda_{1\mu}^{\nu} D_{\nu}}{m})e^{i\vec{q}_1\cdot\vec{x}}+\frac{\Vec{\sigma}\cdot\Vec{q}_2}{2m}e^{-i\vec{q}_1\cdot\vec{x}}f(\frac{\Lambda_{1\mu}^{\nu} D_{\nu}}{m})e^{i\vec{q}_1\cdot\vec{x}}\frac{\vec{\sigma}\cdot\vec{q}_1}{2m}f(\frac{D_{\mu}}{m})\right)N \nonumber\\
&=&e^{i\vec q_2\cdot\vec x}e^{i\vec q_1\cdot\vec x} \left(-\frac{\Vec{\sigma}\cdot\Vec{q}_2}{2m}f(\frac{D_{\mu}}{m})+\frac{\Vec{\sigma}\cdot\Vec{q}_2}{2m}e^{-i\vec{q}_1\cdot\vec{x}}f(\frac{\Lambda_{1\mu}^{\nu} D_{\nu}}{m})e^{i\vec{q}_1\cdot\vec{x}}+\frac{\Vec{\sigma}\cdot\Vec{q}_2}{2m}e^{-i\vec{q}_1\cdot\vec{x}}f(\frac{\Lambda_{1\mu}^{\nu} D_{\nu}}{m})e^{i\vec{q}_1\cdot\vec{x}}\frac{\vec{\sigma}\cdot\vec{q}_1}{2m}f(\frac{D_{\mu}}{m})\right)N,\nonumber 
\end{eqnarray}
where in the last line, only the linear order $q/m$ result is kept. Thus, we find that
\begin{eqnarray}
&&\frac{\Vec{\sigma}\cdot\Vec{q}_2}{2m} \frac{\Vec{\sigma}\cdot\Vec{q}_1}{2m}N  \\
&=&e^{i\Vec{q}_2\cdot\Vec{x}}\left(1+\frac{\Vec{\sigma}\cdot\Vec{q}_2}{2m}f(\frac{\Lambda_{1\mu}^{\nu} D_{\nu}}{m})\right)e^{i\vec{q}_1\cdot\vec{x}}\left(1+\frac{\Vec{\sigma}\cdot\Vec{q}_1}{2m}f(\frac{D_{\mu}}{m})\right)N-e^{i\vec q_2\cdot\vec x}e^{i\vec q_1\cdot\vec x}\left(1+\frac{\vec\sigma\cdot(\vec q_1+\vec q_2)}{2m}f(\frac{D_{\mu}}{m})\right)N\nonumber.
\end{eqnarray}
Note that the symmetric part under  $q_1\leftrightarrow q_2$ in the above equation vanishes in the commutator, thus we obtain the commutation relation
\begin{eqnarray}
 i\frac{\Vec{q}_2\times\Vec{q}_1}{m^2}\cdot\frac{\Vec{\sigma}}{2}N(x) &=& \left[\frac{\Vec{\sigma}\cdot\Vec{q}_2}{2m},\frac{\Vec{\sigma}\cdot\Vec{q}_1}{2m}\right]N \nonumber\\
&=& e^{i\Vec{q}_2\cdot\Vec{x}}\left(1+\frac{\Vec{\sigma}\cdot\Vec{q}_2}{2m}f(\frac{\Lambda_{1\mu}^{\nu} D_{\nu}}{m})\right)e^{i\vec{q}_1\cdot\vec{x}}\left(1+\frac{\Vec{\sigma}\cdot\Vec{q}_1}{2m}f(\frac{D_{\mu}}{m})\right)N(x) \nonumber\\
&&-e^{i\Vec{q}_1\cdot\Vec{x}}\left(1+\frac{\Vec{\sigma}\cdot\Vec{q}_1}{2m}f(\frac{\Lambda_{2\mu}^{\nu} D_{\nu}}{m})\right)e^{i\vec{q}_2\cdot\vec{x}}\left(1+\frac{\Vec{\sigma}\cdot\Vec{q}_2}{2m}f(\frac{D_{\mu}}{m})\right)N(x) \nonumber\\
&=&\left[\left(1+i\frac{\Vec{q}_2}{m}\cdot\Vec{\mathcal{K}}_x\right),\left(1+i\frac{\Vec{q}_1}{m}\cdot\Vec{\mathcal{K}}_x\right)\right]N(x).
\end{eqnarray}

\end{appendix}

\bibliography{ref}

\end{document}